\documentclass[useAMS,usenatbib]{mn2e}

\newcounter{footnew}
\newcommand{\startfoot}{\setcounter{footnew}{0}}

\usepackage{graphicx}
\usepackage{amssymb}
\usepackage{tablefootnote}
\usepackage{pdflscape}
\usepackage{longtable}
\usepackage{journals}
\usepackage{amsmath}
\usepackage{natbib}
\usepackage{color}
\usepackage{supertabular,booktabs}
\usepackage[titletoc]{appendix}

\newcommand{\heinke}{H05}
\newcommand{\edmondsa}{E03a}
\newcommand{\edmondsb}{E03b}
\newcommand{\figbox}[1]{\makebox[0.5\textwidth]{\includegraphics[width=0.5\textwidth]{#1}}} % withoutframe
\newcommand{\twofigbox}[2]{\figbox{#1}\hspace{\stretch{0}}\figbox{#2}\\ \vspace{3pt}}

\begin{document}

\voffset= -1.3cm % ad hoc

\title[CVs in 47\,Tuc]{New Cataclysmic Variables and other Exotic Binaries in the Globular Cluster 47\,Tucanae\thanks{Based on proprietary and archival observations with the NASA/ESA Hubble Space Telescope, obtained at the Space Telescope Science Institute, which is operated by AURA, Inc., under NASA contract NAS $5-26555$.}}
\author[L.E. Rivera-Sandoval et al.]{L. E. Rivera-Sandoval$^{1}$\thanks{E-mail: l.e.riverasandoval@uva.nl}, 
M. van den Berg$^{2,1}$, C. O. Heinke$^{3,4}$, H. N. Cohn$^{5}$,   
\newauthor P. M. Lugger$^{5}$,  J. Anderson$^{6}$,  A. M. Cool$^{7}$, P. D. Edmonds$^{2}$,  R. Wijnands$^{1}$, 
\newauthor N. Ivanova$^{3}$ and  J. E. Grindlay$^{2}$\\
\\
$^{1}$ Anton Pannekoek Institute for Astronomy, University of Amsterdam, Science Park 904, 1098 XH Amsterdam, The Netherlands\\
$^{2}$ Harvard-Smithsonian Center for Astrophysics, 60 Garden Street, Cambridge, MA 02138, USA\\
$^{3}$ Department of Physics, University of Alberta, CCIS 4-183, Edmonton, AB$_{435}$T6G 2E1, Canada\\
$^{4}$ Max Planck Institute for Radio Astronomy, Auf dem Hugel 69, 53121 Bonn, Germany\\
$^{5}$ Department of Astronomy, Indiana University, 727 E. Third St, Bloomington, IN 47405, USA\\
$^{6}$ Space Telescope Science Institute, 3700 San Martin Drive, Baltimore, MD 21218, USA\\
$^{7}$ Department of Physics and Astronomy, San Francisco State University, 1600 Holloway Avenue, San Francisco, CA 94132, USA}

\pagerange{\pageref{firstpage}--\pageref{lastpage}} \pubyear{2016}

\maketitle

\label{firstpage}

\begin{abstract}

We present 22 new (+3 confirmed) cataclysmic variables (CVs) in the non core-collapsed globular cluster 47 Tucanae (47 Tuc). The total number of CVs in the cluster is now 43, the largest sample in any globular cluster so far.  For the identifications we used near-ultraviolet (NUV) and optical images from the \textit{Hubble Space Telescope},
in combination with X-ray results from the \textit{Chandra X-ray Observatory}. 
This allowed us to build the deepest NUV CV luminosity function of the cluster to date.
We found that the CVs in 47 Tuc are more concentrated towards the cluster center than the main sequence
turnoff stars. We compared our results to the CV populations of the core-collapsed globular clusters NGC 6397 and NGC 6752. 
We found that 47 Tuc has fewer bright CVs per unit mass than those two other clusters. That suggests that dynamical interactions in core-collapsed clusters play a major role creating new CVs. In 47 Tuc, the CV population is probably dominated by
primordial and old dynamically formed systems. 
We estimated that the CVs in 47 Tuc have total masses of $\sim1.4$ $M_{\sun}$.
We also found that the X-ray luminosity function of the CVs in the three clusters is bimodal.
Additionally, we discuss a possible double degenerate system and an intriguing/unclassified object.  
Finally, we present four systems that could be millisecond pulsar companions given their X-ray and NUV/optical colors. For one of them we present very strong evidence for being an ablated companion. The other three could be CO- or He-WDs. 

\end{abstract}


\stepcounter{footnote}

\section{Introduction}

Cataclysmic variables (CVs) are binary stars that harbor a white dwarf (WD) that accretes from a low-mass companion. 
These systems have orbital periods between $\sim 75$ mins and 10 h, and the overall distribution of CV orbital periods shows a gap between 2 to 3 h
\citep[see for example][and references therein]{2011-knigge}.

The family of CVs is very heterogeneous. 
However, we can group them in different classes according to four key characteristics. 
1) The way in which accretion occurs.
Depending on the strength of the magnetic field of the WD in the CV, 
the mass accretion can occur through an accretion disk (non-magnetic CVs), 
collimated streams onto the magnetic poles (named polar CVs) 
or a combination of both (so called intermediate polar CVs). 
2) Changes in the accretion disk. Occasional changes to a hotter and more viscous state lead to so called dwarf nova (DN) eruptions. 
3) The ocassional presence of thermonuclear burning on the WD surface, also known as nova eruptions. 
4) The presence or lack of hydrogen in the donor. 

Related to the last point are the AM Canum Venaticorum (AM CVn) stars, 
which are CVs that have no, or very little H left, so Balmer lines are not present in their optical spectra \citep[though see][for a possible exception]{2002-israel}. 
These binaries have orbital periods between
5-65 mins \citep{carter}. These short periods 
indicate that the companion is degenerate \citep[or at least semi degenerate, which causes the period to be shorter than $\sim 75$ min,][]{1986Nelson}.
It is believed that these systems harbor a WD that is accreting mass from an He rich companion,
since spectra of these systems are He dominated \citep[see][and references therein, for a review]{2005-nelemans}.

CVs or likely CVs have been found in different environments, such as the Galactic center, Solar neighborhood, open and globular clusters (GCs).
AM CVn systems have not been identified beyond any doubt in GCs yet \citep[though see ][]{2016Zurek}.
However, the study of CVs in star cluster has some advantages with respect to the study of CVs in other environments. 
Among those advantages are that  
GCs provide samples of these binaries at known distance, metallicity and age. 
Furthermore, the study of CVs in globulars is important, on one hand, for understanding how this population affects the evolution of the clusters 
(for example close binaries can help to prevent cluster core-collapse), and on the other hand, to know how
the stellar interactions in a dense stellar environment, like a cluster, affect the CV evolution.

Numerical studies on the evolution of CVs in GCs have been carried out by, for example
\citet[][2017b]{1988-ver, 1994-stefano, 1997-davies, 2006-shara, hong, 2017-belloni}.
For a GC with about 10$^6$ stars initially, \citet{2006-ivanova} found that after 10 Gyr of evolution, the number of CVs would be more than 200. 
This number is large compared to the number of CVs identified so far in any GC (few tens in massive clusters and only few in the less massive ones). 
However, because of the stellar crowdedness, intrinsic faintness and large distances (compared to the Solar neighborhood)
their identification is not trivial. 
So far, the best way to confirm CVs is to take spectra, which has proven to be a fruitful technique \citep[e.g.][]{1995-grin, 2003-knigge, knigge08}. 
Unfortunately, that is not always possible due to the faintness of these binaries and the stellar crowding in the cores of GCs (where most CVs reside),
even with the use of \textit{Hubble Space Telescope}  {\em (HST)}.
Therefore, the best way to maximize the identification of possible CVs in GCs is by combining different techniques like optical variability, 
blue color (light from accretion is blue compared to light from normal stars), 
H$\alpha$ excess, and X-ray emission \citep[see for example][for a recent review about the production of X-rays in CVs]{2017Mukai}.
X-ray emission is actually a very good tracer of interacting binaries, and thanks to the good spatial resolution of \textit{Chandra}
it is frequently used to identify possible CVs in GCs (see below).
Indeed, most of the CVs in clusters have been identified by their X-ray emission. 
Thus, the available sample of CVs in globulars is X-ray biased compared to the field population \citep[most of which is optically biased, see][]{2013-Preto}.  

CVs are considered low luminosity X-ray sources ($L_X  = 10^{29} -10^{34}$ erg s$^{-1}$). However, there are other accreting binaries that emit in the same range, 
which increases the risk of misclassification or ambiguity. Examples in 47 Tuc are the cases of the X-ray sources W42  (an accreting black hole candidate, \citeauthor{2015-x9} \citeyear{2015-x9})  and W82 (a millisecond pulsar -MSP-, \citeauthor{2006-bogda} \citeyear{2006-bogda}) which were initially classified as CVs by \citet{1992-pare} and 
\citet[][hereafter \edmondsa]{2003-edmonds1}, respectively. 
Although, we note that \citeauthor{2003-edmonds2} \citeyear{2003-edmonds2} (from now on \edmondsb) also suggested that W82 could be an MSP. 
To mitigate the risk of misclassification, the combination of photometric information at different wavelengths is important.

Clusters where multiple identifications of CVs in GCs have been made as counterparts to X-ray sources include 47 Tuc  \citep[\edmondsa; \edmondsb;][]{2014-beccari},
NGC 6752 \citep{2002-pooley, 2012-thom}, NGC 6397  \citep{2001-grinb, cohn}, M30 \citep{2007-lugger}, NGC 2808 \citep{2008-servi}, M71 \citep{2010-huang}, 
M80 \citep{2010-die}, M92 \citep{2011-lu}, NGC 6388 \citep{2012-max},  and $\omega$ Cen \citep{cool}.
These studies have shown the effectiveness of this technique and have revealed different characteristics of CVs in GCs. 
Nonetheless, in order to have a global idea about the properties and behavior of the CVs
in clusters, a comparison among their CV populations is necessary.

Of particular interest is to know how the dynamical interactions in dense environments affect the creation and evolution of CVs.
In this context, a comparison between CVs in core-collapsed and non core-collapsed clusters would help to answer that question. 
Core-collapse is the process that occurs when the core of the cluster has negative heat capacity, getting hotter as it loses energy 
to the outskirts. 
This means that there is a net loss of energy from the core when the dynamical interactions of stars push less massive stars in the cluster from the center to the outskirts. To counteract that effect, the stars in the core lose energy and fall deeper down the potential well, occupying a smaller volume (``collapsing'').
This leads to a dramatic increase in the stellar density of the core.
In terms of observations, a cluster is defined as core-collapsed if its surface brightness profile exhibits a power-law until the limit of observational resolution.
On the other hand, the non core-collapsed Galactic clusters show a clear profile flattening in the central few arcseconds to arcminutes and are well fitted with a King model \citep{king}.

\citet{cohn} have found that in the core-collapsed cluster NGC 6397, 
the population of CVs is divided in two groups, with one group of CVs brighter than the other one, and more centrally concentrated. 
This raises the question whether CVs in non core-collapsed clusters are also divided in two groups or not, 
or whether the radial and brightness distribution of CVs as seen in NGC 6397 is a signature that is typical for dynamically active clusters.
 
47 Tuc is a well known Galactic non-core collapsed cluster that is relatively close ($\sim4.7$ kpc), shows low extinction, is massive and has a high interaction rate (Table \ref{table:clusters}). 
It is also the GC with the most X-ray sources identified so far (300 within the half-mass radius, r$_h$),
reaching a limiting luminosity in the 0.5-6 keV band of $3\times 10^{29}$ ergs s$^{-1}$  \citep[][referred to from now on as H05]{heinke}. 
All these features make 47 Tuc a very good target to study close binaries, increasing the possibilities to identify many CVs. 
Using the high resolution of the Wide Field Camera 3 (WFC3) and the Advanced Camera for Surveys (ACS)
on board the {\em HST}, complemented with results from
\textit{Chandra}, we have identified new CVs in the cluster. This includes CVs below $\sim 10^{30}$ erg s$^{-1}$.
This is the first deep H$\alpha$ study of 47 Tuc with {\em HST} that uses near-simultaneous continuum-band photometry.
We designed an innovative WFC3 program to observe a large field of view of a globular cluster 
core to a high depth in the near-UV for the first time, using the broadest possible UV filters to increase our sensitivity.
In this paper, we identify the largest number of CVs ever identified in a GC through X-ray counterpart identification, 
more than doubling the number of known CVs in 47 Tuc.

\begin{table*}
 \centering
 \caption{Parameters of the globular clusters discussed in this paper that have several cataclysmic variables identified. }
\label{table:clusters}
  \begin{tabular}{l c c c c c c c c c}
\hline
Cluster                                            &  Mass$^a$                            & Distance$^b$                                                     & Age$^c$        &      r$_h^d$                    & [Fe/H]$^c$ & E(B-V)$^e$ & $\rho_0 ^e$     &   $\Gamma^f$     & Core  \\ 
                                                        &($\times10^5$ $M_{\odot}$)   &  (kpc)                                                           &(Gy)  &  ($\arcmin$)              &                    &              &   ($L_{\sun}$/ pc$^3$) & (total)     &collapsed?$^g$\\
\hline 
47 Tuc                                            & 6.45$\pm$ 0.4                        & 4.69 $\pm$ 0.04 $\pm$ 0.13                 & 11.75 $\pm$ 0.25    & 2.79 &   $ -0.76$  &   $0.04$   & 4.88 & 1000 & No\\  
NGC 6397                                      & 1.1 $\pm$    0.1             & 2.2$^{+0.5}_{-0.7}$                                 & 13.00 $\pm$ 0.25    & 2.33 & $ -1.99$  &   $ 0.18$    &5.76 &84.1& Yes\\
NGC 6752                                      & 2.82  $\pm$ 0.4                      & 4.0 $\pm$0.2                                           & 12.50 $\pm$ 0.25    & 1.9 & $-1.55$  &   $0.04$    & 5.04  & 401 & Yes\\

\hline
\end{tabular}
{\flushleft
\startfoot  
\begin{flushleft}
$^a$ Taken from \citet{kimming} for 47 Tuc and from \citet{Heyl} for NGC 6397.\\
$^b$ Taken from \citet{2012-woodley} for 47 Tuc, \citet{Heyl} for NGC 6397, and \citet[][2010 revision]{Harris} for NGC 6752.\\
$^c$ Taken from \citet{VandenBerg}.\\
$^d$ Taken from \heinke \ for 47 Tuc, \citet{cohn} for NGC 6397 and from \citet{2017Lugger} for NGC 6752.\\
$^e$ Taken from \citet[][2010 revision]{Harris}.\\
$^f $ Taken from \citet{2013-bara}.\\
$^g$ Taken from \citet{trager}.\\
\end{flushleft}
}
\end{table*}

\section{OBSERVATIONS}
\label{observations}

In this paper we have used the NUV data set from {\em HST} program GO $12950$ taken with the UVIS channel of the WFC3 on 2013 August 13.
We used a $4\times4$ dithering pattern that includes fractional-pixel offsets.
We obtained 24 images in the filter F300X (U$_{300}$) and 8 images in the filter F390W (B$_{390}$). The total exposure times are
3881 s in  B$_{390}$ and 14\,256 s in U$_{300}$.
The original scale of the WFC3/UVIS images is $0\farcs04$ pixel$^{-1}$. 
We improved that resolution for the DAOPHOT and KS2 analyses (see below) 
by drizzling the images to obtain master frames with a pixel resolution of $0\farcs02$.
All the objects studied here are inside the r$_h$ of the cluster ($2\farcm79$, \heinke).

The optical images used in this work were taken with the Wide Field Channel (WFC) on the ACS under program
GO\,9281 in 2002 September 30, 2002 October 2/3 and 2002 October 11. 
We also used a dither pattern that includes fractional-pixel offsets.
The optical data set consists of 10 images in the filter F435W (B$_{435}$),
20 images in the filter F658N (H$\alpha$), and 22 images in the filter F625W (R$_{625}$). 
The total exposure times are
955 s in B$_{435}$, 1320 s in R$_{625}$, and 7440 s in H$\alpha$.
The native WFC pixel scale is 0$\farcs$05 pixel$^{-1}$, which was improved to 0$\farcs$025 pixel$^{-1}$ after drizzling. 
For more details about the NUV and the optical data sets see \citet{rivera}. 

In this paper we looked for CVs by identifying and classifying counterparts to \textit{Chandra} X-ray sources in the cluster. 
We have used the catalog of X-ray sources in 47 Tuc reported by \heinke. 
That catalog was built using data from \textit{Chandra} in 2002 with the ACIS-S instrument at the focus (281 ks exposure). 
These X-ray observations were scheduled to be simultaneous with the optical data used in this paper. 
A study of correlated X-ray and optical variability will be presented in a separate paper.

\section{DATA REDUCTION}
\label{reduction}

The image reduction and astrometric alignment of the NUV and optical images were done as described in \citet{rivera}. 
The photometric analysis of the NUV and optical images was also done as described in that paper.
Additional to the DAOPHOT and KS2 analysis of the NUV images described there, we also have carried out
a photometric analysis with the DOLPHOT software \citep{dolphin}, which worked on the individual U$_{300}$ exposures. 
We used as a reference frame an image with extension .DRC, 
which is a stacked frame with the native scale of the individual images ($0\farcs04$ pixel$^{-1}$).
It has been corrected for charge transfer efficiency degradation of the UVIS detectors,
and it has also been corrected for geometric distortion. 
We aligned each individual image to the reference frame using the task WFC3FITDISTORT.  
Once the offsets in the X-Y directions were determined, as well as the rotation with respect to the reference frame, 
we ran the DOLPHOT software on the 24 (U$_{300}$) $+$ 8 (B$_{390}$) individual exposures. 
We obtained Vega-mag calibrated and aperture corrected photometric results for all the U$_{300}$ and B$_{390}$ images.  
This photometry was used for the variability analysis of the binaries described in this paper.
The photometric magnitudes obtained with DOLPHOT are in general agreement with the ones obtained with
DAOPHOT and KS2.   

\begin{figure*}
\centering
\vspace{-2cm}
\includegraphics[width=15cm]{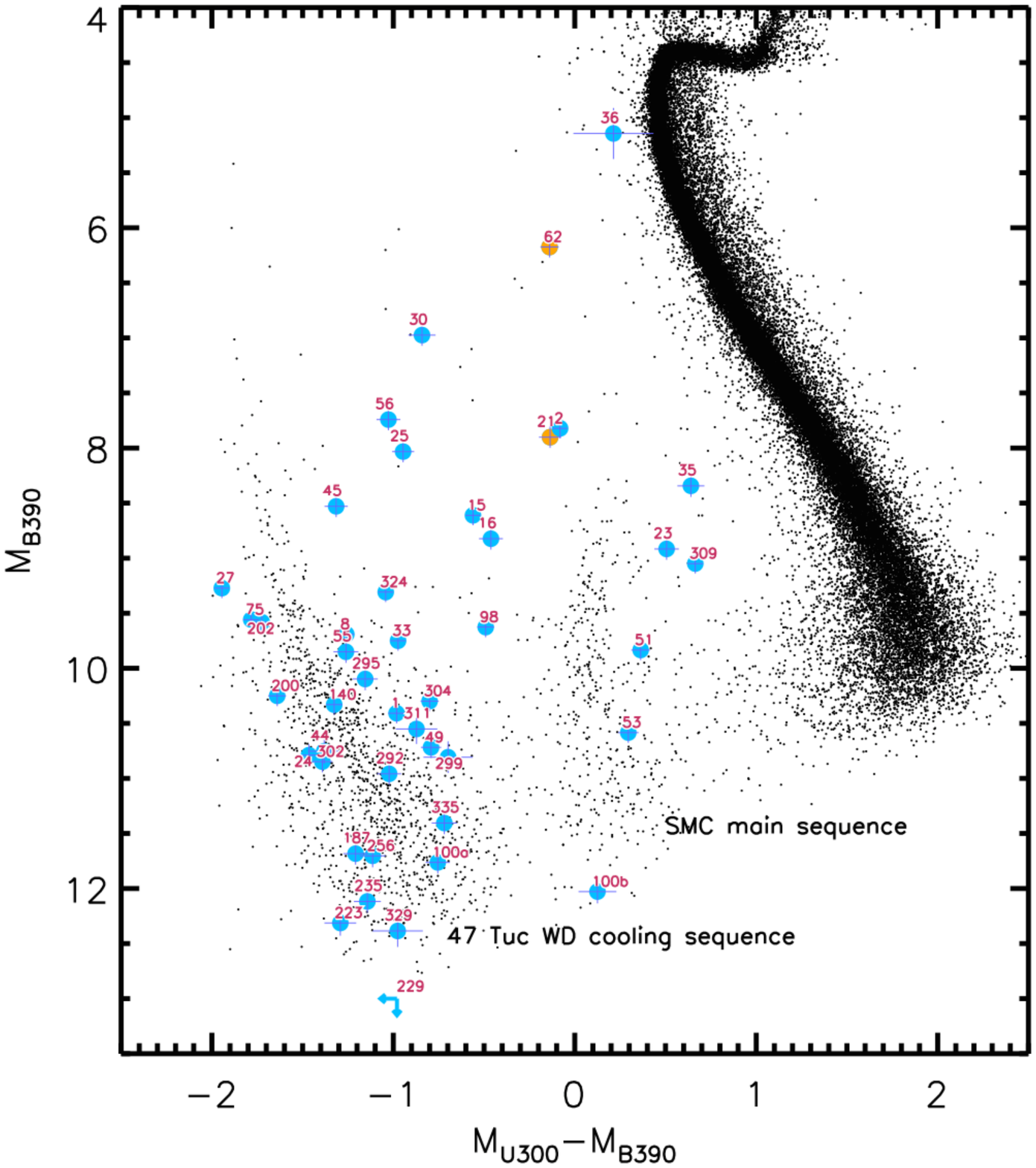}
\vspace{-2cm}\\
\caption{ U$_{300}$ -- B$_{390}$ versus B$_{390}$  CMD based on DAOPHOT photometry of our NUV images. 
Each CV and CV candidate is represented with a blue circle and its respective W number. The
AM CVn candidate W62$_{UV}$ and the unclassified object W21$_{UV}$ are marked with an orange circle.
Apparent magnitudes were converted to absolute magnitudes using a distance to 47 Tuc of
4.69 kpc and reddening E(B-V) = 0.04. Mean magnitudes are given for the variable objects W2$_{UV}$, W15$_{UV}$, W23$_{UV}$  and W27$_{UV}$.
The CV candidate W229$_{UV}$ was just detected in the U$_{300}$ band and thus is plotted in this CMD using an upper limit in B$_{390}$. }
\label{fig:NUV} 
\end{figure*}

\begin{figure*}
\centering
\includegraphics[width=18cm]{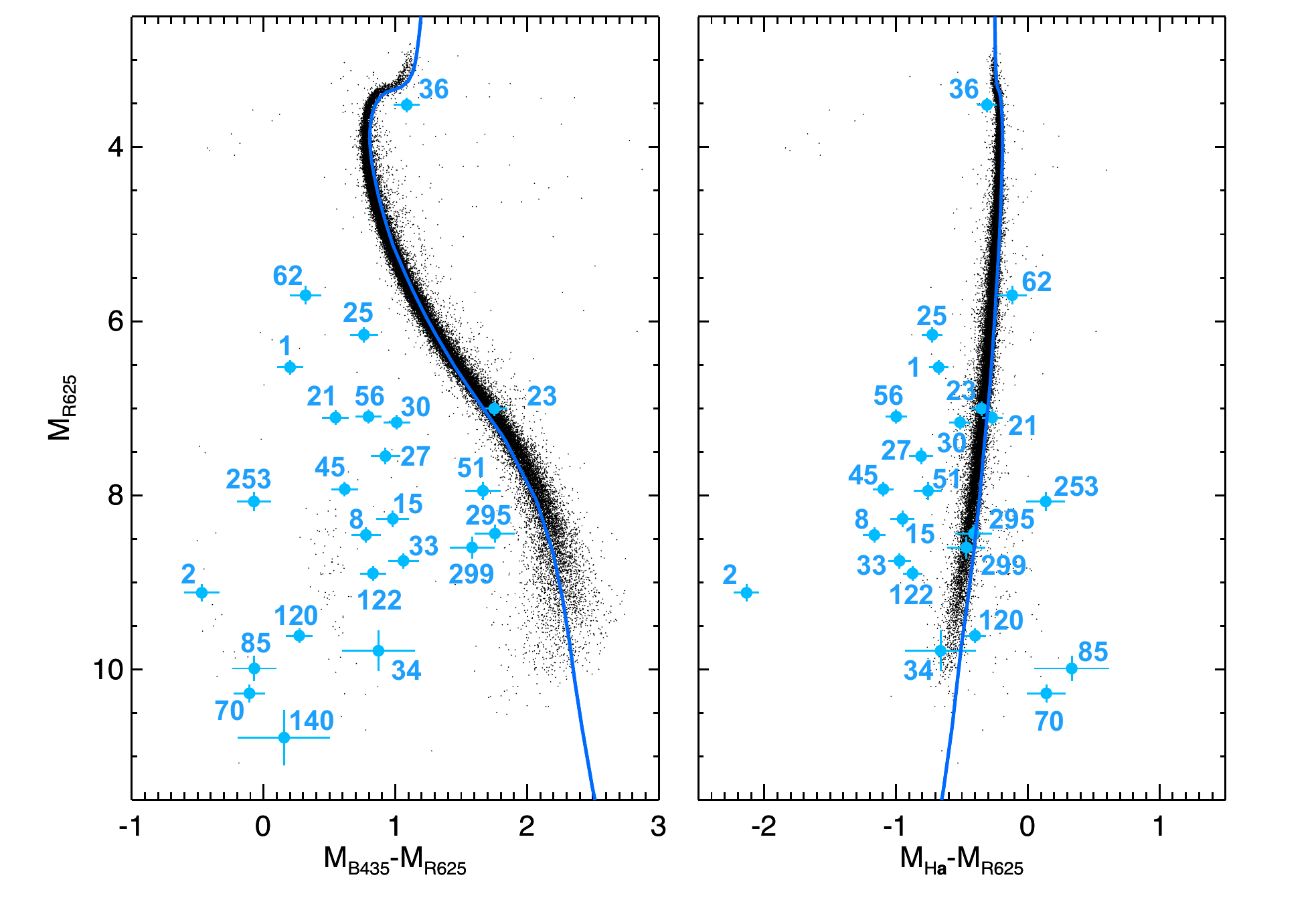}
\vspace{-0.5cm}\\
\caption{ B$_{435}$ -- R$_{625}$ versus R$_{625}$ (left) and H$\alpha$ -- R$_{625}$ versus R$_{625}$ (right) CMDs. Counterparts discussed in this paper are plotted with blue circles. Calibrated Vega-mag magnitudes were converted to dereddened absolute magnitudes using a distance to 47 Tuc of 4.69 kpc 
and reddening E(B --V) = 0.04. The mean magnitude is given for W2$_{UV}$. 
Overplotted are isochrones for an age of 11 Gyr and Z$=0.004$ for the corresponding {\em HST} filters \citep{2012-bressan}. }
\label{fig:BR} 
\end{figure*}

The absolute magnitudes obtained with the three software packages mentioned above were computed using a distance of $4.69\pm0.04\pm0.13$ kpc \citep{2012-woodley}. 
The two terms correspond to the random and the systematic errors, respectively. 
We also have used the reddening value towards 47 Tuc, viz. E(B--V) = 0.04 $\pm$ 0.02 \citep{2007-salaris} together with the UVIS Exposure Time Calculator
of  {\em HST} to obtain the extinction values in the corresponding NUV filters. We obtained A(U$_{300}$)=0.26 $\pm$ 0.13 and A(B$_{390}$)=0.18 $\pm$ 0.09. 
For the optical filters we used the conversions given in \citet{siriea05} which result in A(B$_{435}$)=0.16 $\pm$ 0.08, A(H$\alpha$)=0.10 $\pm$ 0.05, and A(R$_{625}$)=0.11 $\pm$ 0.05.

\section{RESULTS}
\label{results}

Out of the 300 X-ray sources found by \heinke \ within the r$_h$ of the cluster, 
238 are in the field of view (FOV) of the NUV images and 244 are in the FOV of the optical images. 

As previously mentioned, we have identified potential CVs in 47 Tuc by looking for counterparts to \textit{Chandra} X-ray sources in our {\em HST} NUV/optical images. 
We found an NUV/X-ray boresight offset of --0\farcs023$\pm$0\farcs008 in right ascension and 
--0\farcs003$\pm$0\farcs009 in declination in both filters \citep[see][for more information about the NUV/X-ray alignment]{rivera}.
We searched within $3\sigma$ from the X-ray positions. 
The size of each of these $3\sigma$ error circles takes into account the {\em Chandra} positional error of the corresponding source (see \heinke), 
the boresight error (see above), and the {\em HST} astrometric errors \citep[see][]{rivera}.
For each of the {\em HST} sources within the error circle, we examined their NUV and optical colors. 
A source is considered blue if it is located off the left side of the main sequence (MS) of the cluster in the 
NUV or in the B$_{435}$--R$_{625}$ versus R$_{625}$ color magnitude diagrams (CMDs; Figures \ref{fig:NUV} and \ref{fig:BR}).

Based on our photometry in the filters B$_{435}$, H$\alpha$ and R$_{625}$ we have constructed a color-color diagram (CCD; Figure \ref{fig:color-color}).
We consider a star to have H${\alpha}$ excess emission if it lies above the MS in that CCD (which most of the times coincides with the source being 
off the left side of the MS in the H$\alpha$--R$_{625}$ versus R$_{625}$ CMD). 

If a star inside the $3\sigma$ error circle has a blue color, shows an H$\alpha$ excess and/or is optically variable (variability will be discussed below), 
then it is classified as a CV. If a star is blue but it was not detected in the H$\alpha$ images and/or it does not show variability, 
that star is classified as a CV candidate. 

We have applied the same criteria to the previously suggested CVs by other authors 
(see for example \citeauthor{knigge02-1} \citeyear{knigge02-1}, \edmondsa, \heinke, \citeauthor{2014-beccari} \citeyear{2014-beccari}).
In that case if the star shows H$\alpha$ excess and/or is variable in our images, then it is still considered as a CV. 
On the other hand, if the star is blue and visible in the H$\alpha$ images, but it does not show an H$\alpha$ excess and/or variability, 
then an alternative explanation for this behavior is considered. 

Using our CCD, we estimated the H$\alpha$ equivalent width (EW) of our CVs and CV candidates by comparing their dereddened H$\alpha$--R$_{625}$ color 
to the expected H$\alpha$--R$_{625}$ color for an MS star of the same B$_{435}$--R$_{625}$ color. The latter were computed using the Synphot package\footnote{$www.stsci.edu/institute/software\_hardware/stsdas/synphot$}.
We observe that all the CVs and CV candidates have EW(H$\alpha$) $\lesssim -5$ \AA, which are consistent with EWs found for CVs in the field \citep{2009-vandenberg, 2011-szkody}.

To complement our analysis, we have also studied variability of the objects discussed here as CVs and CV candidates. 
This is an important tool to identify CVs, since variability is expected from these binaries due to flickering, eclipses or ellipsoidal variability.
We have used the root main square (rms) deviation about the U$_{300}$ median magnitude obtained with DOLPHOT as a measure of variability. 
To reduce the possible scatter due to photometric issues in the 24 measurements for each object, we have only taken into account photometric 
measurements labeled by DOLPHOT as good quality measurements (flag equal zero).
The possible influence of outliers\footnote{Measurements that result in magnitudes smaller or larger than $m \pm \ 3\sigma$, where m is the median magnitude and $\sigma$ is the standard deviation.} 
has been reduced by using an iterative 3$\sigma$ clipping algorithm (which removes most of the measurements affected by noise, cosmic rays, hot pixels, etc.). 
To quantify the level of variability for each CV and CV candidate, we have used the percentile technique. 
Thus we have drawn lines along the well defined sequence of non-variables (or very-small-amplitude variables) in this diagram. 
They represent the levels above which $95\%$ and $99\%$ of the stars at a given magnitude are found. 
If an optical source is on or above the $95\%$ confidence level line, then we consider it variable.  

It is possible that periodic behavior is not observed (even when present) for many objects because of the short time coverage (few hours) of our observations.
However, we detected an eclipse of the known CV AKO 9 \citep[][see Section \ref{sec:CVs47}]{1989ortolani}.
In order to investigate the presence of outbursts or eclipses we have also constructed light curves in the filter U$_{300}$. 
We thus detected four CVs that exhibit large-amplitude variability at the time of our observations (see Section \ref{sec:outburst}).

Using the methods mentioned above, we classified 43 X-ray sources in the cluster as CVs and CV candidates (with likely non degenerate donors). 
Of all those, 22 are new identifications. 
All of these CVs and CV candidates are X-ray sources with luminosities $L_{X(0.5-6 \ keV)}\sim 5\times10^{29}-2\times10^{32}$ erg s$^{-1}$ (\heinke).
In the NUV CMD, most of these CVs and CV candidates lie on or near the primary WD cooling sequence of the cluster, while
few others fall on the Small Magellanic Cloud\footnote{The outskirts of the SMC are behind 47 Tuc, leading to some contamination 
of the 47 Tuc CMD by SMC stars around U$_{300}$--B$_{390} \sim 0.1$ and M$_{B390}\gtrsim8.5$
in Figure \ref{fig:NUV} and 
B$_{435}$--R$_{625} \sim 0.5$ and M$_{R625}\gtrsim8$
in Figure \ref{fig:BR}.}(SMC) MS. For the objects that lie on the SMC MS, we think that it is unlikely that they actually belong to that galaxy
because only a small fraction of SMC stars fall inside our field; 
the chance of finding superpositions with X-ray sources with luminosities of the order  $10^{32}-10^{34}$ erg s$^{-1}$ (at the distance of the SMC, $\sim 60$ kpc),
is pretty low. Also, these X-ray luminosities would be very high for being isolated MS or active binary (AB; $L_{X}<10^{31}$ erg s$^{-1}$) stars (for instance, W51 would have an X-ray luminosity of $\sim10^{34}$ erg s$^{-1}$). 
We note that the primary source of spurious identifications to our sample would be single WDs that happen to land in our error circles. 
These WDs would lie on the WD sequence, but will not show H$\alpha$ excess or variability.

The apparent magnitudes of the 43 CVs and CV candidates, together with their astrometric results are given in Table \ref{table:results}.
Finding charts for some of them are shown in Appendix \ref{finding_charts}. 
In Section \ref{AM} we discuss two objects with optical (specifically H$\alpha$) properties that distinguish them from the CVs and CV candidates discussed in 
Section \ref{sec:CVs47}, viz. a candidate AM CVn system and another intriguing, but unclassified object.
in Section \ref{hidden} we present four candidate counterparts to {\em Chandra} sources that show striking similarities to known MSP optical counterparts; 
the difference is that no radio source has been detected at the position of these four sources, yet. Finally in Section \ref{match}
we discuss in detail the probability for chance coincidences between {\em Chandra} sources and optical sources with CV-like properties. 

From now on we use the subscript ``UV'' to refer to the NUV counterpart to the corresponding ``W''  X-ray source from the catalog of \heinke , e.g. W1$_{UV}$
as the counterpart of W1. We also use the abbreviation ``NUV CMD'' for the U$_{300}$-B$_{390}$ versus B$_{390}$ CMD, 
``optical CMD'' for the B$_{435}$-R$_{625}$ versus R$_{625}$ CMD, and
``H$\alpha$ CMD'' for the H$\alpha$-R$_{625}$ versus R$_{625}$ CMD. 

\subsection{Cataclysmic variables in 47 Tuc}
\label{sec:CVs47}

\subsubsection{Previously identified CVs and CV candidates}
\label{previous_CVS}  
 
The nineteen CVs and CV candidates associated with the X-ray sources W1, W2, W8, W15, W21 (see Section \ref{AM}), W25, W27 (V3), W30, W33, W35, W36 (AKO 9), W44,
W45, W49, W51, W53, W56, W120 and W122 were previously suggested by
\citeauthor{1992-pare} (\citeyear{1992-pare}), \citeauthor{1994-Pare} (\citeyear{1994-Pare}), \citeauthor{1996-shara} (\citeyear{1996-shara}), 
 \citeauthor{2001-grindlay} (\citeyear{2001-grindlay}) and \edmondsa \ to be CVs.
Where clearly detected in our data, these CVs are blue in our NUV and optical CMDs, and have an H$\alpha$ excess (Figures \ref{fig:NUV} and \ref{fig:BR}).
Exceptions are W120$_{UV}$ and W122$_{UV}$ which are not in the FOV of our NUV images,
W35$_{UV}$  that is surrounded by bright stars preventing R$_{625}$ and H$\alpha$ photometry, 
as well as W53$_{UV}$ and W44$_{UV}$ in which diffraction spikes affect the optical photometry.  
The object W49$_{UV}$ was detected neither in H$\alpha$ nor R$_{625}$. 
Two additional sources with WD-like colors were found within the $2\sigma$ error circle of W49. 
For one of them optical photometry was derived but due to contamination by a nearby saturated star the photometry is not reliable. 
The other WD-like object was not detected in the optical bands. Among the three possible candidates, only one (W49$_{UV}$) shows variability (Figure \ref{fig:rms}).

\begin{figure}
\centering
\includegraphics[width=9cm]{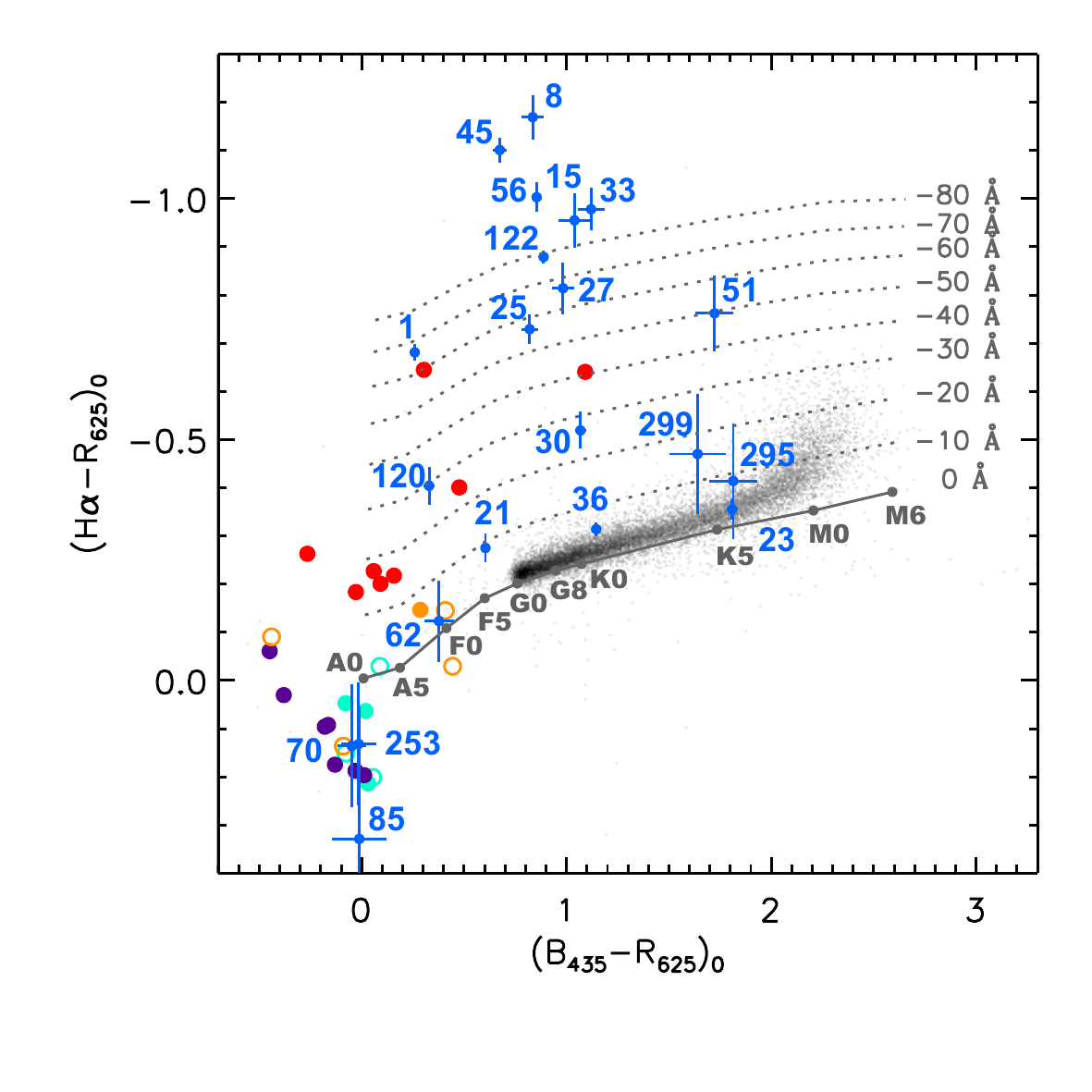}
\vspace{-1.3cm}
\caption{Color-color diagram of 47 Tuc that shows the optical counterparts to \textit{Chandra} X-ray sources in the cluster
discussed in this paper (small blue circles with error bars). 
The photometry has been dereddened using E(B--V) = 0.04. 
Synthetic colors of various spectral types were computed with the Synphot package using low-metallicity (log Z = --2.5) 
Castelli-Kurucz model spectra, and are shown as dark gray circles. 
The solid gray line corresponds to EW(H)=0 \AA \ and the dotted lines show
increments of $\Delta$EW(H$\alpha$)=--10 \AA, which result from adding a gaussian emission line with FWHM of $25$ \AA.
Orange, cyan and red circles are colors obtained from AM CVn, DA WD and CV spectra from the Sloan Digital Sky Survey (SDSS) by \citet{carter}, respectively. 
The purple circles are colors obtained for DA WDs model spectra with Synphot.
For some SDSS systems a small fraction of the B$_{435}$ filter
bandpass is not covered by the spectra, which means that the actual
B$_{435}$--R$_{625}$ colors of the SDSS binaries are slightly bluer than plotted in this figure (open circles).
Filled circles correspond to good B$_{435}$ coverage. 
Other stars in the cluster are denoted with black dots. }
\label{fig:color-color} 
\end{figure}

We also note that W120$_{UV}$ lies to the right of the H$\alpha$ -- R$_{625}$ MS
as some faint CVs in the cluster NGC 6397 \citep{cohn} do. 
However, considering its blue color (Figure \ref{fig:BR} left), the CCD (Figure \ref{fig:color-color})
shows an H$\alpha$ excess that corresponds to a line with EW(H$\alpha$) of --30 \AA, which is consistent with the EW(H$\alpha$) of typical CVs.

The objects W1$_{UV}$, W8$_{UV}$, W27$_{UV}$, W33$_{UV}$, W44$_{UV}$, W45$_{UV}$ and W49$_{UV}$ lie on the WD 
cooling sequence of the NUV CMD, which suggests a large contribution from the WD and/or accretion disk to the NUV spectrum. 
On the other hand, the colors of the CVs W15$_{UV}$, W25$_{UV}$, W30$_{UV}$, W35$_{UV}$, W36$_{UV}$, W51$_{UV}$, W53$_{UV}$
and W56$_{UV}$ suggest a larger contribution from the companion star, 
though it is also possible that some of them (like W25$_{UV}$ and W56$_{UV}$) have emission dominated by hot, relatively low-mass WDs. 
In the case of W36$_{UV}$ (AKO 9), which harbors a subgiant secondary star, the NUV emission is dominated by the (edge-on) accretion disk \citep{2003-knigge},
and it is the brightest NUV object among the CV and CV candidates discussed in this paper. 

Among the stars that show variability at a confidence level equal or higher than $95\%$ in U$_{300}$ (Figure \ref{fig:rms}) are 
W2$_{UV}$, W8$_{UV}$, W15$_{UV}$, W25$_{UV}$, W27$_{UV}$, W30$_{UV}$, W33$_{UV}$, W35$_{UV}$, 
W36$_{UV}$, W45$_{UV}$, W49$_{UV}$, W51$_{UV}$ and W56$_{UV}$. 
Given that the expected number of variable blue objects or variable WDs associated by chance within these error circles is small (Section \ref{match}),
it is likely that they are the real counterparts to the respective X-ray sources.   
The objects W1$_{UV}$, W44$_{UV}$ and W53$_{UV}$, do not show variability in U$_{300}$. 
However, \edmondsb \ (see for instance their Figure 9) found that they display long term variability in the filters F814W and/or F300W, which also suggests that they are the correct X-ray counterparts. 
The objects W120$_{UV}$ and  W122$_{UV}$ are outside of the NUV FOV. But they are also clearly variable in other bands (\edmondsb), 
supporting their classification as CVs. 

For AKO 9 the variability is related to eclipses (a partial light curve of this 1.1 day accreting binary is shown in Figure \ref{fig:AKO9}). 
For the other CVs the nature of variability
appears more stochastic and could be due to flickering.  
The variability of W2$_{UV}$, W15$_{UV}$ and W27$_{UV}$ is discussed in more detail in Section \ref{sec:outburst}.

\subsubsection{Confirmation of previously suggested CVs}

The stars W16$_{UV}$, W55$_{UV}$ and W140$_{UV}$ were previously mentioned as possible CVs (\edmondsa). 
These objects were not firmly identified as CVs due to the stellar crowdedness in the optical images of these authors. 
Our results support the classification of these objects as CVs. 
These three objects have no H$\alpha$ information (there are saturated stars nearby or very close neighbors) but they were clearly identified in the NUV images. 
In the NUV CMD, W16$_{UV}$ lies between the primary WD cooling sequence and the MS of the SMC, while
W55$_{UV}$ and W140$_{UV}$ lie on, or near the primary WD cooling sequence.
The objects W16$_{UV}$ and W140$_{UV}$ display variability at a confidence level higher than $99\%$, which strongly supports their classification as CVs. Variability for W55$_{UV}$ was not detected. 

\subsubsection{Unidentified previously suggested CVs}

We note that for the X-ray sources W10 and W20, previously cataloged as CV 
candidates based on the presence of saturated/bright stars and X-ray properties (\edmondsa, \edmondsb, \heinke), we did not find 
signals of any blue object within the 3$\sigma$ 
error circle in the NUV images that could indicate the presence of an accreting WD. 
We obtained the same results from the analysis at optical wavelengths.
Both W10 and W20 have relatively bright (giant, subgiant or MS) stars in the X-ray error circles, 
which may prevent the identification of true, faint counterparts (\edmondsa), 
such as CVs or MSPs. Alternatively, the bright stars may be the true counterparts, 
in which case they are likely RS CVn-type coronally active binaries.
Apart from W10, W20, W32 and W71 (the latter two reclassified as likely ABs), all the other 
CVs and CV candidates mentioned by \edmondsa, \edmondsb \ and \heinke \ are presented in our paper.

\subsubsection{New CVs and CV candidates}
\label{sec:new_cvs}

In this work we have classified 22 X-ray sources as new CVs and CV candidates (most of which were previously unclassified X-ray sources).
The respective counterparts are W23$_{UV}$, W24$_{UV}$, W75$_{UV}$,  W98$_{UV}$,  W100$_{UV-\text{a or --b}}$, 
W187$_{UV}$,  W200$_{UV}$,  W202$_{UV}$,  W223$_{UV}$,  W229$_{UV}$,  W235$_{UV}$,  
W256$_{UV}$, W292$_{UV}$, W295$_{UV}$ W299$_{UV}$,  W302$_{UV}$,
 W304$_{UV}$,  W309$_{UV}$,  W311$_{UV}$,  W324$_{UV}$,  W329$_{UV}$, and W335$_{UV}$.
All of them show blue colors in the NUV except W229$_{UV}$,
which was only detected in the U$_{300}$ stacked image, and therefore we were not able to investigate its variability in that band.
The non-detection of W229$_{UV}$ in the B$_{390}$ filter indicates it is a blue object.   
Most of the counterparts to these 22 X-ray sources lie on the WD cooling sequence of the NUV CMD, 
indicating the strong contribution from the accreting WD to the NUV light. 
Because of intrinsic faintness, stellar crowding, bright saturated stars nearby and 
diffraction spikes, it was not possible to get optical photometry, including H$\alpha$, for all of the 23 optical stars (counting W100$_{UV-a}$ and W100$_{UV-b}$ separately).
In the filter B$_{435}$ photometry was derived for fourteen objects. 
H$\alpha$ information was obtained for three stars 
(W23$_{UV}$, W295$_{UV}$ and W299$_{UV}$), which supports their identification as CVs. 
They show EW(H$\alpha$)  $\lesssim$--5 \AA \ (up to --30 \AA \ for W299$_{UV}$), 
also their different location in the optical and NUV CMDs with respect to the MS or WD sequence 
show that they are not single-temperature WDs, and thus most likely contain WDs and donor stars. 
Variability at the confidence level of $95\%$ or higher was detected for nine out of the 23 stars
(W23$_{UV}$, W200$_{UV}$, W256$_{UV}$, W292$_{UV}$, W295$_{UV}$ W299$_{UV}$, W304$_{UV}$,  W309$_{UV}$ and W311$_{UV}$). 
Because of that, 
they have only a small chance of being random coincidences compared to the blue stars found in the 3$\sigma$ error circles of the other thirteen sources
that are newly classified as candidate CVs. 
Eclipses or ellipsoidal variations were not found in any of the new 23 (counting W100$_{UV-a}$ and W100$_{UV-b}$ separately) systems.
For the non variable counterparts, it is possible that they are variable in other bands different than U$_{300}$, like the previously discussed cases of 
W1$_{UV}$, W44$_{UV}$ and W53$_{UV}$. 
Another possibility is that those systems have long term variations that were missed because of our short time coverage.
It is also possible that due to the $\sim$1.5 hr observational gap (between both visits due to Earth occultation), 
orbital variations, like brief eclipses, were missed. 
We note that unlike the rest of CVs and CV candidates in the cluster, which display a blue color in the optical CMD, 
the object W23$_{UV}$ lies on the MS of the optical CMD. Using the isochrones computed by \citet{2012-bressan} 
we determined that W23$_{UV}$ could have an optical companion with mass $0.55-0.6 \ M_{\sun}$.
This object is also highly variable in the NUV and thus will be further discussed in Section \ref{sec:outburst}.

\begin{figure*}
\centering
\vspace{-2cm}
\includegraphics[width=13cm]{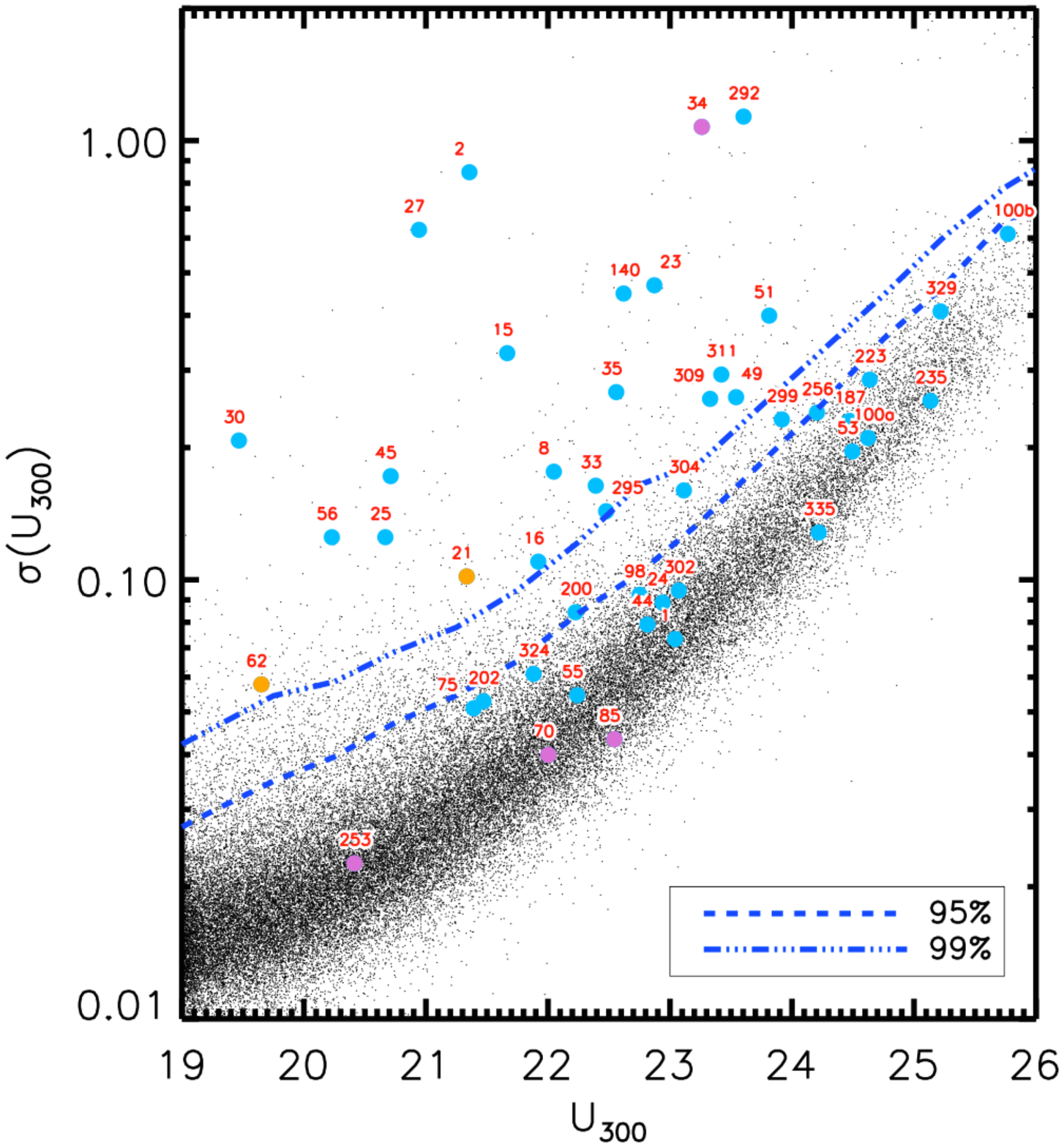}
\vspace{-2.5cm}\\
\caption{Variability diagram in the filter U$_{300}$, showing the rms variability in the (at most 24) U$_{300}$ 
magnitudes extracted by DOLPHOT from the individual exposures, versus the average magnitude. 
Outliers were filtered by using 3$\sigma$ clipping algorithm. 
The CVs and CV candidates are represented with blue circles.
The blue dashed lines represent the $95\%$ and $99\%$ confidence level of variability, as computed by the percentiles in the $\sigma$(U$_{300}$) distribution below which 95\% and 
99\% of the stars are found (percentiles computed in bin widths of 0.5 mag). Other stars in the cluster are represented with black dots. 
The AM CVn candidate W62$_{UV}$ and the unclassified object W21$_{UV}$ show variability at a confidence level higher than $99\%$ and are marked with orange circles. The candidates to MSP companions are market with pink circles. 
The eclipsing CV AKO 9 is not shown because its brightness makes it fall outside the frame. 
W120$_{UV}$ and W122$_{UV}$ are outside of the NUV FOV, thus there are no NUV results for them. The CV candidate W229$_{UV}$
was just detected in the U$_{300}$ stacked image, and so it is not included in this diagram.}
\label{fig:rms} 
\end{figure*}

The X-ray sources W23, W24, W75, W100, W200, W202, W256, W299, W302, W304 and W335 are also 
among the X-ray sources for which we found multiple potential counterparts, and more information about them is given below (Section \ref{sec:multiple}).

\subsubsection{Confused X-ray sources and X-ray sources with multiple potential counterparts}
\label{sec:multiple}

The high stellar crowding in 47 Tuc can lead to multiple potential NUV/optical counterparts within a single {\em Chandra} error circle. This may be due to chance coincidence of one or more non-X-ray-emitting stars in the error circle, and/or due to X-ray sources lying too close to each other to be resolved by  {\em Chandra}. \heinke \ suggest that of order 14\% of X-ray sources in the core brighter than $\sim10^{30}$ erg/s are not detected, due to being ``confused" with similar or brighter X-ray sources. This is consistent with the evidence that 3 of 22 MSPs (47 Tuc F, 47 Tuc G, and 47 Tuc O) are confused with other sources (\heinke), two of which are other MSPs.
Such ``confused" X-ray sources are also more difficult to analyze in the NUV and optical. 
This is because they have larger error circles.
However, most of the NUV/optical sources in the error circles have colors that are compatible with those 
expected from MS stars, which suggests that they are aligned by chance with the X-ray source. 

For twelve X-ray sources we found more than one interesting NUV or optical source inside the X-ray error circle, of which at least one is a CV or CV candidate. 
These are W23, W24, W35, W75, W100, W200, W202, W256, W299, W302, W304 and W335. 
Six of these exhibit NUV variability. 
The sources W24, W35, W75, W202 and W302 are among the list of X-ray ``confused" sources in \heinke \
and are therefore likely source blends. 
For statistical purposes we have just given upper limits for the X-ray luminosities of these CV candidates
(except for W100; Table \ref{table:results}), 
since an accurate and reliable splitting of the luminosities among the number and kind of X-ray sources identified for each of these W sources
is impossible given the available information. Results about the other possible counterparts (that are not CVs) 
for these X-ray sources will be shown in a separate paper. 
We stress that in this work we focus on the results about CVs and CV candidates.

From our NUV analysis we found that all the X-ray sources mentioned above have at least one blue object within the 3$\sigma$ error circle. 
An exception is W75, for which the WD-like object identified lies outside 3$\sigma$. 
A counterpart at 2.5$\sigma$ has been proposed by \edmondsa, but other counterparts outside 3$\sigma$ have also been associated with that X-ray source
(e.g. \citeauthor{knigge02-1} \citeyear{knigge02-1}, \citeauthor{knigge08} \citeyear{knigge08}).

The sources W23, W24, W256, W299 and W304 were previously classified as ABs (\citeauthor{albrow01} \citeyear{albrow01}, \edmondsb, \heinke). 
For all these sources we identified the corresponding ABs based on published finding charts, or on the coordinates of the associated optical variables. 
However, for the first time we also identified a blue object inside the 3$\sigma$ error circles (all but W23$_{UV}$ lying on the WD sequence in the NUV CMD), which 
in the cases of W23, W24 and W304 are closer than the ABs to the center of the error circle.
On the other hand W200, W202, W302 and W335, were until now unclassified X-ray sources.
For all of them (including W35) we found two possible counterparts, a blue object and an AB.
An exception is W200, which apart from the presumable CV (at $2\sigma$ from the center of the X-ray circle) 
there is a blue straggler star (at 0.43\arcsec or 4$\sigma$ from the center of the X-ray error) that could contribute to the X-ray flux.

The source W100 is a peculiar case for which we found two blue objects within the 3$\sigma$ error circle (see Appendix \ref{finding_charts}). 
They lie at 0.25\arcsec or $1.5\sigma$ (W100$_{UV-a}$) and 0.18\arcsec or $1\sigma$ (W100$_{UV-b}$) from the center of the circle. 
Only W100$_{UV-a}$  was detected in the filter B$_{435}$. 
Based on the available information it is impossible to tell which one is the correct counterpart to W100, 
or whether W100 is a blended X-ray source with two CV counterparts (though this would be less likely).

\subsection{CVs with large-amplitude NUV variability}
\label{sec:outburst} 

In this section, we present results for the counterparts to the X-ray
sources W2, W15, W23 and W27, which display large-amplitude
NUV variability during the time of our NUV observations. In Figure \ref{fig:DNe}
we show their DOLPHOT light curves; these are in good agreement with
those obtained with KS2. Information about the NUV and optical colors
of these four objects were discussed in Sections \ref{previous_CVS} (W2, W15,
W27) and \ref{sec:new_cvs} (W23). 
Based on an exquisite {\em HST} WFPC2 data
set of more than 600 $V$ and $I$ exposures obtained over an 8.3 d
period, \edmondsb \  performed a variability analysis of many optical
counterparts to {\em Chandra} sources in 47 Tuc, 
the results of which we compare with more recent 
results and our findings on a case-by-case basis below.

\begin{figure}
\centering
\includegraphics[width=8cm]{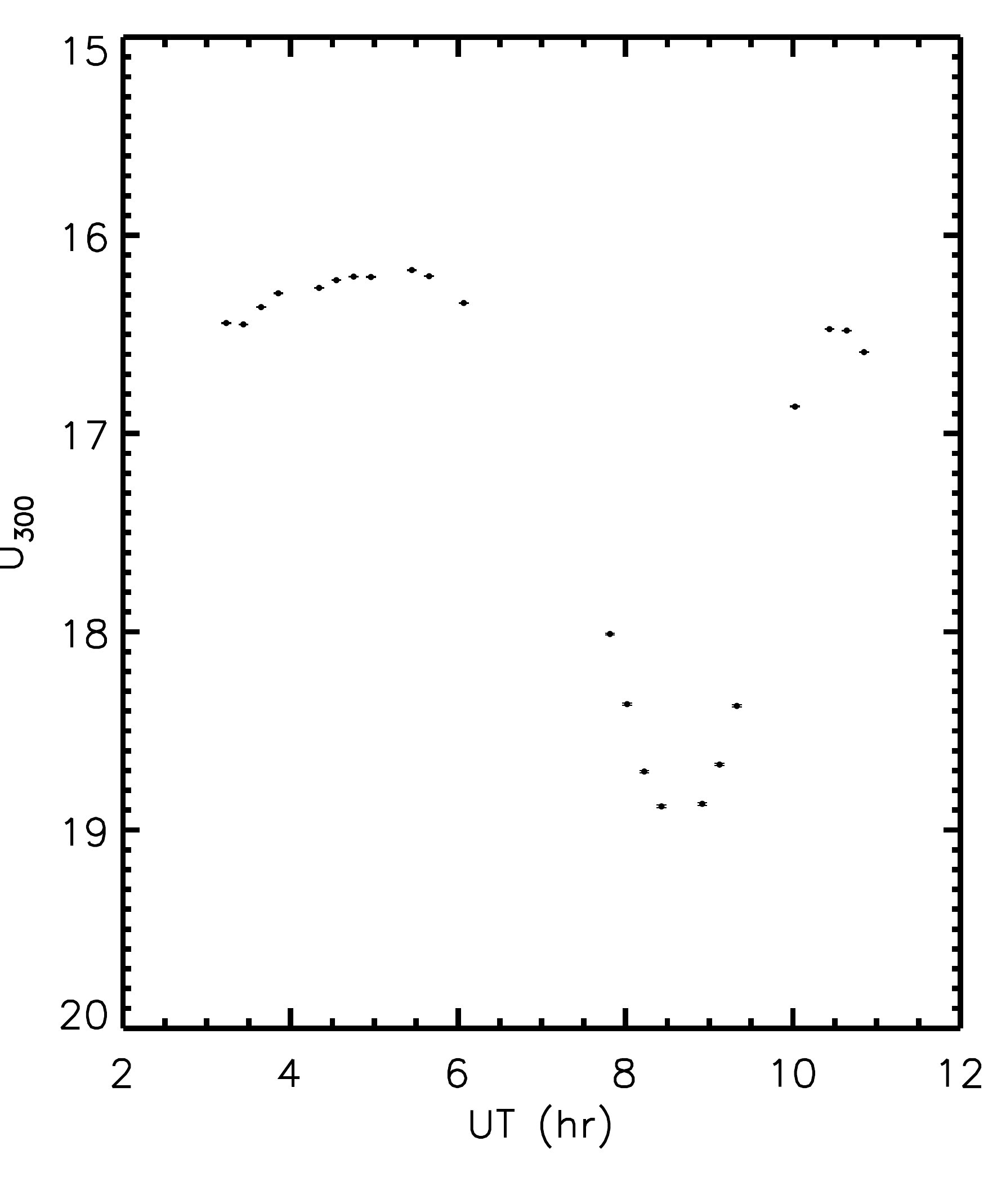}
\caption{Dereddened partial light curve in the filter U$_{300}$ of the of 1.1 day orbital period eclipsing CV AKO 9. 
Only photometric measurements that were labeled by DOLPHOT as good quality measurements are plotted.}
\label{fig:AKO9} 
\end{figure}

For the CV W2$_{UV}$, no indication of periodicity was found when
folding the U$_{300}$ light curve on any of the possible optical
periods (2.2, 5.9 and 8.2 hr) or X-ray period (6.287 hr) suggested by
\edmondsb. However, recently \citet{2016-israel} used a larger {\em
Chandra} data set to do a period search and found that this system
shows eclipses with a period of 2.4025 h. Using that period, we did
not find a periodic pattern in our NUV light curves; the optical data
on the other hand, do show periodic behavior and deep, eclipse-like
features, most notable in the R$_{625}$ filter (see Figure  \ref{fig:W2R} for the
folded R$_{625}$ and U$_{300}$ light curves). Interestingly, the amplitude of the
variations observed by \edmondsb \ in their light curves from 1999 is small
(about 0.5 mag in $V$) compared to our findings. 
W2 is also matched to the miscellaneous blue variable WF2-V48 in \citet{albrow01}, 
who already noted that this star could be a CV. 
Their $V$ light curve, which may be affected by a nearby diffraction spike, 
shows deep sharp dips, resembling the behavior of our R$_{625}$ light curve.
Eclipses that come
and go on a time scale of years could be due to long-term changes in
the system's configuration with respect to the observer, for example
due to precession of the accretion disk, or changes in the thickness
of the disk. Despite the absence of a periodic NUV signal, the
U$_{300}$ light curve shows magnitude variations of $\sim3$ mag
(Figure \ref{fig:DNe}) within the time span of our observations ($\sim8$
hr). These variations cannot be attributed to contamination by
neighboring stars, since W2$_{UV}$ is the only star inside the
3$\sigma$ error circle. Non-periodic, large-amplitude variability has
been observed for highly magnetic CVs,
and is associated with changes between high and low accretion states. State switches can occur on a
time scale of about one or two weeks \citep[see for example HS 0943+1404;][]{2005-rodriguez}
but large amplitude variations in time-scales of tens of minutes have also been observed for magnetic CVs
\citep[like in AE Aqr e.g.][]{2012-zama}; 
if W2 is an intermediate polar (the presence of a disk is required to account for the observed/unobserved eclipses at different epochs), 
we could have caught the system during
a high/low state transition.

W15$_{UV}$ is an eclipsing CV with an optical period of 4.238 hr
(\edmondsb). In the U$_{300}$ we did not find the eclipse for this
object when folding the light curve using that period. 
However, we observe photometric variations of $\sim$2.5 mag in
the same band (see Figures \ref{fig:DNe} and \ref{fig:15fold}). 
This contrasts with the optical light curve presented by
\edmondsb \ which shows ellipsoidal variations, and does not show a large
decay or increase on a time scale of hours like we observe. Visual
inspection of the individual images suggests that the variations
obtained in the NUV are authentic. Like is the case of W2$_{UV}$, this
change in behavior could be due to long-term changes in the system's
configuration; or the component that dominates the NUV light is very
different from the component that dominates the optical (e.g. disk
versus mass donor).

The light curve of W23$_{UV}$ shows variability up to $\sim2$ mag on
a time scale of hours, as well (Figure \ref{fig:DNe}). Examination of the
individual images suggests that the observed variations are real. The
NUV light curve of this object looks like it is declining back to a
quiescent level, perhaps after a strong flare lasting at least several
hours. Alternatively, we could have observed W23 right at the end of
a DN outburst, which are recurrent brightenings of 2--5
mag observed in non-magnetic CVs. DN eruptions have already been
detected in several 47\,Tuc CVs (AKO 9, \citeauthor{1989ortolani}, \citeyear{1989ortolani};
W30$_{UV}$, \citeauthor{1994-Pare}, \citeyear{1994-Pare}; and possibly W51$_{UV}$ and
W56$_{UV}$, \edmondsb, \heinke). Unfortunately the time coverage of the
$W23_{UV}$ light curve is short and no more information can be
obtained.

W27$_{UV}$ has already been noted as a CV displaying large-amplitude
variability \citep{1996-shara} and has been classified as a magnetic
CV (\heinke) based on its periodic X-ray light curve. 
\citet{2001-grindlay} and \edmondsb \ proposed a possible orbital period of 3.83 hr and
∼3.7 hr, respectively for this object (though Hattawi et al. -in
preparation- have found a period of ∼ 6.7 hr). Our U$_{300}$
phase-folded light curve using these periods does not show signs of
periodicity. However, variations of $\sim2.6$ mag in amplitude are
visible in the light curve, which are similar to the ones observed in
W2$_{UV}$.

Flickering variations in the optical light of CVs occur on time scales
of $\sim10$ min, and the corresponding brightness fluctuations around
the mean value are not of the order of a few magnitudes but rather a
few tenths of a magnitude. Flickering is therefore not a plausible
explanation for the observed behavior of the systems discussed above.
In Figure \ref{fig:DNe} we show our U$_{300}$ light curve for W30$_{UV}$, in which
the amplitude variations are likely due to flickering from the hot
spot or accretion disk.

\begin{figure*}
\centering
  \begin{tabular}{@{}ccc@{}}
    \includegraphics[width=.33\textwidth]{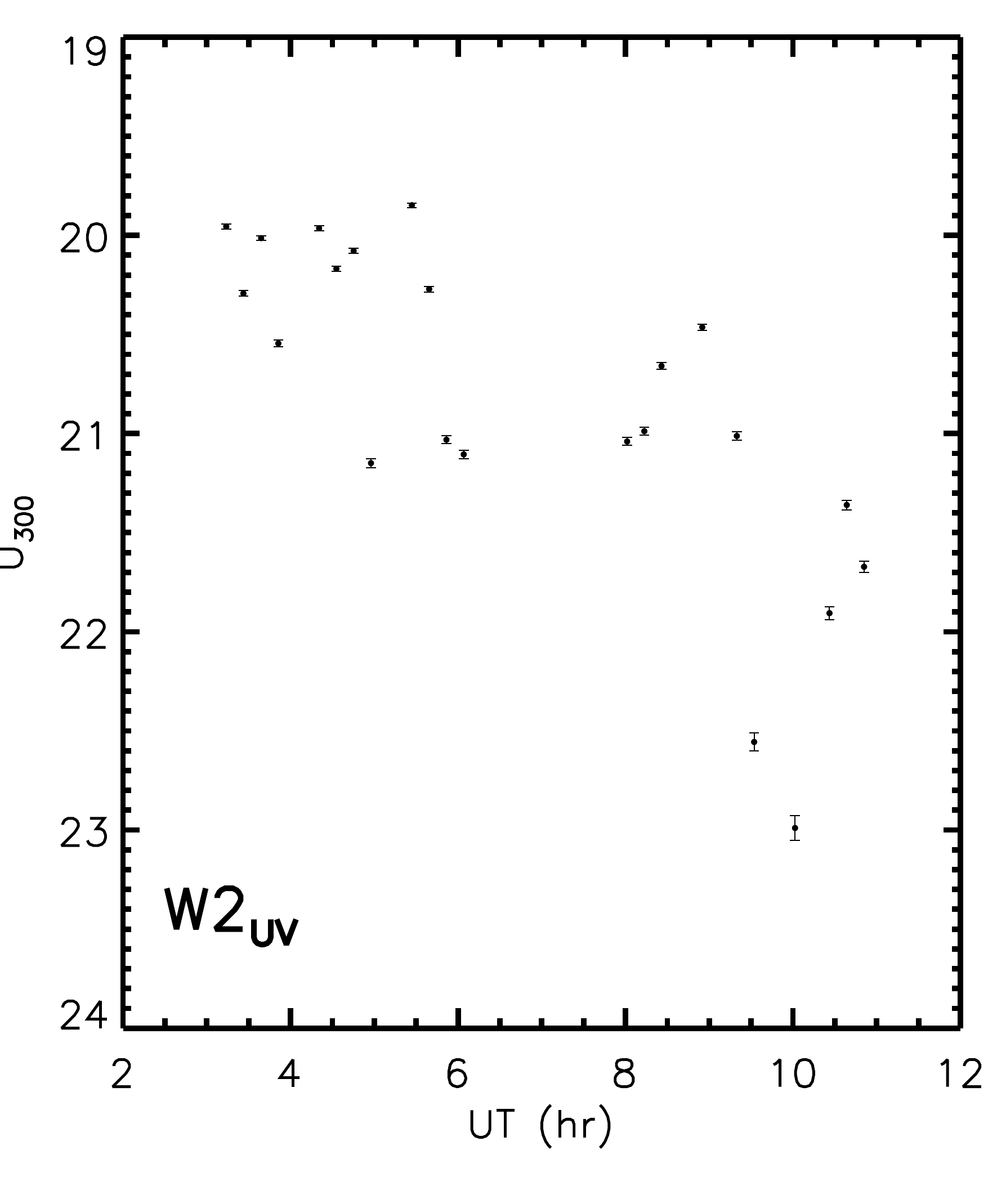} &
    \includegraphics[width=.33\textwidth]{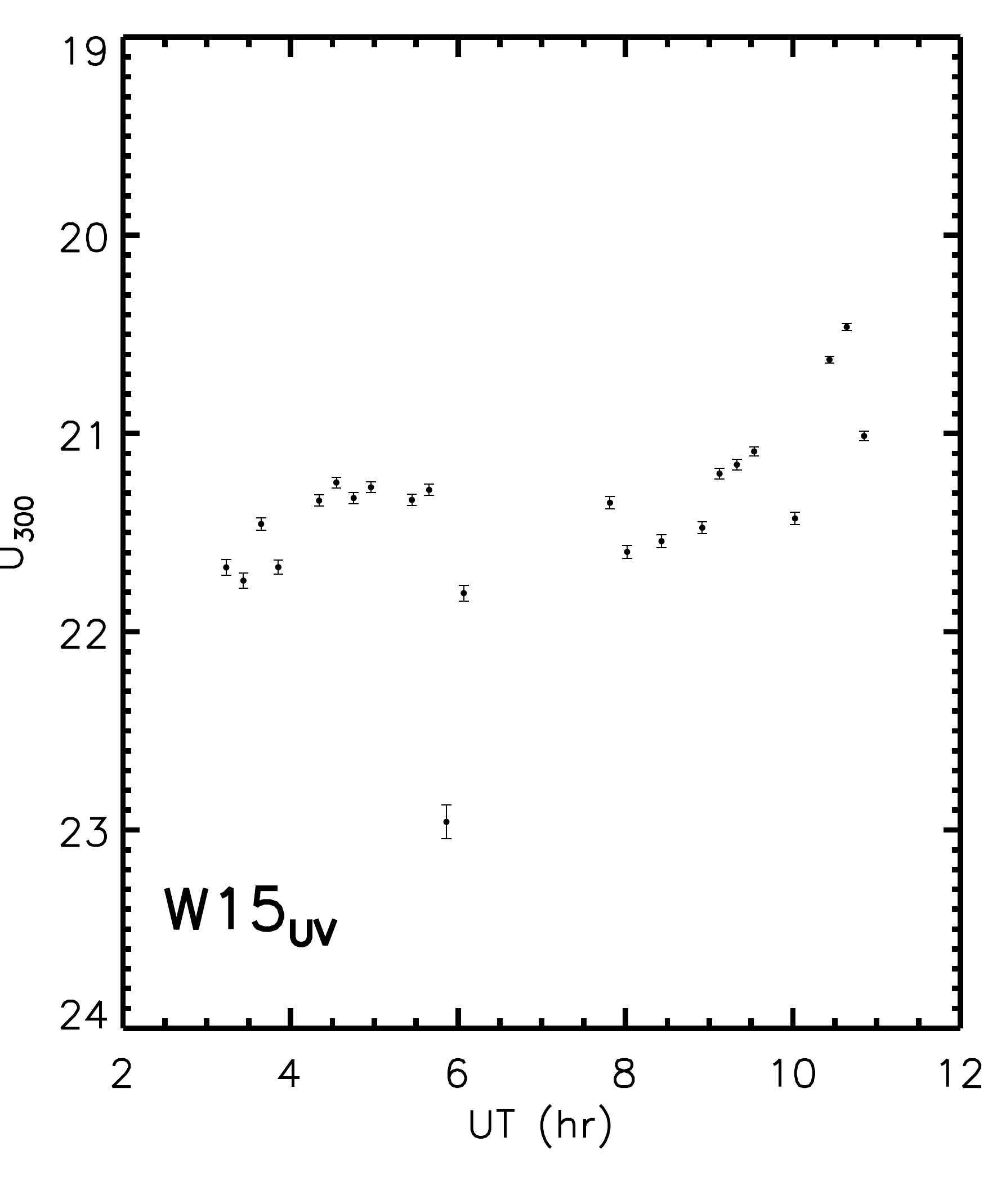} &
    \includegraphics[width=.33\textwidth]{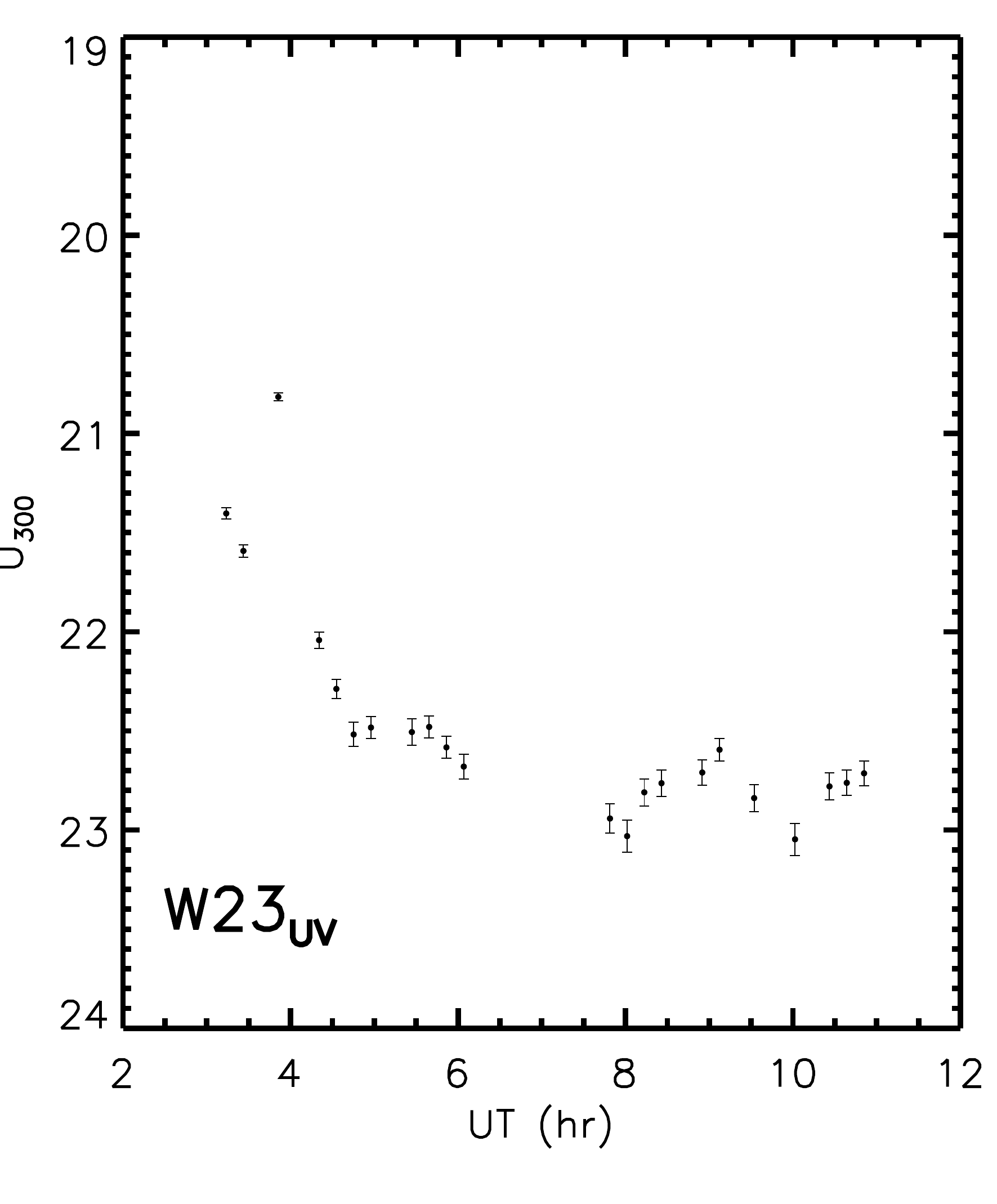} \\
  \end{tabular}
\begin{tabular}{@{}cc@{}}
   \includegraphics[width=.33\textwidth]{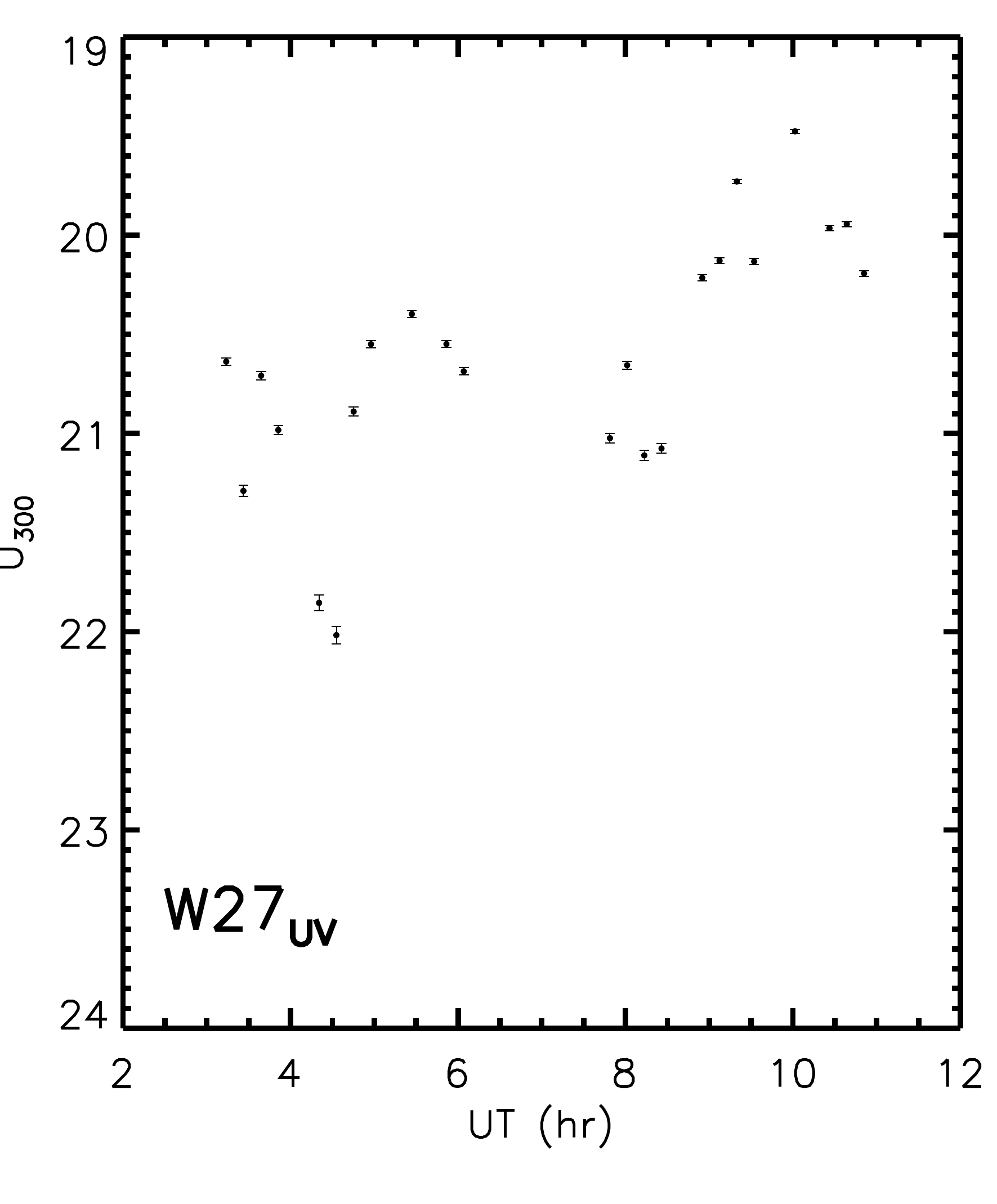} &
    \includegraphics[width=.33\textwidth]{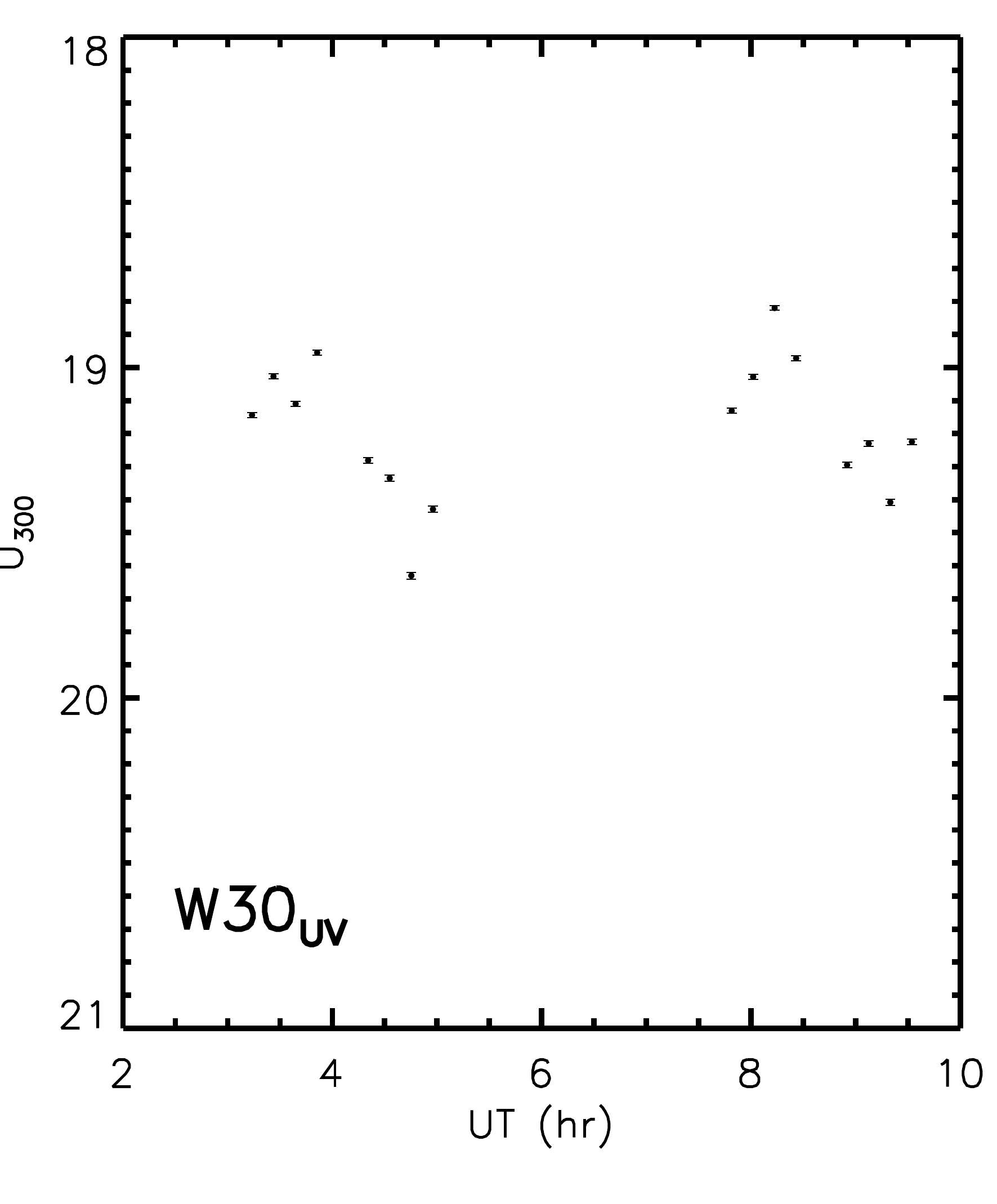} \\
  \end{tabular}
 \caption{Dereddened light curves in the filter U$_{300}$ of the CVs W2$_{UV}$, W15$_{UV}$, W23$_{UV}$, W27$_{UV}$ and W30$_{UV}$. 
The last is an already known dwarf nova system, added just for comparison matters. 
Large amplitude variations (more than 1 mag) are observed in the light curves of W2$_{UV}$, W15$_{UV}$, W23$_{UV}$ and W27$_{UV}$
in time-scales of hours. 
Only photometric measurements that were flagged by DOLPHOT as good quality measurements are shown.}
\label{fig:DNe} 
\end{figure*}

\subsection{A possible AM\,CVn system (W62) and an intriguing object (W21)}
\label{AM}

The counterparts to the \textit{Chandra} X-ray sources W21 and W62 
are blue in the NUV and optical CMDs. Both objects have variability at a confidence level higher than 99$\%$ (Figure \ref{fig:rms}). 
Also, both objects lie very close to, or even slightly ($\lesssim$0.15 mag in H$\alpha$-R$_{625}$) to the right side of the H$\alpha$-R$_{625}$ MS. 
Compared to W120$_{UV}$, which also lies on that region, the inferred H$\alpha$ emission for W21$_{UV}$ is much weaker, and
perhaps even absent for W62$_{UV}$ (Figure \ref{fig:color-color}).
\citet[][]{2006-whil} found that typical CVs show H$\alpha$ emission lines with $|EWs|> 10$ \AA, with DN in outburst the ones that 
usually have smaller EWs. 
However, the absence of large (several magnitudes) amplitude variations in the optical data of W21 and W62, which were
taken at different epochs, argues against a DN nature. 
An intermediate polar nature could be a possibility, though spectroscopic observations of these kind of objects 
show that usually they exhibit H$\alpha$ emission lines with EWs from tens to hundreds of \AA \
\citep[e.g. RX J0558+53 $|$EW(H$\alpha$)$|$ = 604 \AA,  IGR J19267+1325  $|$EW(H$\alpha$)$|$ = 120 \AA, 
V2306 Cyg $|$EW(H$\alpha$)$|$ = 35 \AA ;][]{1999emilios, 2008stee, 2006-whil}
which contrasts with that estimated for W62$_{UV}$. Therefore, below we explore alternative explanations for W21$_{UV}$ and W62$_{UV}$. 

W21$_{UV}$ is a periodic variable with an orbital period of 52 or 104 mins (\edmondsb).
The power spectrum and amplitude variations of this object favors the former period.
However the folded light curve using the longer period is asymmetric, which could be due to ellipsoidal variations. 
If indeed the period is 52 min, 
W21$_{UV}$ should have a fully or partially degenerate secondary 
\citep[the minimum period for CVs with a non degenerate donor is $\sim$ 75-80 mins;][]{2009-gan}.
H-rich CVs with periods below 75 mins have been detected, examples are 485 Cen \citep[59 min period,][]{1996-augus} and 
1RXS J232953.9+062814 \citep[64 min period,][]{2002-Thor}. However, these CVs have EW(H$\alpha$)$\sim$ --30 \AA \ and  --25 \AA, respectively.
Which are considerably larger than the determined EW(H$\alpha$) for W21$_{UV}$.
Moreover, these two short period CVs are $\sim1$ absolute magnitude fainter than W21$_{UV}$ in V and R, respectively.
These characteristics argue against a very short period H-rich CV nature for W21$_{UV}$.
On the other hand, we think that the possibility for W21$_{UV}$ of being a system with a high accretion mass rate (nova-like) is low, 
since these systems are usually significantly brighter \citep[M$_{V} \lesssim 6.4$ mag,][]{1987-Warner}.

The object W62$_{UV}$ is the second brightest binary discussed in this paper (Figure \ref{fig:NUV}).
There are no observational constraints on the orbital period.
The EW(H$\alpha$) inferred from Figure \ref{fig:color-color} is small: EW(H$\alpha$) $\sim -3 \pm6$ \AA, consistent with no H$\alpha$ emission line.

\begin{figure}
\centering
\includegraphics[width=8cm]{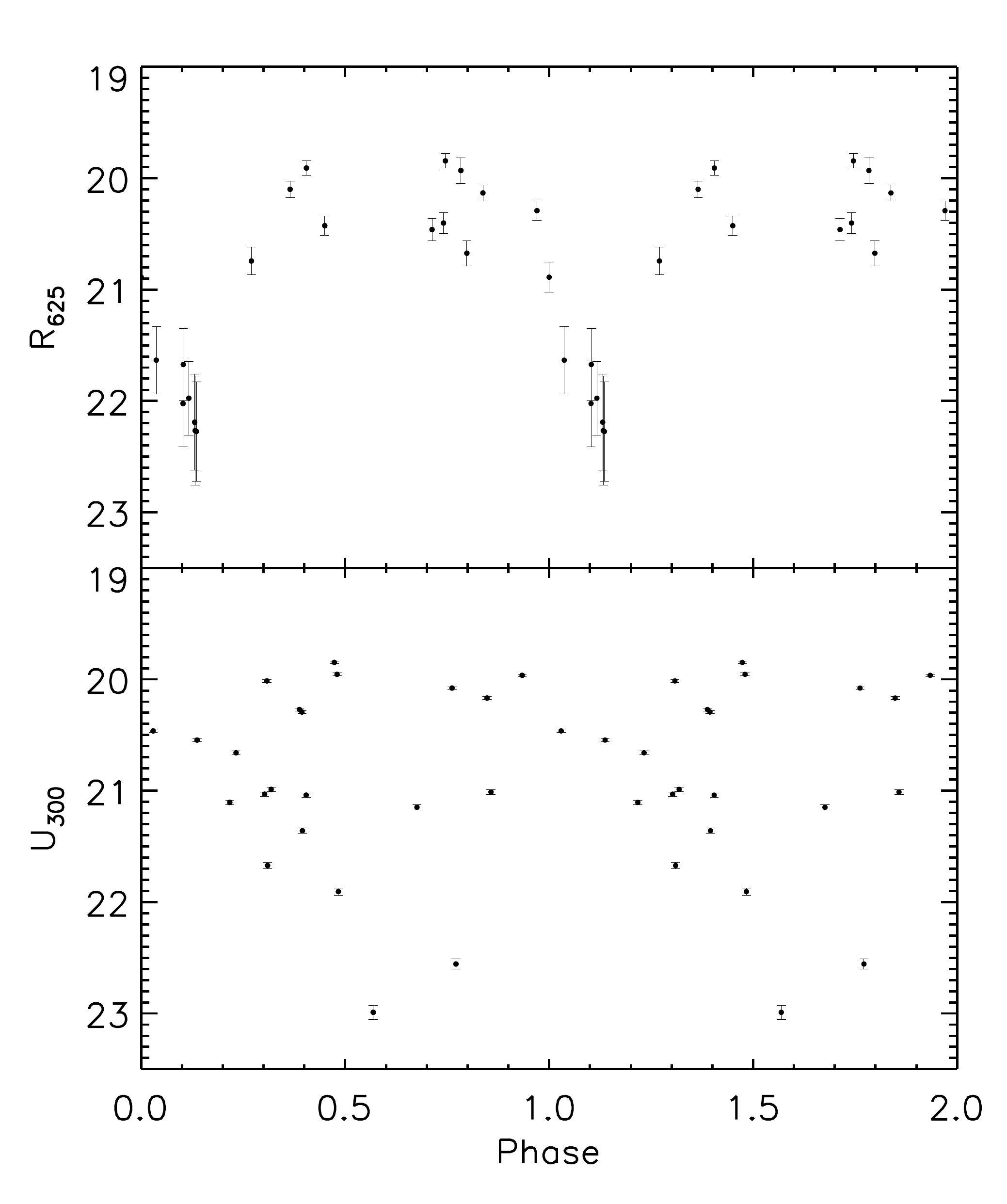}
\vspace{-0.3cm}\\
\caption{Phase-folded light curve in the R$_{625}$ and U$_{300}$ bands for the eclipsing CV W2$_{UV}$. 
We have used the period of 2.4 h given by \citet{2016-israel}.
The light curves have been dereddened but not barycentered.
We used the same initial time when folding.
The eclipse is clearly visible in R$_{625}$ but not in U$_{300}$. 
Note that the observations in U$_{300}$ were taken at a different epoch than the observations in R$_{625}$.}
\label{fig:W2R} 
\end{figure}

\begin{figure}
\centering
\includegraphics[width=8cm]{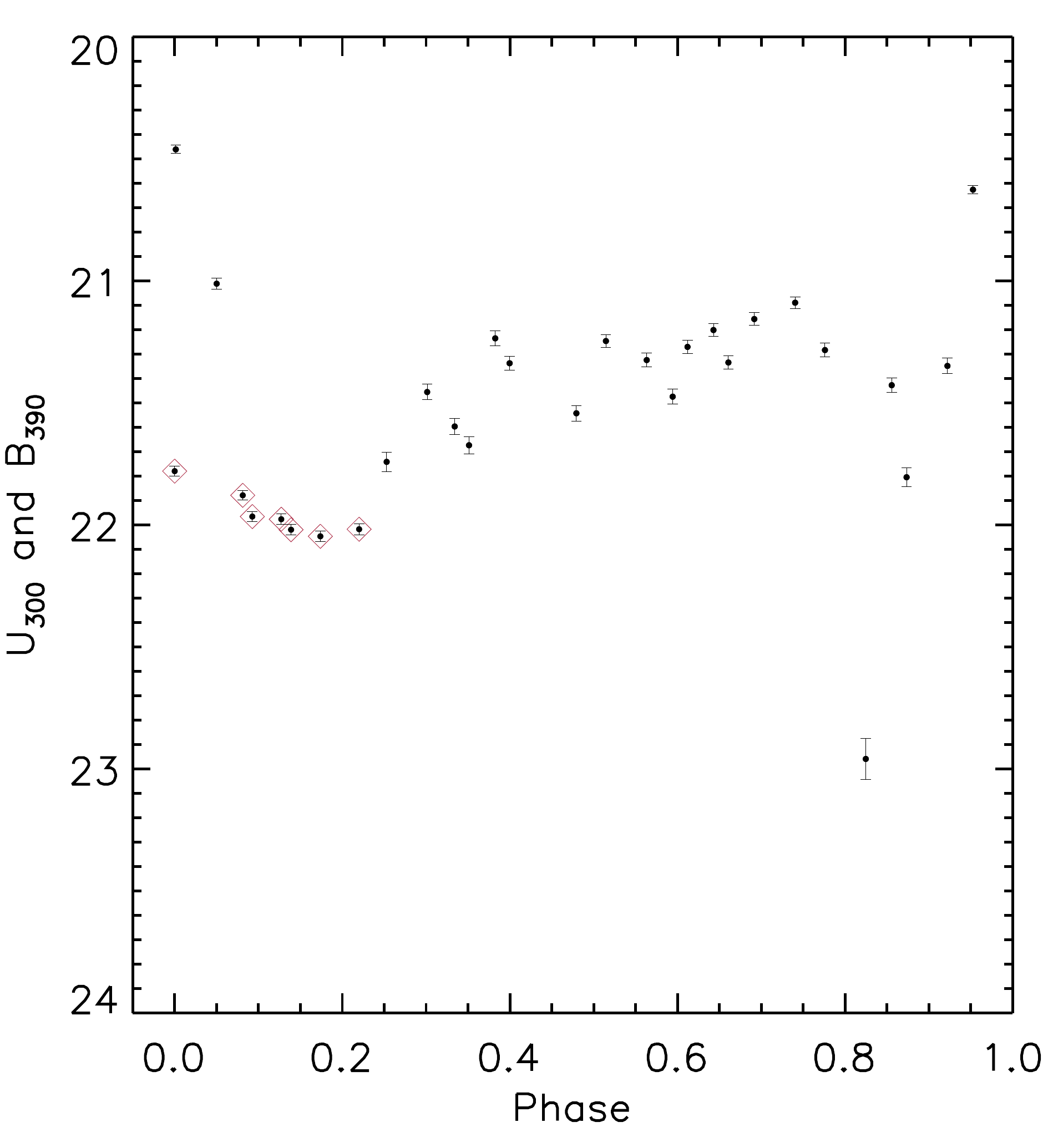}
\caption{Dereddened phase-folded light curves in the filters U$_{300}$ and B$_{390}$ for the eclipsing CV W15$_{UV}$. 
We have used the period of 4.238 h given by \edmondsb.
Red diamonds correspond to the observations in the filter B$_{390}$.
The eclipse is not visible in U$_{300}$, instead variations of $\sim$2.5 mag are observed. 
Only photometric measurements that were labeled by DOLPHOT as good quality measurements are plotted.}
\label{fig:15fold} 
\end{figure}

We explored the possibility that W21$_{UV}$ and W62$_{UV}$ are double degenerate systems. In fact we found that
the colors of W62$_{UV}$ are compatible with those obtained from AM CVn spectra (Figure \ref{fig:color-color}) published by \citet{carter}.
Moreover, if we estimate the period of W62$_{UV}$ using the magnitude-period relation for AM CVns given by \citet{bil}, 
we find that W62$_{UV}$ would have a period
of $\sim$ 6 mins. Very similar to that of the known AM CVn system HM Cnc \citep{2002-israel}. 
Therefore, if W62$_{UV}$ would be confirmed as an AM CVn, 
then it is the first of that type detected in 47 Tuc and it is one of the few with an orbital period below 10 mins.
The NUV color of W62$_{UV}$ (U$_{300}$-B$_{390}$ $\approx 0.15$) is consistent with those of AM CVn binaries \citep[NUV--u color goes from --0.8 to 0.7;][]{carter},
and the same is true for its X-ray luminosity \citep[$L_X \sim 10^{30}-10^{32}$ erg s$^{-1}$ for AM CVn binaries;][]{2006-ramsay}. 
We note that W21$_{UV}$ has an absolute V magnitude ($\sim$7.4 mags; \edmondsa) 
that is not consistent ($\sim 5.5$ mag brighter) with the one expected for an AM CVn system with an orbital period of 52 mins (the 104 min period would automatically exclude  W21$_{UV}$ from being a double degenerate).
According to \citet{bil}, an AM CVn system with that period should have M$_{V}$ $\approx$ 13 mag \citep[as it has been observed for systems with similar periods,][]{Sole}.

Comparing the colors of W62$_{UV}$ and W21$_{UV}$ with those of He WD (like in Figure \ref{fig:msps}),
they lie slightly to the right of the least massive cooling track.
Together with the lack of radio emission, this gives evidence against a possible He WD MSP companion nature (but see Section \ref{hidden}). 
The probability of these objects of being black widow or redback systems is low, 
since the confirmed systems of that type in GCs have positions in the CMDs 
that are not consistent with those of W62$_{UV}$ or W21$_{UV}$ \citep[see for example][]{2002-Edmonds, cadelano71}.
This is because black widow or redback systems lack an accretion disk and a hot, accreting WD. 
Instead, the optical flux comes from a (irradiated) secondary star, which causes the binary not to be 
as blue as W62$_{UV}$ or W21$_{UV}$ are. 
We discard a possible quiescence low mass X-ray binary (qLMXBs) nature for both objects, on account of their X-ray to ultraviolet flux ratio.
Since CVs lack the deep potential wells of qLMXBs, they should generate a much smaller ratio of X-ray to UV flux from material passing through their accretion disks.
We found that the known qLMXBs in 47 Tuc have log$(F_X/F_{300})$ between 1 and 3, while W21$_{UV}$ has a value of around $ -0.1$
and W62$_{UV}$ $\sim -1.5$. Also, the positions of these two binaries in the NUV CMD do not match with those of qLMXBs in the cluster
\citep[][van den Berg et al. in preparation]{2002-Edmonds}.
Given all the arguments mentioned above, the nature of the binary W21$_{UV}$ remains uncertain. 

\subsection{Millisecond pulsar companions?}
\label{hidden}

\begin{figure}
\centering
\vspace{-1cm}
\includegraphics[width=8.5cm]{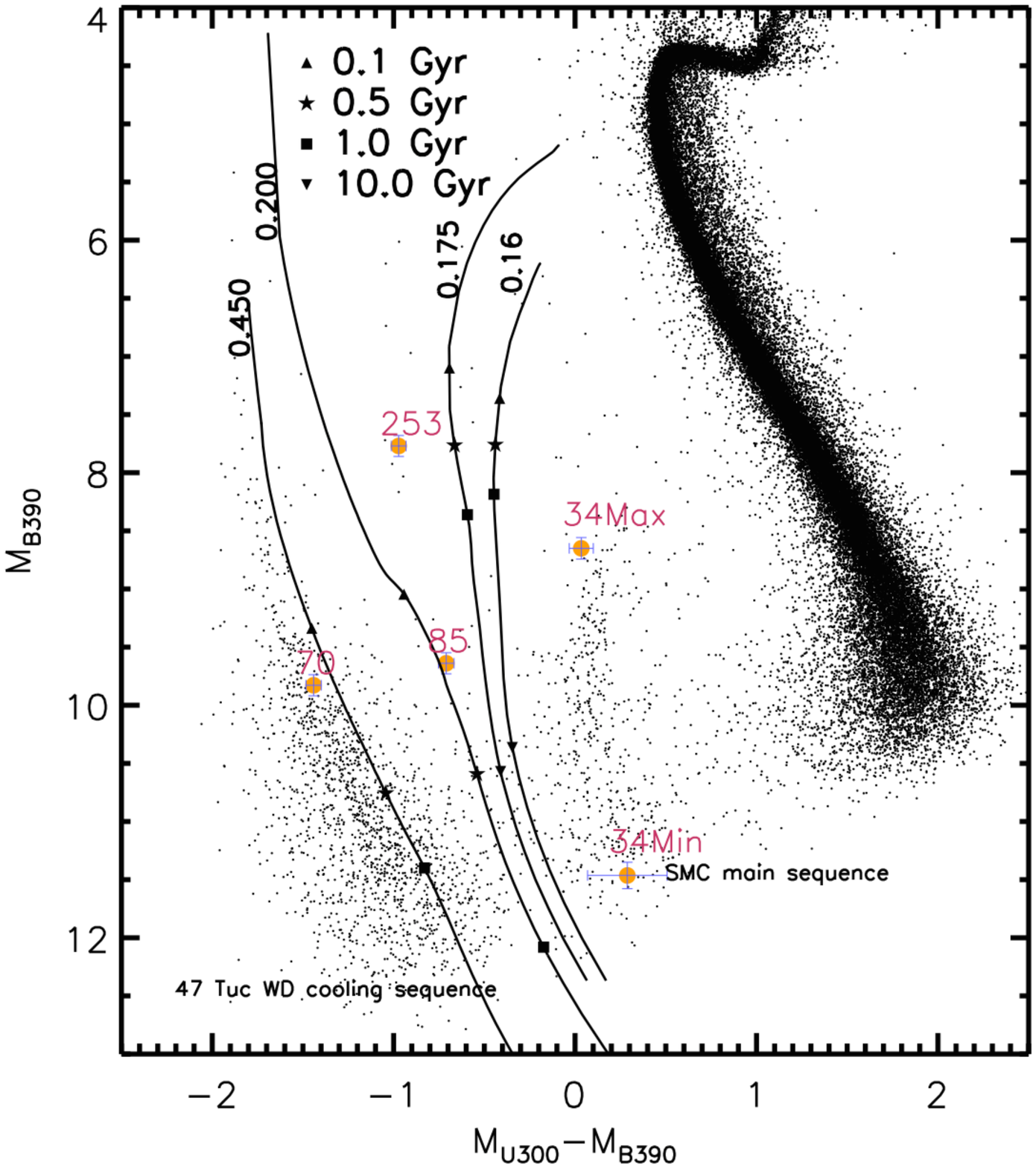}
\vspace{-2cm}\\
\caption{$U_{300}-B_{390}$ versus $B_{390}$ CMD based on DAOPHOT photometry of our NUV
  images. Each candidate MSP companion is denoted with the corresponding number and a filled orange circle. 
Calibrated magnitudes were converted to absolute magnitudes
  using a distance to 47\,Tuc of $4.69$ kpc and
  reddening $E(B-V)=0.04$. The lines correspond to
  cooling models for He WDs with thick H envelopes as in \citet{2002-serenelli}. 
They were computed for an initial metallicity of $Z=0.003$. 
The masses for which each track were calculated are indicated on the top of each track.
The units are $M_{\sun}$.}
\label{fig:msps} 
\end{figure}

We place in this group the counterparts to the X-ray sources W34, W70, W85 and W253. 
The first three objects were previously suggested to be either CVs or MSPs \citep[][\edmondsa, \edmondsb]{2002-Edmonds}. 
However, our photometric results (especially the H$\alpha$) argues against the CV nature (see Figure \ref{fig:color-color}). 
The X-ray source W253 is classified for the first time in this paper, and has been
included in this section because of the similarities with W70 and W85.
Figure \ref{fig:msps} shows the position of these four binaries in the NUV CMD, 
and Table \ref{table:photomsp} contains their
photometric and astrometric results. 

\begin{table*}
\centering
\caption{Astrometric and photometric results for the MSP companion candidates. 
Units of right ascension are hours, minutes, and seconds, and units of declination are degrees, arcminutes, and arcseconds.
The photometric values are calibrated to the Vega-mag system. 
Photometric errors are given in parenthesis. The projected distance from the cluster center (r) was determined using the coordinates of the center of 47 Tuc
as given by \citet{2010-Golds}. }
\label{table:photomsp}
  \begin{tabular}{c c c c c c c c c }
Source & R.A. & Decl (J2000) & U$_{300}$ & B$_{390}$ & B$_{435}$ & H$\alpha$& R$_{625}$ & r (\arcmin)\\
\hline
 34$_{UV}^*$ & 00:24:05.216  &   -72:04:46.72   & 23.8(2)  & 23.59(9) &24.1(1)  &  22.5(1)  & 23.2(2)  &0.10\\
 70$_{UV}$ & 00:24:11.874  &  -72:04:44.21    & 22.003(6) & 23.36(1) & 23.69(3) &23.8(1) & 23.73(6) & 0.49 \\
 85$_{UV}$ & 00:23:59.387  &  -72:04:38.57    & 22.545(9) & 23.17(1) & 23.44(5) & 23.7(2) & 23.4(1) &  0.54 \\
253$_{UV}$ & 00:24:10.983 &  -72:05:04.99    & 20.414(3) & 21.306(4)& 21.52(3)** &21.6(1)** & 21.53(7)** & 0.45\\
\end{tabular}
{\flushleft
\startfoot  
\begin{flushleft}
*  \  \ Mean photometric values.\\
** Photometry may be slightly affected by the presence of a bad column. 
\end{flushleft}
}
\end{table*} 

W70 and W85  were previously discussed by \edmondsa, \edmondsb \, 
who found both objects with blue colors in the U--V vs U  and F300W -- F555W vs F300W CMDs, respectively. 
However, a lack of I magnitude was mentioned by those authors for W70$_{UV}$, together with a mismatch of 2 pixels in the position
of this object between the V and the U images. This mismatch is the result of the presence of a very close and bright neighbor at only 0.12\arcsec \
from the real counterpart. Since in the NUV images both stars are clearly visible, reliable photometry for W70$_{UV}$ was obtained. The resulting
magnitudes place this X-ray binary on the primary WD cooling sequence of the NUV and optical CMD. 
Its NUV, optical and H$\alpha$ colors, indicate that W70$_{UV}$ could be a CO WD. 
Additionally its X-ray softness and lack of variability in the NUV, optical or X-ray emission are consistent with a 
(low state) magnetic CV, a CO WD companion (it seems too blue to be an He WD) to a so-far-undetected MSP, or a chance coincidence of an isolated WD within the X-ray error circle (see Section \ref{match}). 

The objects W85$_{UV}$ and W253$_{UV}$ show blue colors that place them between the primary WD sequence and the SMC MS.
There are no indications of variability in any wavelength for these objects.
In the optical CMD, W85$_{UV}$ appears on the primary WD sequence. 
This slight position change in the CMDs suggests that this binary is not an isolated WD matched by chance with the X-ray source. 
The object W253$_{UV}$ is between the WD sequence and the SMC MS in the optical CMD. 
However, we note that the presence of a bad column nearby might slightly affect the optical photometry.
From the fitting of the X-ray spectra, \heinke \ found that W253 is more suggestive of an MSP than a thermal plasma spectrum typical of CVs or ABs. 
Additionally, W253$_{UV}$ is almost at the center of the 3$\sigma$ error circle, which significantly reduces the probabilities of being a chance coincidence. 

Given the results described above, we considered the possibility that W85$_{UV}$ and W253$_{UV}$
are faint CVs like W120$_{UV}$. 
However, from our Figure \ref{fig:color-color} we see that the colors of both binaries 
are indicative of the presence of prominent H$\alpha$ absorption lines that are not typical of CVs, in contrast to the findings of W120$_{UV}$. 
Instead, their colors are in very good agreement with those of DA WDs 
(WDs with H-dominated atmospheres that show deep Balmer absorption lines) obtained from spectra published by \citet{carter}.
We note that the lack of variability in W85$_{UV}$ and W253$_{UV}$ argues against an AM CVn nature.
However, our results for these two objects are consistent with those described by \citet{rivera} for He WD companions to MSPs in the cluster
\citep[see][for more examples of (DA) WDs companions to pulsars]{1996-van, 2006-bassa}. 
The difference is that radio emission has not been reported either for W85$_{UV}$ or W253$_{UV}$.

Since W70$_{UV}$, W85$_{UV}$ and W253$_{UV}$ have similar properties to the MSP companions,
in Table \ref{table:modelshidden} we show estimated masses, effective temperatures, surface gravities, luminosities, cooling ages, and X-ray properties
for these three binaries assuming that they are indeed MSP companions. 
These parameters were obtained by comparison of He WD cooling models from \citet{2002-serenelli} to our NUV photometry. 
These models were computed for a metallicity $Z=0.003$, as appropriate for 47 Tuc. 
Using the absolute magnitudes in the filter B$_{390}$ as independent variable, we have estimated the corresponding parameters
by interpolating linearly along the two cooling tracks that bracket the location of W253$_{UV}$ and W85$_{UV}$. 
For W70$_{UV}$ the values were
obtained by interpolating to the closest cooling track (that one for an He WD with 0.45 $M_{\sun}$).
As mentioned before, this object could be consistent with a CO WD with mass $> 0.45 \ M_{\sun}$.

\begin{figure*}
\centering
\vspace{-1.5cm}
\twofigbox{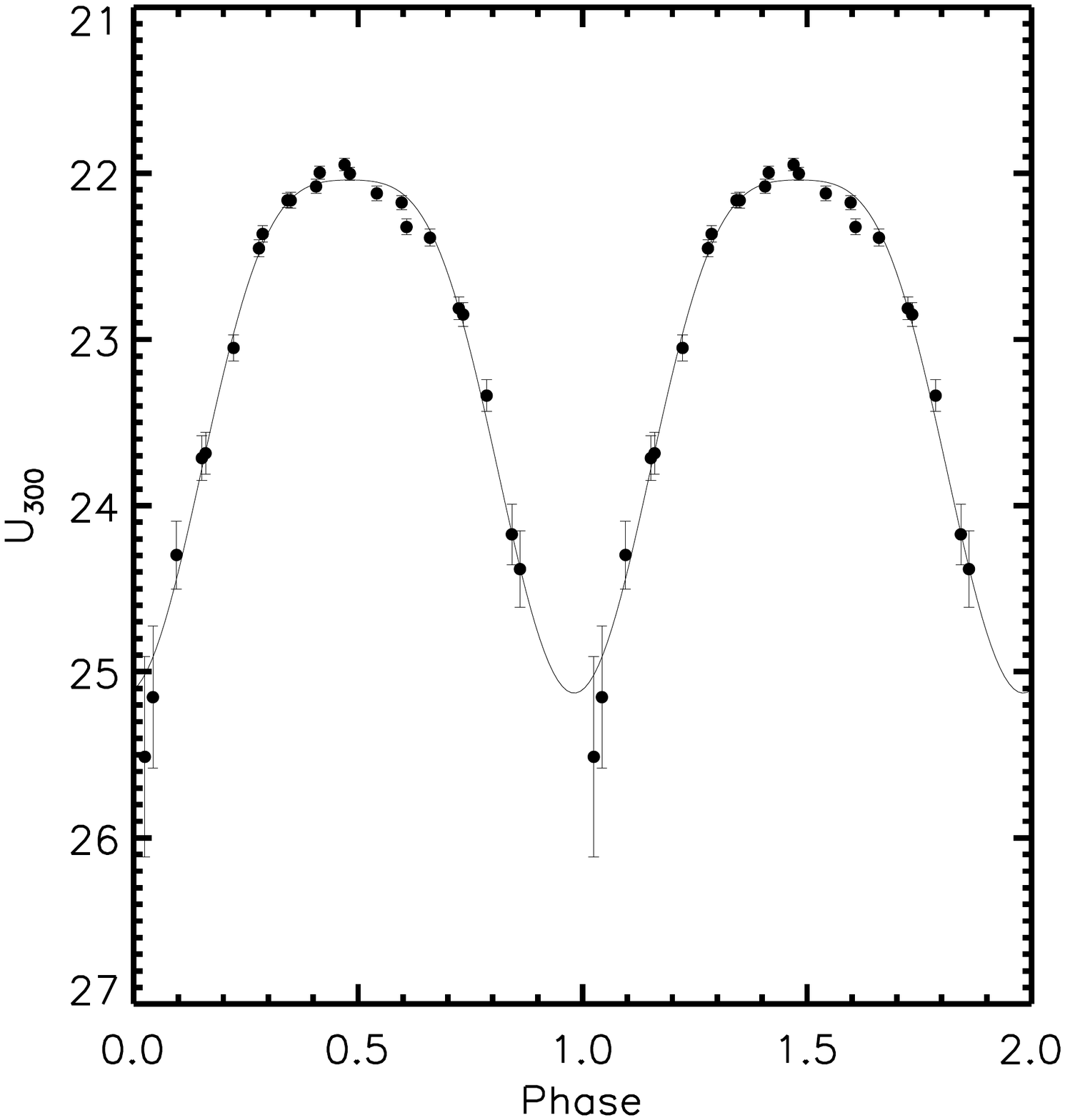}{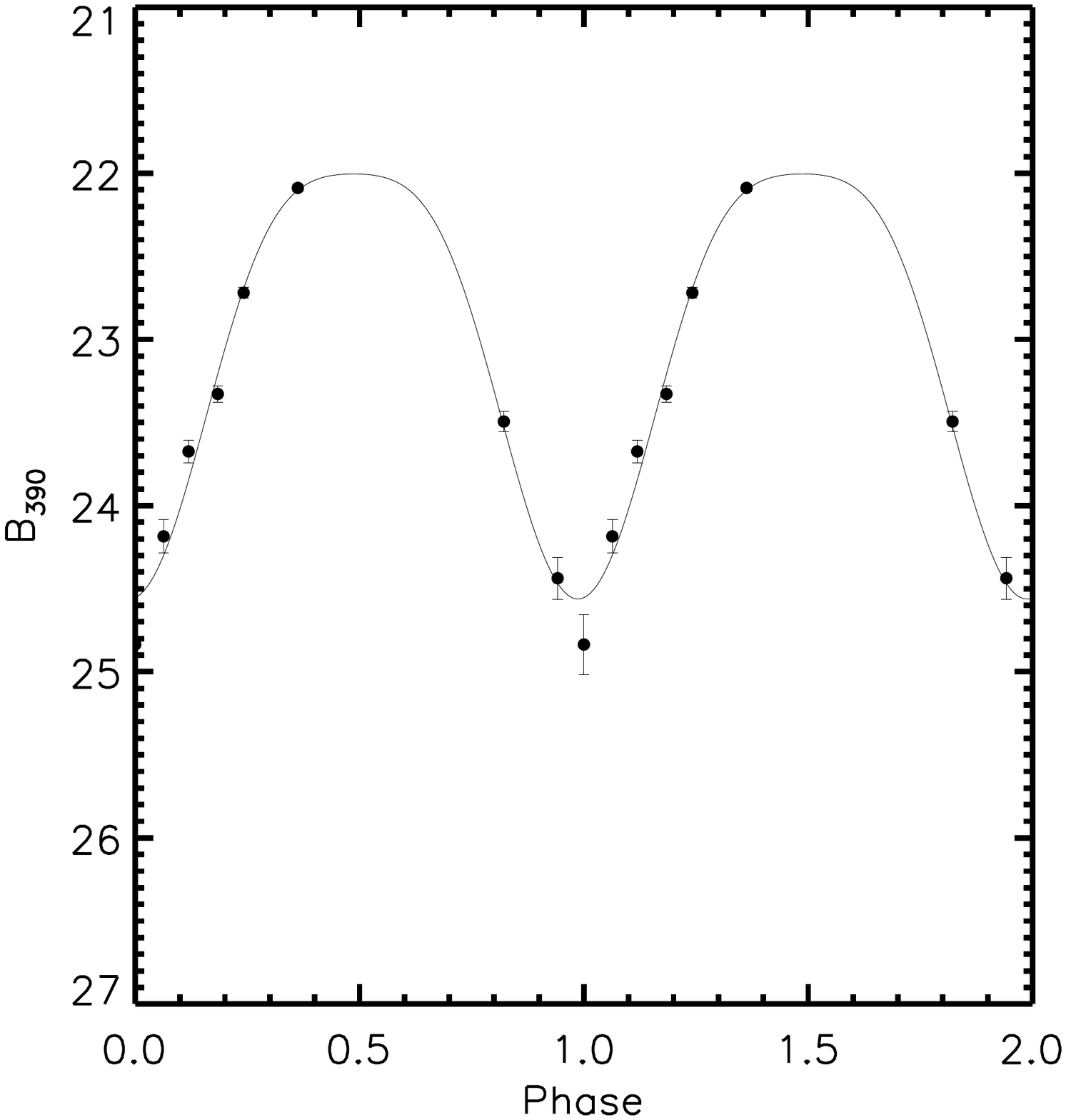}
\vspace{-0.9cm}
\caption{Folded light curves of W34$_{UV}$ in the filters U$_{300}$ and B$_{390}$. We have used a period of 1.624 hr.
For clarity, the phase is plotted twice. The best fit is overplotted. The light curves are not similar to light curves of CVs, 
but they resemble more the light curves of systems in which the companion is strongly irradiated. The light curves are dereddened but not barycentered.}
\label{fig:34}
\end{figure*}

The counterpart to the X-ray source W34 is a short period variable with a period of 1.624 hr \citep{2002-Edmonds}. 
Light curves in the filters U$_{300}$ and B$_{390}$ folded on that period are shown in Figure \ref{fig:34}, 
where we can observe the non sinusoidal behavior of W34$_{UV}$,
with amplitude variations of $\sim 2.8$ magnitudes in the filter B$_{390}$ and $\sim 3$ magnitudes in the filter U$_{300}$. 
Only measurements labeled by DOLPHOT as good quality measurements were used. 
We note that the light curve of W34$_{UV}$ is not the typical light curve of a CV system. 
In fact, the light curve of W34$_{UV}$ matches in shape and amplitude those of radio MSPs with strongly irradiated companion stars in short orbits
(such as 47 Tuc W \citeauthor{2002-Edmonds} \citeyear{2002-Edmonds}; 
PSR J1653-0158,  \citeauthor{2014-romani} \citeyear{2014-romani}; 
XSS J12270-4859, \citeauthor{2017muv} \citeyear{2017muv}; 
and especially PSR B1957+20, \citeauthor{2007Rey} \citeyear{2007Rey}).
W34$_{UV}$ displays a blue color in the NUV and optical CMDs (Figures \ref{fig:BR} and \ref{fig:msps}). 
It shows small H$\alpha$ excess, 
but the photometric errors in that band are considerable due to the presence of bright neighbors, thus
it remains the possibility that this binary lies on the MS of the H$\alpha$ CMD.  
\citet{2002-Edmonds} have already pointed out optical similarities between W34$_{UV}$ and the black widow system associated
to W29 (47 Tuc W), with the difference that W34 has not been reported in radio wavelengths. 
Moreover, the strong heating from the light curves of W34$_{UV}$ proves that the intrinsic color of the companion is redder than the observed color. 
This rules out any kind of WD companion nature 
(as W34$_{UV}$ lies far to the right of the least massive He WD cooling track in Figure \ref{fig:msps}),
suggesting that the companion is almost certainly a low-mass MS star. 
We also noted that this binary lies on the SMC MS. However, we discard a possible membership to that galaxy because 
at the distance of the SMC the source
would be very bright (2.6$\times 10^{33}$ erg s$^{-1}$), 
and it would have an orbital period far too short for being a SMC MS binary star. 

\begin{table*}
 \centering
  \caption{Estimated masses (M$_c$), cooling ages, effective temperatures (T$_{eff}$), surface gravities (log g), bolometric luminosities (L) and X-ray luminosities ($L_X$)  
for the possible counterparts to hidden MSPs in 47 Tuc. These parameters were obtained by comparing our NUV photometry to 
He WD cooling models with metallicity $Z=0.003$ \citep{2002-serenelli}. }
\label{table:modelshidden}
  \begin{tabular}{c c c c c c c c c c c}
Source  &M$_c$                  & T$_{eff}$      & Log g & Log L               &  Cooling age$^a$        & $L_X$ 0.5-6.0$^b$              & $L_X$ 0.5-2.5$^b$ \\
             &(M$_{\sun}$)& (kK)  &          &($L_{\sun}$)    &  (Gyr)      &(10$^{30}$ ergs s$^{-1}$) & (10$^{30}$ ergs s$^{-1}$) & \\
\hline
 34$_{UV}$ & $<$ 0.16       &  --  & --   & --  & --   & 16.5$^{+1.3}_{-1.1}$         & 10.5$^{+0.8}_{-0.7}$\\
 70$_{UV}$ &  $>$ 0.45    & $>$18.3   & $>$7.6  & $>$ -1.4           & $>$ 0.18    &1.6$^{+0.4}_{-0.3}$         & 1.5$^{+0.3}_{-0.2}$ \\
 85$_{UV}$ &  0.2 -- 0.175 & 11.4 -- 9.2 & 6.6 -- 6.3& -1.7 -- -1.8      & 0.2 -- 3.8   & 2.9$^{+0.6}_{-0.4}$        & 1.7$^{+0.3}_{-0.3}$ \\
253$_{UV}$  &  0.2 -- 0.175 & 11.7 -- 9.9 & 5.9 -- 5.6 & -1.0 -- -1.1     & 0.5    &  2.7$^{+0.6}_{-0.4}$       & 1.7$^{+0.3}_{-0.3}$ \\
\end{tabular}
{\flushleft
\startfoot  
\begin{flushleft}
$^a$ Ages represent the time between the point of highest T$_{eff}$ on the cooling track until the observed point along the track.\\
$^b$ Values taken from \heinke.
\end{flushleft}}
\end{table*} 

As already mentioned, radio emission for all the systems discussed in this section has not been reported. 
That could be due to different reasons. The simplest one is that the direction of the radio beam is not pointed towards Earth.
An alternative explanation is that the corresponding radio MSPs show disappearances (probably caused by clouds of material in the binary system)
for several orbits that coincide with the time when the radio observations of the cluster have been taken, as occurs for PSR 1744-24A \citep[e.g.][]{1992-nice}. 
Another explanation could be that they are faint radio sources compared to the other pulsars in the cluster, so they have been missed in the radio data analysis. 
Indeed, \citet{2016-pan} recently found two new isolated radio MSPs in 47 Tuc by reprocessing existing radio data with new 
methods. That opens the possibility of future detection (probably by more sensitive telescopes) of radio pulsations from the systems here discussed. 

\subsection{Random coincidence test}
\label{match}

In a crowded cluster such as 47 Tuc it is important to take into account that some matches between X-ray sources
and optical counterparts can be chance coincidences. In order to know how reliable the associations discussed 
in this paper are, we have carried out a test that computes the expected number of random coincidences between the X-ray sources and the NUV counterparts. 

We first distinguish two regions in our NUV images: 
the core, limited by the core radius \citep[$r_c = 24\arcsec$;][]{2000-howell}, and the 
region surrounding it but inside the NUV images. Of the 238 X-ray sources in our NUV FOV, 88 are in the core and 150 are outside
the core but within the r$_h$ of the cluster (see Figure \ref{fig:mosaic} of Appendix \ref{stacked}).

For the analysis, we followed the definition of blue object given at the beginning of Section \ref{results}. 
However we now call blue-gap object those stars found between the WD sequence and the MS of 47 Tuc ($\sim1450$ objects). 
In the same context, we define as WD an object that falls on the WD cooling sequence of the NUV CMD (Figure \ref{fig:NUV}).
Additionally, if an object shows NUV variability at a confidence level equal or higher than $95\%$, we call it variable.
Given that the number of blue-gap objects is smaller than WDs, 
we deduce that the probability of finding a blue-gap object within an X-ray error circle is smaller than finding a WD. Additionally, we identified around $620$ sources in the NUV FOV that show evident Ha excess ($\geq$ 99\% confidence level) and are not associated to X-ray sources in the cluster.

We determined the density (stars arcsec$^{-2}$) of variable (or non-variable) blue-gap objects inside and outside of the core radius of the cluster.
We also determined the density of variable (or non-variable) WDs inside and outside of the core radius.
Thus, we have four densities for variable objects and four densities for non variable objects. 
We have used these densities to calculate the total expected number of false matches within the 3$\sigma$ X-ray 
error circles of the 238 sources in our NUV FOV.
 
We expect $\sim0.2$ false matches for variable blue-gap objects and the same number for variable WDs outside the core radius.
On the other hand, we expect $\sim0.8$ false matches for variable blue-gap objects and a similar value for variable WDs inside the core radius. 
Such values become much smaller if instead we consider the areas of the 1$\sigma$ and 2$\sigma$ X-ray error circles, where many of the counterparts that we propose are. 

For the non variable objects, we expect 10 (and 14) false matches for blue-gap objects (and WDs) outside the core radius ($\sim35\%$ of the detected matches).
On the other hand, we expect 13 (and 15) false matches for blue-gap objects (and WDs), inside the core radius ($\sim50\%$ of the detected matches).
These relative large numbers are not surprising, since within the 3$\sigma$ error circles of  X-ray sources for which we have secure optical counterparts 
we have also found blue-gap or WD objects that are likely spurious alignments. 
Also, the total numbers of false matches are largely influenced by the sources that have large X-ray error circles (e.g. many of the sources discussed in Section \ref{sec:multiple}). It is possible that few of the non variable (and not detected in H$\alpha$) CV candidates reported in this paper are part of these chance coincidences.
These objects are W24$_{UV}$, W55$_{UV}$, W75$_{UV}$, W98$_{UV}$, W100$_{UV-\text{a and --b}}$, W187$_{UV}$, W202$_{UV}$, W223$_{UV}$, W229$_{UV}$, W235$_{UV}$, W302$_{UV}$, W324$_{UV}$, W329$_{UV}$ and W335$_{UV}$.
However, the objects W24$_{UV}$ and W302$_{UV}$ are WDs that fall well inside the 1$\sigma$ error circle and are inside the core of the cluster, 
where the expected number of false matches among the 88 X-ray sources is $1.2$. 
The counterparts to W229 and W253 (a proposed binary MSP also non-variable in NUV) are also within 1$\sigma$ but they lie outside the core of the cluster. The objects W98$_{UV}$, W100$_{UV-\text{a and --b}}$, W223$_{UV}$, W235$_{UV}$ and W324$_{UV}$, are inside the 1.5$\sigma$ error circle.

As discussed in previous sections, the counterparts to W1, W44 and W53 are not variable in U$_{300}$, but they actually are in other wavelengths (\edmondsb). This suggests that most likely they are the correct counterparts to their X-rays sources and so the chance of being spurious matches is reduced. Among the MSP counterparts proposed (W34$_{UV}$, W70$_{UV}$, W85$_{UV}$ and W253$_{UV}$), we think that W70$_{UV}$ has the largest chances to be a match coincidence. It was found at 3$\sigma$ and the counterpart does not show position changes in our different CMDs that could indicate that it is not a single-temperature WD. However, some of the confirmed MSP companions (also non-variable in NUV) by \citet{rivera} present similar characteristics to W70$_{UV}$ and were also found at 3$\sigma$ from the X-ray source. 

\section{Comparison to other globular clusters: core-collapsed vs non core-collapsed}
\label{sec:comparison}

Given the results obtained in this paper, 
47 Tuc is now the GC with the most CVs and CV candidates identified so far. 
It is followed in number of identified CVs and CV candidates by the cluster $\omega$ Cen (27 systems - \citeauthor{cool} \citeyear{cool}), 
NGC 6752  \citep[16 systems -][]{2017Lugger} and NGC 6397 (15 systems - \citeauthor{cohn} \citeyear{cohn}). 
Characteristic parameters for these last two clusters are given in Table \ref{table:clusters}, and 
a list of the CVs and CV candidates found in them is given in Tables \ref{table:otherclusterscvs_1} and \ref{table:otherclusterscvs_2} of Appendix \ref{fCVs_other}.  
The approximate X-ray limiting luminosities of these clusters are few times $10^{29}$ erg s$^{-1}$. 

In the following sections we will compare the results found in 47 Tuc with the available results of the GCs 
NGC 6752 and NGC 6397 to obtain the best comparison of the CV population among the Galactic clusters. 
We exclude $\omega$ Cen from our comparison 
due to its remarkable differences, like
its mass, its orbit and its velocity dispersion measure which suggest that 
$\omega$ Cen could be the disrupted core of a dwarf galaxy instead of a GC \citep[see for example][]{2000-omega,2008-eva}. 
However, a comparison between the X-ray counterparts of $\omega$ Cen and the ones in NGC 6397, including the CV population, 
has been reported by \citet{cool}. They found a notable difference between the samples of both clusters.
$\omega$ Cen (with a mass of $40.5 \times 10^5 M_{\odot}$) has $\sim3$ times more CV candidates in the range $L_X =  10^{30} -  10^{31}$ erg s$^{-1}$
than NGC 6397, but similar number of  CVs in the range $L_X = 2 \times 10^{31} - 2\times 10^{32}$ erg s$^{-1}$.
This suggests that unlike NGC 6397, the CV sample in $\omega$ Cen could be dominated by primordial CVs.
We refer the interested reader to that paper for more details. 

It is important to mention that in our comparison we only include the CVs and CV candidates discussed in Subsections \ref{sec:CVs47} and \ref{sec:outburst}. 
This is because the characteristics of the binaries W34$_{UV}$, W70$_{UV}$, W85$_{UV}$ and W253$_{UV}$ do not match those of typical CVs, 
instead they match much better those of MSP companions. 
The nature of the sources W21$_{UV}$ and W62$_{UV}$ is not confirmed, and thus they have been excluded from the comparison as well.
By excluding these sources, we can make a fairer comparison\footnote{Note that NGC 6397 and NGC 6752 have not been studied in NUV.}
of the CVs with likely non degenerate donors between cluster's samples,
since none of the CVs and CV candidates in NGC 6397 and NGC 6752
were suspected to be either AM CVn or MSP companions. 
For practicality, from now on we will refer as CVs to
all CVs and CV candidates in the three clusters.

\subsection{X-ray and optical properties of the CV population}
\label{properties}

\begin{figure}
\centering
\vspace{-0.8 cm}
\includegraphics[width=8.5cm]{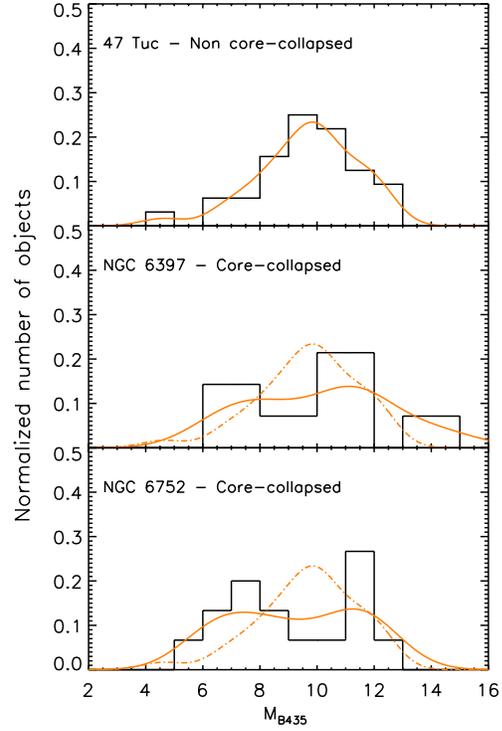}
\vspace{-0.5 cm}
\caption{Normalized luminosity functions in the filter B$_{435}$ for the CVs in 47 Tuc, NGC 6397 and NGC 6752.
In the case of 47 Tuc, only CVs for which we derived photometry in the B$_{435}$ band are plotted. 
An exception is W100 which was excluded because the counterpart is ambiguous. 
The integral under the histogram is equal to 1. 
In each plot, the orange line is the probability density function
estimated by the KDE method using as kernel a Gaussian function.
The dashed curves in all three plots represent the probability density function of 47 Tuc.  
The distribution function of 47 Tuc is unimodal, while the ones of NGC 6397 and NGC 6752 are bimodal.}
\vspace{-0.1 cm}
\label{fig:b435}
\end{figure}

Studies of the X-ray sources in the core-collapsed GCs NGC 6397 and NGC 6752 
show that these clusters have two populations of CVs: a bright population (with apparent magnitudes R$_{625}\lesssim21.5$) that is more
concentrated towards the core than a faint population (R$_{625}>21.5$), for which the WD dominates the brightness  \citep{cohn,2017Lugger}.
When we compare the optical CMDs of these clusters with the optical CMD of the non core-collapsed cluster 47 Tuc,
we find that in the last one we can not distinguish two groups of CVs as in NGC 6397 and NGC 6752  (Figure \ref{fig:BR}).
For a better comparison, in Figure \ref{fig:b435} we show normalized histograms of the absolute B$_{435}$ magnitudes of the CVs
with respect to the total number of identified CVs in each cluster. 
It is important to mention that the results shown in this histogram and in all the other histograms in this paper
do not depend on the binning size. 
From Figure \ref{fig:b435} we see that there are clear differences in the B$_{435}$ luminosity distributions of the two types of clusters.
We derived the probability density function of the absolute B$_{435}$ magnitudes in the three clusters
by carrying out a kernel density estimation \citep[KDE,][]{parzen}
on our samples. Again, the shape of the KDE estimation does not depend 
on the binning size, but purely on the data. 
We chose a Gaussian function as kernel and a bandwidth given by the optimization approximation of Silverman \citep{silverman}. 
We found that 47 Tuc has an unimodal distribution, reaching the 
peak at absolute magnitude M$_{B435}$ = 10. 

\begin{figure}
\centering
\includegraphics[width=8.5cm]{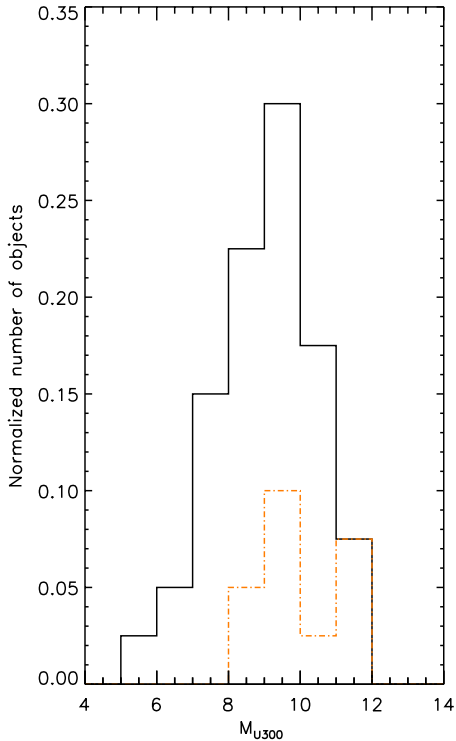}
\vspace{-1.3cm}
\caption{Normalized luminosity function in the filter U$_{300}$ for the CVs in 47 Tuc.
This cluster shows an unimodal distribution. Almost all the CVs discussed in this paper are included in this plot.
Exceptions are the counterparts to W100, W120 and W122. 
The first one has an ambiguous counterpart. The last two are outside of the NUV FOV. 
The integral under the solid line histogram is equal to 1. 
The dashed line histogram represents the CVs for which M$_{B435}$ was not obtained.}
\label{fig:300}
\end{figure}

\begin{figure}
\centering
\vspace{-0.6 cm}
\includegraphics[width=8.5cm]{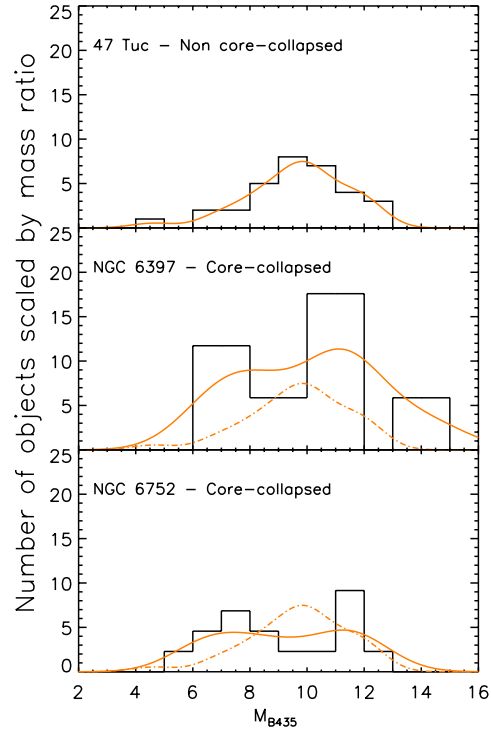}
\vspace{-0.6cm}
\caption{Mass scaled luminosity functions in the filter B$_{435}$ for the CVs in 47 Tuc, NGC 6397 and NGC 6752. 
The corresponding scale is the ratio between the mass of 47 Tuc 
and the mass of these clusters. In the case of 47 Tuc, only CVs for which we derived photometry in the B$_{435}$ band are plotted. An exception
is W100 which counterpart is ambiguous and so it was excluded. 
In each plot, the orange line is the probability density function
estimated by the KDE method using as kernel a Gaussian function.
The dashed curves in all three plots represent the probability density function of 47 Tuc. 
The distribution function of 47 Tuc is unimodal, while the ones of NGC 6397 and NGC 6752 are bimodal.
Note that core-collapsed clusters have a larger number of CVs with absolute magnitudes M$_{B435}\le8$ than 47 Tuc. }
\label{fig:bymass}
\end{figure}

In Figure \ref{fig:300} we show the normalized luminosity function for 47 Tuc in the filter U$_{300}$, which 
is more complete, since for some CVs the B$_{435}$ magnitude could not be obtained. 
We observe that in the filter U$_{300}$ the distribution is very similar to that in the filter B$_{435}$.
Since the shape of the distribution does not depend on the filter used, 
this suggests that 47 Tuc has an intrinsic unimodal distribution of CVs. 
On the other hand, the bimodal luminosity functions of NGC 6397 and NGC 6752 suggest that 
the core-collapsed clusters actually harbor more bright CVs than the non core-collapsed GC 47 Tuc. 
The difference between the distributions for non core-collapsed and core-collapsed clusters becomes 
more evident when we scale the number of CVs in NGC 6397 and NGC 6752 by the ratio between the mass of 47 Tuc 
and the mass of these clusters (Figure \ref{fig:bymass}).

In all the figures discussed above we just excluded the counterpart to W100 since there 
is an ambiguity about which object is the correct counterpart (but even considering W100$_{UV-a}$ or W100$_{UV-b}$ the conclusions remain unchanged). 
Also, we noted that by excluding the counterparts to the X-ray sources W23, W24, W35, W75, W200, W202,
W256, W299, W302, W304 and W335, which are classified as confused X-ray sources and X-ray sources with
multiple potential counterparts (Section \ref{sec:multiple}), 
the general results above mentioned do not change.

\begin{figure}
\centering
\includegraphics[width=8.5cm]{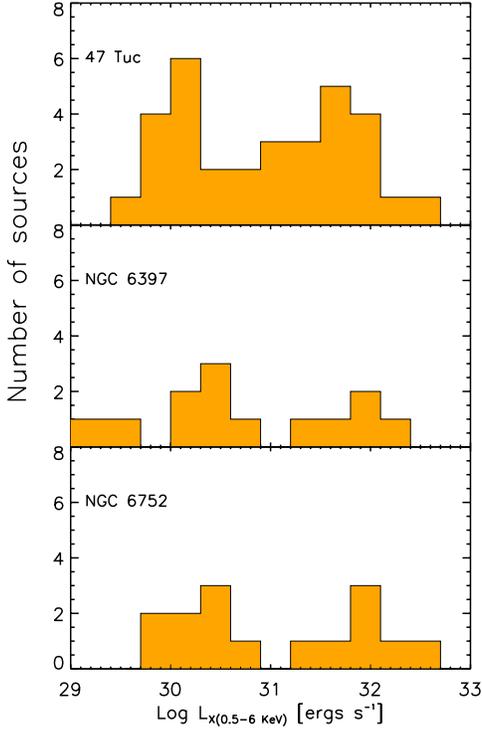}
\vspace{-0.6cm}
\caption{X-ray luminosity function for the CVs in the globular clusters 47 Tuc, NGC 6397 and NGC 6752.
All the clusters show a bimodal distribution. 
For 47 Tuc, only X-ray sources with one possible NUV/optical counterpart are plotted. 
The three clusters have an X-ray limiting luminosity of $\sim 10^{29}$ erg s$^{-1}$.}
\label{fig:histo_xray}
\end{figure}

For the three clusters, we also compared the X-ray luminosity distributions of the CVs in the band $0.5-6$ keV (Figure \ref{fig:histo_xray}). 
In order to avoid ambiguities in this comparison, 
we have removed eleven out of twelve X-ray sources that could be a blend of multiple nearby X-ray sources from the CV sample of 47 Tuc (see Section \ref{sec:multiple}). 
However, we have included W100 in this analysis because we consider that that X-ray source is the result of only one emitter
(if we remove it, the gap of the X-ray CV distribution of 47 Tuc in Figure \ref{fig:histo_xray} becomes slightly more visible). 
For this cluster we have used the X-ray luminosities reported by \heinke. 
In the case of NGC 6752 we have transformed the X-ray luminosities given by \citet{2014-forestell} in the band 0.5-8 keV to 
luminosities in the band 0.5-6 keV by multiplying for the appropriate factor (0.95) assuming a 2 keV bremsstrahlung model. 
For NGC 6397 we have converted the X-ray fluxes reported in \citet{2010-bogda} into X-ray luminosities assuming a 
distance of 2.2 kpc. In Figure \ref{fig:histo_xray} we show a histogram of X-ray luminosities 
of the CVs. We see that all the clusters present a bimodal distribution: 
a population of faint X-ray CVs ($L_X\lesssim 9 \times 10^{30}$ ergs s$^{-1}$) and another population of bright X-ray CVs ($L_X> 9 \times 10^{30}$ ergs s$^{-1}$). 
This could be explained by the mass transfer rates in CVs, the higher the mass transfer rate, the higher the X-ray luminosity. 
The gap observed in all the clusters could be related to the CV period gap, since objects below the period gap would have lower mass transfer rates compared to CVs above it \citep{2001-howell}.

\subsection{Spatial distribution}
\label{spatial}

\begin{figure}
\centering
\vspace{-1cm}
\includegraphics[width=8.5cm]{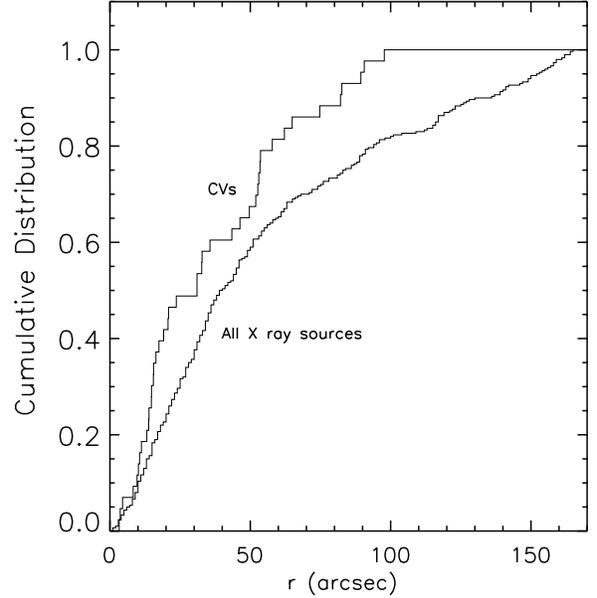}
\vspace{-1cm}
\caption{Cumulative radial distribution of the CVs and all the X-ray sources in the catalog of \heinke. 
All the CVs detected are within $100\arcsec$, where $\sim80\%$ of all the X-ray sources detected by \heinke \ reside.
X-ray sources with $r > 100\arcsec$ are more consistent with background objects.}
\label{fig:all_xray}
\end{figure}

\begin{figure}
\centering
\includegraphics[width=8.5cm]{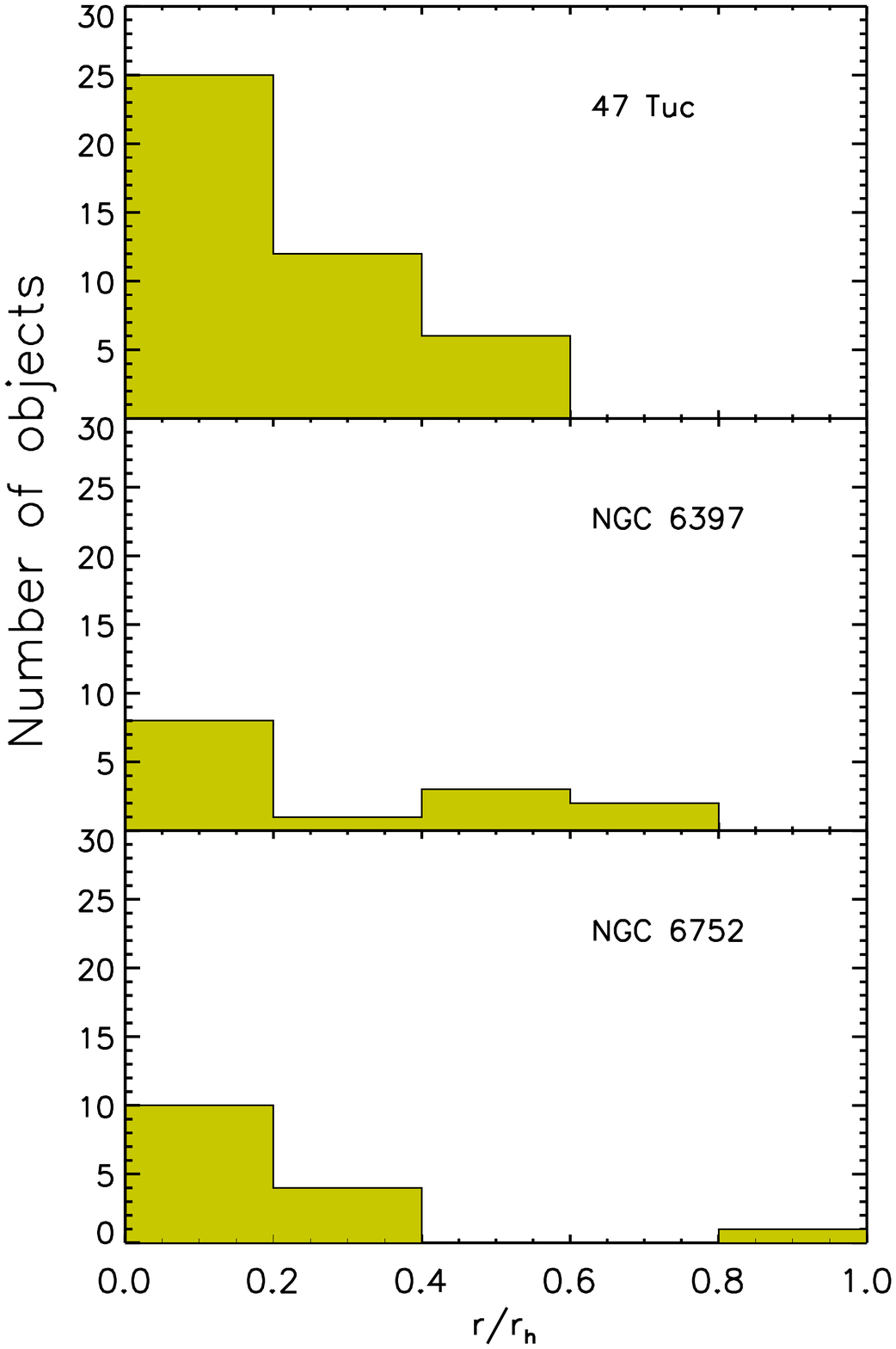}
\vspace{-0.6cm}
\caption{Histogram of the normalized radial distance of CV identifications in the globular clusters 47 Tuc, NGC 6397 and NGC 6752. 
The last two clusters show a second group at larger radii. 
For each cluster, the projected radial distance (r) has been normalized by the corresponding half-mass radius (r$_h$).
The green color denotes the total number of objects per bin. }
\label{fig:histo_rh}
\end{figure}

Assuming that the center of 47 Tuc corresponds to that reported by \citet{2010-Golds}, 
which is R.A.$= 00$:$24$:$05.71$, Decl. $= -72$:$04$:$52.7$ in epoch J2000, we have calculated the 
projected radial distance to each CV in the cluster using that reference point. 
Results of the projected distance are given in column 4 of Table \ref{table:results}.  
In Figure \ref{fig:histo_rh} we have a histogram of the distribution of CVs with respect to the corresponding center of 47 Tuc, 
NGC 6752 and NGC 6397. The projected radial distributions have been normalized by the r$_h$ of each cluster. 
For 47 Tuc, NGC 6752 and NGC 6397 the respective r$_h$ values are 2$\farcm$79, 1$\farcm$91 and 2$\farcm$33.
Our NUV images of 47 Tuc cover a radial distance equivalent to $50\%$ of its r$_h$ (see Appendix \ref{stacked}). 
At that distance we cover around $80\%$ of the total number of X-ray sources detected within the r$_h$ (see Figure \ref{fig:all_xray}).
However, \heinke \ showed that at radial distances larger than $\sim100\arcsec$, most of the X-ray sources identified are consistent with background objects. 
This means that if we discard these sources, our NUV images cover $\sim95\%$ of the X-ray objects in 47 Tuc within r$_h$ (see Figure 8 of \heinke).

From Figure \ref{fig:histo_rh} we observe that apart from the different number of CVs detected in each cluster, 
the three clusters also show different radial distributions. 
In 47 Tuc, the number of CVs decreases gradually as we get far from the core. 
A different radial distribution is observed for the core-collapsed clusters NGC 6752 and NGC 6397 for which two groups
are distinguishable: a large group of CVs more concentrated in the cores of these clusters, and a small group at larger radii,
as has been already pointed out by \citet[][]{cohn} and \citet{2017Lugger}. 
We carried out a Kuiper test (which is as sensitive in the tails as at the median of the distribution; \citeauthor{kuiper} \citeyear{kuiper}) 
to know which is the probability that the radial distributions of NGC 6397 and NGC 6752 are drawn from
the same radial distribution as 47 Tuc. We obtained a probability of $7.2\%$ between 47 Tuc and NGC 6397, 
and a probability of $28\%$ between 47 Tuc and NGC 6752.
This relatively large percentage could be due to the small CV sample at $r/r_h> 0.4$ in NGC 6752.
The probability obtained between NGC 6397 and NGC 6752 is $74\%$.

\begin{figure}
\centering
\includegraphics[width=8.5cm]{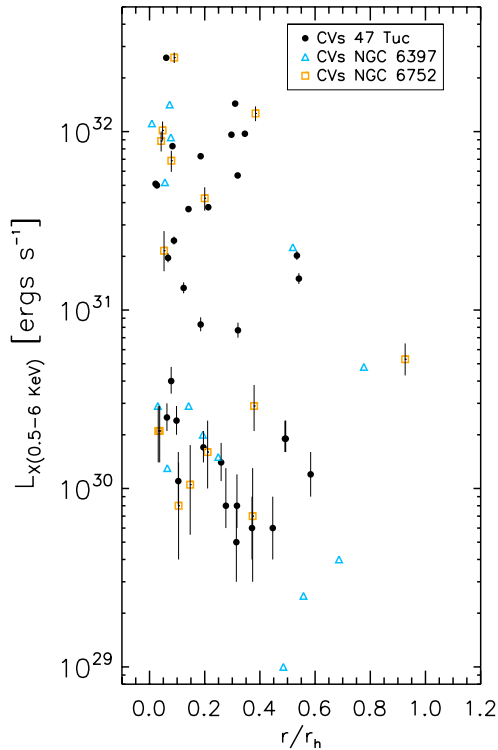}
\vspace{-1.3cm}
\caption{Normalized radial distribution vs X-ray luminosity for the CVs in the globular clusters 47 Tuc, NGC 6397 and NGC 6752.
For 47 Tuc, only X-ray sources (plus W100) that are not part of the X-ray sources with multiple NUV/optical counterparts are plotted. 
In all the clusters we see that at $r/r_h \sim 0.5$ the CVs have X-ray luminosities below  $2 \times 10^{31}$ erg s$^{-1}$.
X-ray luminosities for the CVs in NGC 6397 were derived from \citet{2010-bogda}. However,
errors are not specified by those authors. }
\label{fig:rh_xray}
\end{figure}

In Section \ref{properties} we showed that the CVs in the three clusters have a bimodal distribution in X-rays. 
In Figure \ref{fig:rh_xray} we show a plot of $L_{X\ (0.5-6 keV)}$ vs $r/r_h$ for the three clusters. 
We note that at $r/r_h \gtrsim 0.5$ most of the CVs in all the clusters have X-ray luminosities below $2 \times 10^{31}$ erg s$^{-1}$.
This can be explained by faint and old CVs that have been ejected to larger radii due to stellar interactions in the clusters.

For 47 Tuc, we also analyzed the radial distribution of the CVs following the
procedures described in \citet{cohn} and \citet{2017Lugger} for
fitting a parameterized surface-density profile to a main-sequence
turnoff (MSTO) sample and subsamples of the CV population. This
produces an estimate of the characteristic mass for all of the CVs
taken together, and for the bright - and faint - CV subsamples.
The KS2 photometry was used for this analysis. We
first determined the cluster center by iterative centroiding in a
$30\arcsec$ aperture, using a MSTO sample defined by $17.7 \leq
B_{390} \leq 20.2$. The resulting center coordinates of R.A. $= 00$:$24$:$05.76$, Decl.$= -72$:$04$:$52.6$ agree
well with the definitive center determination by \citet{2010-Golds},
differing by only 0\farcs2.

We next determined the cumulative radial distribution of a MSTO
sample that extends out to the nearest edge of the mosaic frame, which
is 80\arcsec\ from the cluster center. We fitted the
cumulative radial distribution of the MSTO group with a ``generalized
King model'', which we have also called a ``cored power law''.  As
discussed by \citet{cohn}, this takes the form,
\begin{equation}
\label{eqn:Cored_PL} 
S(r) = S_0 \left[1 + \left({r \over r_0}\right)^2 \right]^{\alpha/2},
\end{equation}
with the core radius related to the scale parameter $r_0$ by,
\begin{equation}
r_c = \left(2^{-2/\alpha} -1 \right)^{1/2} r_0\,.
\end{equation}

The result of the maximum-likelihood fit of this profile to the MSTO
sample is given in Table \ref{t:Cored_PL_fits}. We note that the
best-fit core radius value, $ 25\farcs1 \pm 0\farcs6$ agrees well
with the value of $24\farcs0 \pm 1\farcs9$ determined by
\citet{2000-howell}. The best-fit value of the slope parameter, $\alpha =
-1.6$, falls close to the value of $-1.8$ that we previously found to
give a good fit to the central profile of a \citet{king} model with
a concentration parameter of $c=2$ \citep{1995-Lugger}. This supports the
conclusion by \citet{2000-howell} that 47 Tuc has a large, normal core
and is not presently undergoing core collapse.

We next examined the radial profile of the CVs, taken both as a single
sample and as separate bright and faint samples. We adopted a separating
magnitude between the bright and faint samples of M$_{B390} =
9.0$. The limit was chosen to be consistent with the magnitude limits found observationally in NGC 6397 and NGC 6752.
The cumulative radial distributions of these samples are plotted in Figure \ref{fig:cum_msto}. 
From that figure we see that the population of CVs in the cluster is much more centrally
concentrated than the MSTO population, in agreement with numerical models of \citet{hong}.
This concentration of CVs in the cluster could be due either to their production in the core 
and/or to their higher average mass compared to other stars. 
However, the fact that the brighter CVs (which must be relatively young) are not substantially more centrally concentrated than the fainter CVs, 
indicates that the central concentration of CVs is caused more by their higher total masses than by their place of formation.
This may indicate that both the bright and faint CVs are old enough to have relaxed to their equilibrium radial distribution.
We then quantified the CV masses by carrying out maximum-likelihood fits of the cored-power-law
model to three samples (all, bright and faint CVs). As discussed by
\citet{cohn}, this procedure provides an estimate for the
characteristic mass of an object in each group relative to the MSTO
mass.  As in \citet{cohn}, we adopt the approximation that the mass
groups above the MSTO mass are in thermal equilibrium.  In this case,
the surface-density profile for a mass group with mass $m$ is given by
Eqn. \ref{eqn:Cored_PL}, with a slope parameter $\alpha$ related to the
turnoff-mass slope $\alpha_{\rm to}$ by,

\begin{equation}
\alpha = q\, (\alpha_{\rm to} - 1) + 1
\end{equation}
where $q = m/m_{\rm to}$, with $m_{\rm to}$ the average mass of the MSTO stars and $m$ is the mass of the binaries (in this case of the CVs). 

\begin{table*}
 \centering
  \caption{Cored-Power-Law model fits to 80\arcsec from the cluster center. the parameter N  is the number of stars in each sample within 80\arcsec\ of cluster center, 
$q$ is the mass ratio, $r_c$ is the core radius, $\alpha$ is the slope parameter, 
$m$ is the mass of the MSTO stars or of the CVs, $\sigma$ is the significance of mass excess above MSTO mass, 
and finally the parameter p is the K-S probability of consistency with the MSTO group.}
\label{t:Cored_PL_fits}
  \begin{tabular}{c c c c c c c c  }

Sample    &     N   &   q          &   r$_c$\ ($\arcsec$)            & $\alpha$             &     $m\ (M_{\sun})$  &  $\sigma$ & p  \\
\hline
MSTO       & 34358 &   1.0            & $25.1 \pm 0.6$ & $-1.58 \pm 0.04$ & $0.85 \pm 0.05$ & - & - \\
all CVs    &    36 &  $1.63 \pm 0.20$ & $15.6 \pm 1.5$ & $-3.21 \pm 0.51$ & $1.39 \pm 0.17$ & 3.0     & 0.067\% \\
bright CVs &    11 &  $1.70 \pm 0.36$ & $15.1 \pm 2.6$ & $-3.39 \pm 0.94$ & $1.45 \pm 0.31$ & 1.9     & 9.7\% \\
faint CVs  &    25 &  $1.60 \pm 0.24$ & $15.8 \pm 2.0$ & $-3.14 \pm 0.62$ & $1.36 \pm 0.20$ & 2.5     & 1.0\% \\
\end{tabular}
\end{table*}

\begin{figure}
\centering
\includegraphics[width=8cm]{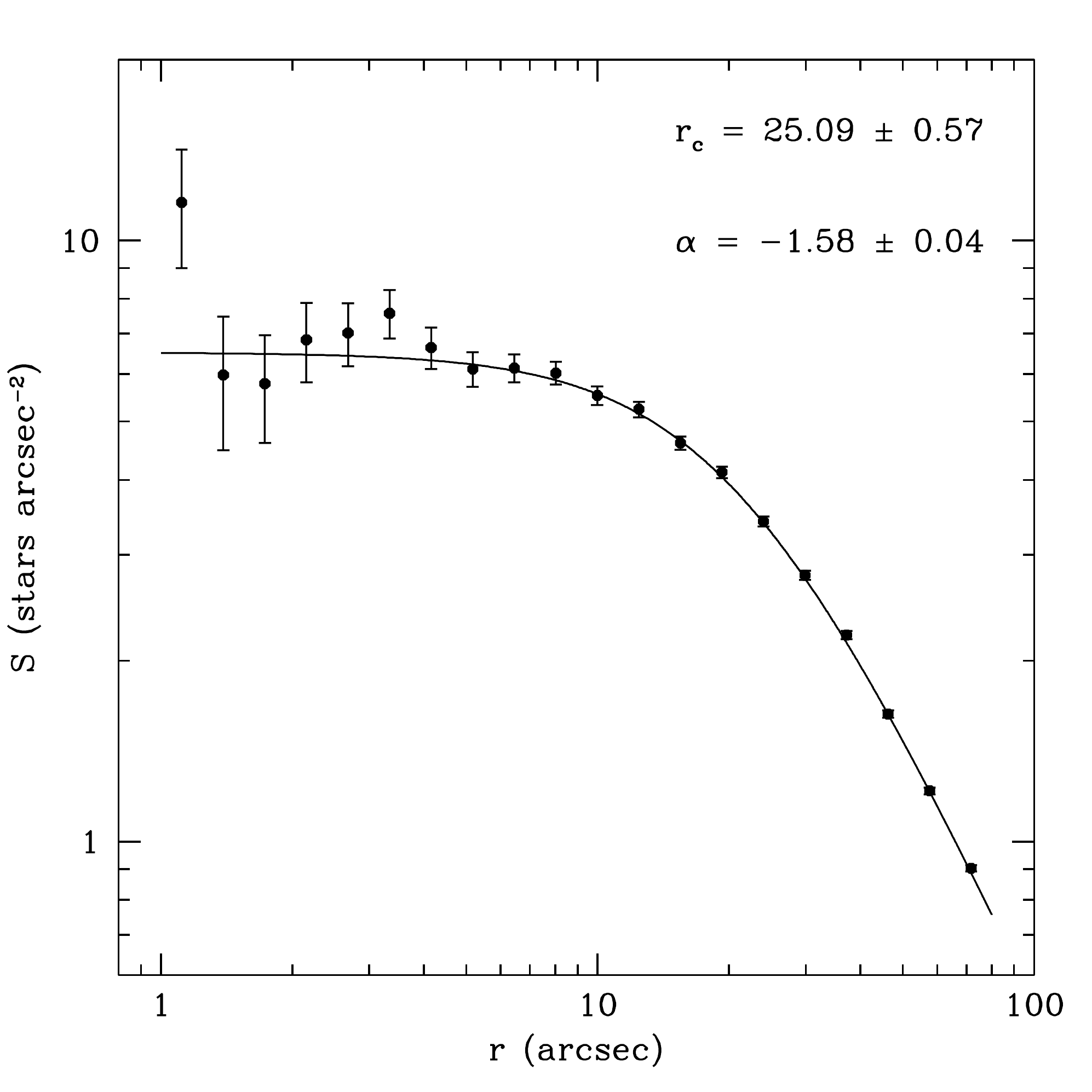}
\caption{Binned radial surface-density profile for MSTO group with cored-power-law fit. The model gives a good fit to all but the
innermost bin, which is the most subject to small-number statistics. }
\label{f:radial_profile}
\end{figure}

\begin{figure}
\centering
\includegraphics[width=8cm]{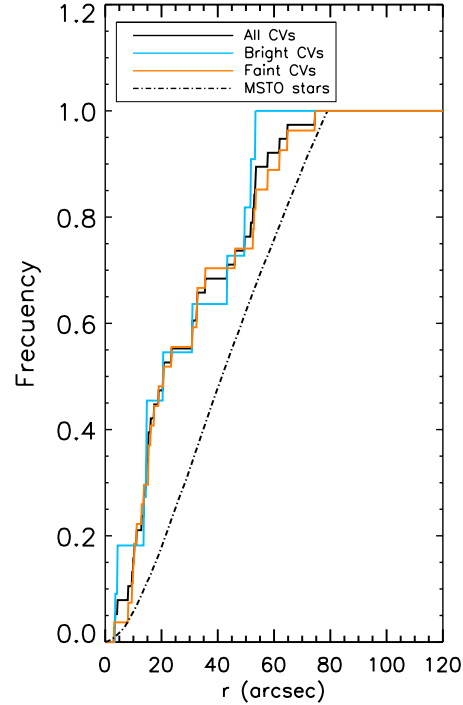}
\vspace{-0.8cm}
\caption{Cumulative radial distributions for selected stellar groups, within $80\arcsec$ of the cluster center. Note that the all-CV, bright-CV,
and faint-CV groups all show central concentration relative to the MSTO group. }
\label{fig:cum_msto}
\end{figure}

Table \ref{t:Cored_PL_fits} gives the results of maximum-likelihood
fits of Eqn. \ref{eqn:Cored_PL} to the samples formed by the
turnoff-mass stars, all CVs, bright CVs, and faint CVs. In all cases,
we only consider objects that lie within $80\arcsec$ of the cluster
center, to exclude incomplete annuli that extend into the corners of
the mosaic. Figure \ref{f:radial_profile} shows a binned radial profile
for the MSTO group with the cored-power-law fit. As can be seen from 
the table, the $q$ values for the three CV samples exceed unity, 
indicating that the characteristic masses indeed exceed the turnoff mass. 
We assume a MSTO mass of $0.85 \pm 0.05 \ M_{\sun}$, 
based on the study of \citet{2016-ferraro}. For the all-CV
and faint-CV samples, the excesses are significant at the
$\sim 3\sigma$ level, while for the smaller bright-CV sample, the
significance of the excess is nearly $2\sigma$.
As found in the analysis described above, there is no apparent
difference between the characteristic masses of the bright and faint
CVs. 

The inferred characteristic mass for the all-CV sample of $1.4 \pm 0.2
M_\odot$ is similar to what has been found in the core-collapsed
clusters NGC 6397 \citep{cohn} and NGC 6752 \citep{2017Lugger}. For 47 Tuc, the median absolute magnitude of the
bright CVs is about $M_R \approx 7.0$. Assuming that most of this flux
is contributed by the main sequence secondary, the mass of the
secondary is about $0.55\,M_\odot$, based on stellar evolutionary
models of \citet{1997_Bara} (this is consistent with the mass
derived for the object W23$_{UV}$ in Section \ref{sec:new_cvs}). This implies a WD
mass of $0.9 M_\odot$. We note that this is consistent with the mean
WD mass of $0.83 \pm 0.23 M_\odot$ obtained by \citet{2011_zoro}
for field CVs in the Sloan Digital Sky Survey (SDSS).  For the faint
CVs, in which the optical flux is dominated by the WD and
accretion disk, the secondary mass is likely to be $\lesssim 0.2
M_\odot$, assuming that the secondary is a main sequence star.  In
this case, the inferred typical WD mass is $\sim 1.2
M_\odot$. This suggests a scenario in which there is a net increase in
WD mass in CVs over the combined action of accretion and nova
outbursts, as has been investigated by \citet{2015_wijnen}. 
Alternatively, the high
system masses that we infer for the faint CVs in 47 Tuc may, at least
in part, represent a selection effect. If we are most sensitive to
the most X-ray bright sample of a larger population of faint CVs, then
we may be biased towards detecting the high-WD-mass end of the
population, since larger WD masses (and concomitantly smaller radii) 
lead to deeper potential wells that generate more X-rays for a given mass accretion rate.

To complement this analysis, we performed the Kolmogorov-Smirnov (K-S) comparison of the CV
samples to the MSTO sample, restricting the
samples to objects that are within 80\arcsec\ of the cluster
center. The resulting cumulative distributions for the present analysis
are shown in Figure \ref{fig:cum_msto} and the K-S
probability values are given in Table \ref{t:Cored_PL_fits}. The
results of this KS analysis are similar to the results of the profile
fitting. Both the all-CV and faint-CV samples are found to have
significantly more centrally concentrated distributions than that of
the MSTO sample. 

\subsection{The influence of stellar interactions}

Our results and comparisons suggest that core-collapsed GCs, in terms of CVs, behave optically different  
than 47 Tuc which is non core-collapsed. 
Of course, more identifications of CVs in both types of clusters are necessary in order to have better statistics. 
However, our comparison gives for the first time suggestions about the observational differences between the two kind of clusters.

For a long time it has been debated whether the core-collapse process in GCs affects the creation and destruction of CVs.
Observations of the core-collapsed clusters NGC 6397 and NGC 6752
are consistent with an evolutionary scenario in which dynamical interactions have a stronger effect in the creation of these 
binaries near the center of the cluster. 
This explains the brighter and more concentrated population of CVs.
When these CVs become old and faint they scatter to larger radii due to interactions with other stars in the cluster, in agreement with 
the observed less bright population \citep{cohn,2017Lugger}.

The population of CVs in 47 Tuc is dominated by faint systems, suggesting
that there are more primordial and/or old dynamically formed CVs in comparison to
NGC 6397 and NGC 6752. The bright CVs in 47 Tuc were probably created a few Gyrs ago, but clearly in less amount than in the core-collapsed clusters. 
This is consistent with evolutionary models of CVs which indicate that, since dynamically-formed CVs generally have relatively massive secondaries, 
brighter CVs have generally been produced more recently than fainter ones (e.g \citeauthor{2017-bellonib}, 2017b). 

\begin{figure}
\centering
\includegraphics[width=9cm, trim=1.5cm 1.0cm 0.0cm 5cm]{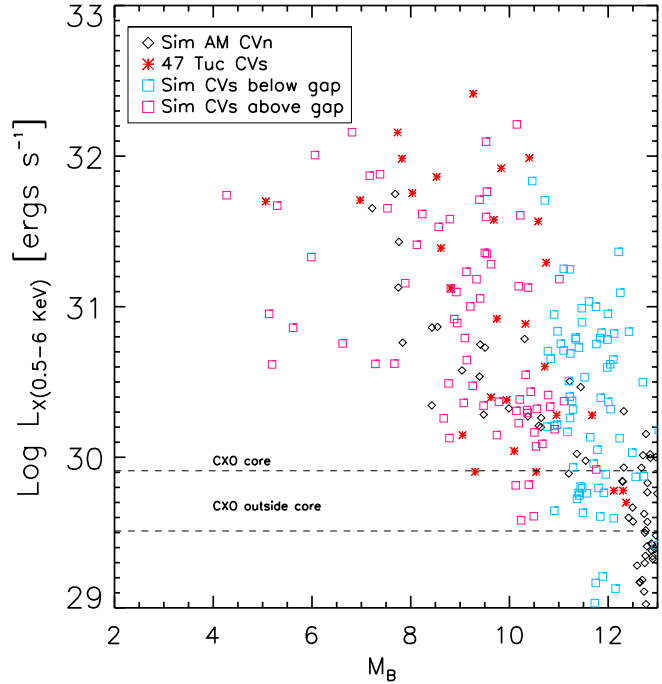}
\vspace{-0.5cm}
\caption{M$_B$ vs log L$_X$ for CVs and AM CVn systems (sim CVs and sim AM CVn) obtained with the models of \citet{2006-ivanova}.
For comparison, we have plotted the CVs discussed in this paper, excluding those mentioned in Section \ref{sec:multiple} and the counterparts to W100, W120, W122 and W229. The dashed lines indicate the depth of the $Chandra$ X-ray observations in 47 Tuc inside and outside the core.} 
\label{f:sim_cvs}
\end{figure}

According to the CV evolutionary models of \citet{2006-ivanova} that are relevant for 47 Tuc (results of their Table 3 multiplied by a factor of 1.3, considering a core mass for 47 Tuc of 6.5$\times 10^{4} \ M_{\sun}$), $\sim$276-287 CVs should currently\footnote{At an age of 11-12 Gyr, approximate age of 47 Tuc.} be present in the cluster. Of these CVs, 53-66 are predicted to be “detectable”, for which \citet{2006-ivanova} required that $L_X>3\times10^{30}$ erg s$^{-1}$, and the donor stellar luminosity $>0.06\ L_{\sun}$, such that the CV could be detected by both \textit{Chandra} and \textit{HST}. 
In Section \ref{sec:multiple} we mentioned that due to the high stellar crowding in 47 Tuc, 
some X-ray emitters can lie too close to each other and therefore remain unresolved by {\em Chandra}. 
\heinke \ estimated that around 10\% of the X-ray sources detected within the r$_h$ could be blends. 
This implies that a similar percentage of CVs detected through X-ray emission could be missing. 
It is also possible that some CVs could have landed on bright stars, thus being undetected. 
Additionally, our limiting magnitudes in X-rays ($3\times 10^{29}$ erg/s in the band 0.5-6 keV)
and in NUV/optical ($\sim27.5$ in NUV) prevent us to detect CVs that are fainter than these limits.
However, the predicted number of "detectable" CVs by \citet{2006-ivanova} is in good agreement with the 43 CVs that we have detected in 47 Tuc, 
though note that neither the X-ray nor the optical/UV flux can be accurately predicted for real (highly variable) CVs, 
so the \citet{2006-ivanova} model predictions should be taken as rough indicators \citep[Figure \ref{f:sim_cvs} shows a comparison between our detections and the models of][]{2006-ivanova}.
\citet{2006-ivanova} also found that at 11-12 Gyrs, 30-34\% of the CVs (corresponding to $\sim$15 of the identified CVs in 47 Tuc), 
were descendants of primordial binaries (though a significant fraction of these primordial CVs have been affected by encounters, e.g. by increasing their eccentricity).  
Also, 35-39\% of CVs were formed through binary exchange encounters, the largest channel for production of CVs. 
Binary collisions (13\%), tidal captures (3\%), and direct collisions of stars with red giants (15\%) offer alternative dynamical formation mechanisms for CVs in these models. 

\citet{2006-shara}, and \citeauthor{2016-belloni} (\citeyear{2016-belloni, 2017-belloni}; 2017b) also found that the creation of exchange binaries is more common in dense environments. 
These exchange interactions are likely to make CVs with MS star masses $<0.8$ M$_{\sun}$  (\citeauthor{2017-bellonib}, 2017b), which tend to be brighter and shorter-lived than CVs in non-cluster environments. \citet{2006-ivanova} pointed out that the formation of CV progenitor binaries in high-density environments happens a couple of Gyrs before they appear as CVs, and that immediately after a substantial increase in the density of the cluster core, some existing CVs are dynamically destroyed while CV progenitors that are just being created are not yet bright. Thus, the relatively bright CVs in each cluster are a tracer of the density of the cluster core a couple of Gyrs ago. The relatively high number of bright CVs in NGC 6397 and NGC 6752 with respect to 47 Tuc, scaled by mass, indicates that these clusters have higher production rates of CVs per unit mass than 47 Tuc, as might be expected given their higher core densities (though NGC 6752's core density is only slightly higher).
These clusters also have a relatively high number of bright CVs compared to 47 Tuc, when scaled by stellar encounter rate. 
This suggests that NGC 6397 and NGC 6752 have been at even higher densities at some time in the past couple Gyr, during which a large number of CVs with massive secondaries were generated, perhaps through episodes of deep core collapse \citep[see e.g.][]{2003-Frege, 2009-heggie, 2012Hurley}.  If so, we might expect that other dense clusters have been observed at times of maximum collapse, and show relatively few bright CVs for their stellar encounter rate. Alternatively, the difference could be due to other unmodeled dynamical or evolutionary effects. 
In contrast, the differences between the fainter CV populations are less strong, and in particular the faint CV populations in NGC 6752 and 47 Tuc are roughly comparable (when scaled by stellar encounter rate, or indeed by mass). The faint CV population likely reflects a combination of primordial CVs and CVs produced more than 1-2 Gyrs ago (as found in the models of \citeauthor{2017-bellonib}, 2017b)

Using Monte Carlo simulations, \citet{hong} studied the effect of stellar interactions on the CV population in GCs. 
They distinguished two groups of CVs, one group formed by primordial binaries (P), which have experienced some changes in their orbital parameters
but which keep the same initial components during the evolution of the cluster, and another group (E) in which the primordial binaries have suffered 
strong dynamical interactions leading to the replacement of one of the components. Group E has also more massive and younger companions than group P. 
These authors found that the group E is radially more concentrated than the group P. 
This is in agreement with findings for core-collapsed clusters like NGC 6397 and NGC 6752. 
However, in 47 Tuc we do not see such radial distribution (see Section \ref{spatial}). 
This suggests that there is not such a clear distinction between the older and younger CV populations 
in 47 Tuc as seen in the core-collapsed clusters, strengthening the case for a recent episode of close binary formation in the core-collapsed clusters.

We appear to see a higher ratio of older to younger CVs in 47 Tuc than in NGC 6752 and NGC 6397. 
This suggests that the core-collapse process may have formed many more young CVs in the last few Gyrs in NGC 6397 and NGC 6752. 
At the same time, that process may be more efficient in destroying\footnote{Among the `destruction' channels of CVs are unstable mass transfer, strong dynamical interactions 
and cluster escape \citep{2017-belloni}.} and/or ejecting older CVs and chromospherically active binaries than creating them, 
as core-collapsed clusters seem to have a deficiency of X-ray binaries overall \citep{cohn, 2013-bara},
which would explain the bimodal CV luminosity distributions and different radial distributions of the CVs in core-collapsed clusters.

\section{summary}
\label{suma}

We can summarize the findings of this paper as follows:

\begin{itemize}
\item{In this work we report the identification of 43 CVs and CV candidates in 47 Tuc using NUV and optical data from the {\em HST}.
All of them are {\em Chandra} X-ray sources.
Of these 43 systems, 22 are for the first time reported in this work. 
We found variability at a confidence level equal or higher than $95\%$ for 9 out of these 22 objects. 
This suggests that these systems are indeed the real counterparts to the corresponding X-ray sources. 
With 43 CVs and CV candidates, 47 Tuc is the GC with the most CVs identified so far.} 
\item We have made use of the first deep H$\alpha$ study of 47 Tuc with {\em HST} that uses near - simultaneous continuum - band photometry, 
which reduces the risk of obtaining misleading results since variability is accounted for.
\item All the objects classified as CVs and CV candidates in this paper have a blue color in the NUV and in the optical CMDs
due to emission from the accretor (the WD) and the accretion disk. 
\item For eighteen CVs out of the 43 CVs and CV candidates in the cluster we have identified H$\alpha$ emission likely coming from the accretion
disk or stream, and contributions from the companion star. Unfortunately, because of stellar crowding, faintness, presence of saturated stars, 
objects outside of the optical FOV and bleeding trails, H$\alpha$ photometry was not 
obtained for all the systems here reported. 
\item We have identified four CVs that have large amplitude variations (up to $\sim$ 3 magnitudes) in time-scales of hours in the filter U$_{300}$.
Flickering does not seem to be a plausible explanation for these variations. 
The dramatical changes of the eclipsing variable W2$_{UV}$ 
could be due to precession of the disk, or changes in the thickness of the accretion disk.
Alternatively they could be due to low and high state switches, like happens in magnetic CVs. 
\item For the first time we compared the CV population in 47 Tuc (a non core collapsed cluster), to those of NGC 6397 and NGC 6752 (core-collapsed clusters). 
\item We found that unlike the luminosity functions in these two core-collapsed clusters, the NUV and optical luminosity functions of CVs in 47 Tuc are unimodal.
\item 47 Tuc has many more faint CVs than those core-collapsed clusters. 
This suggests that the CV population in 47 Tuc is a combination of primordial CVs and dynamically CVs formed a long time ago
\item If we scale the masses of  NGC 6397 and NGC 6752  to that of 47 Tuc, 
the core-collapsed clusters have many more bright CVs.
This seems to indicate that the dynamical interactions play a major role in core-collapsed clusters to create CVs. 
\item We showed that 47 Tuc and the core-collapsed clusters NGC 6397 and NGC 6752 have a bimodal distribution of CVs in X-rays. 
This could be related to the orbital periods of the systems and the donor mass transfer rates (the higher the mass transfer, the higher the X-ray luminosity).
CVs below the period gap would have smaller transfer rates than CVs above the gap.
\item The CVs in 47 Tuc are more concentrated in the core of the cluster than the MSTO stars. 
Given that faint and bright CVs are equally segregated, 
it means that they are more massive. In fact we found that CVs with M$_{B390}\le 9$ could have
$M_{\rm WD} \sim 0.9 M_{\sun}$ with a MS secondary of mass $\sim 0.55 M_{\sun}$. At the same time CVs with M$_{B390} > 9$ 
could have secondary masses $\lesssim 0.2 M_{\sun}$ and  $M_{\rm WD} \sim 1.2 M_{\sun}$. 
These massive WDs in faint CVs suggest a scenario in which there is a net increase in mass over the combined action of accretion and nova
outbursts.
\item In this paper we also present the first AM CVn candidate in 47 Tuc (W62$_{UV}$). 
That object is blue in the NUV and optical. 
Its colors are consistent with those of confirmed AM CVns. 
It shows NUV variability at a confidence level higher than $99\%$.  
It is also an X-ray source with $L_{X (0.5-6 \ keV)} = 2.6 \times 10^{30}$ erg s$^{-1}$ and
it seems to have an EW(H$\alpha$) emission line of $\sim -3\pm6$ \AA, consistent with zero.
The V absolute magnitude of the system is consistent with that of an AM CVn with an orbital period of $\sim 6$ mins. 
All these characteristics would make it very similar to the known AM CVn system HM Cnc.
If confirmed, it would be one of the few AM CVn systems with a period shorter than 10 mins.
Also, It would help understanding the formation and evolution of AM CVn systems in dense stellar environments. 
\item The nature of W21$_{UV}$ is intriguing and uncertain since we have discarded possible explanations given its photometric behavior. 
\item Additionally, we discuss four counterparts to our X-ray sources that show NUV and optical colors similar to those 
found for MSP counterparts. The colors of three of these objects (W70$_{UV}$, W85$_{UV}$ and W253$_{UV}$) are consistent with having deep H$\alpha$ absorption lines, which are compatible with those of DA WDs. 
The other system (W34$_{UV}$) shows extremely strong heating (varying by $\sim3$ mags in the NUV)
which cannot be provided by the weak X-ray source, requiring irradiation by unobserved processes from an accretor star (e.g. particle heating from pulsar winds). 
The photometric results obtained for this object strongly suggest that this system could be an ablated (black widow or redback system) MSP companion. 
None of the four discussed objects have been reported as radio sources yet. 
\end{itemize}

\section*{Acknowledgments}

The authors appreciate the comments by the anonymous referee, which improved this paper. 
LERS acknowledges support from NOVA and a CONACyT (Mexico)
fellowship. COH acknowledges support from an NSERC Discovery Grant and a Discovery Accelerator Supplement.
The authors would like to thank Dr. Andrew Dolphin who kindly answered questions about the usage of the
DOLPHOT photometry package, as well to Dr. Paulo Freire for comments on the paper.

\onecolumn
\begin{landscape}
\begin{longtable}{c c c c c c c c c c  c c  c c}
\caption{Results for the CVs and CV candidates in 47 Tuc.}\\
W source &R.A. & Decl. & r ($\arcmin$)& $d_X$ & U$_{300}$ & B$_{390}$ & B$_{435}$ & H$\alpha$& R$_{625}$ & $L_{X (0.5-6 \ keV)}$  & Type & Notes    \\
counterpart & (J2000)  & (J2000)  & &$(\arcsec/\sigma_X)$ & & & & & (10$^{30}$ erg s$^{-1}$)  &\\
\hline
1                & 00:24:16.969  & -72:04:27.20   & 0.964 & 0.03/0.4     &  23.04(1) &  23.94(2)   & 20.25(1)   &  19.31(1)     &     19.99(1)                              & 97.3$^{+2.4}_{-2.2}$                       & CV  & p-v*  \\
2$^1$        & 00:24:15.887  & -72:04:36.38   & 0.828 & 0.009/0.1   & 21.356(4) &  21.357(4)   & 22.17(8)  & 20.44(4)       &   22.58(5)                            & 96.1$^{+2.4}_{-2.3}$                      & CV & p-v \\
8                & 00:24:10.753  & -72:04:25.71   & 0.594 & 0.03/0.3     &   22.048(7) &  23.22(1)   & 22.75(4)  & 20.75(3)  & 21.92(3)                                     & 37.7$^{+1.6}_{-1.5}$                   & CV  & p-v \\
15$^1$      & 00:24:08.479  & -72:05:00.31   & 0.247 & 0.005/0.06 &  21.666(6) &  22.147(8)   & 22.77(6) & 20.78(4)  & 21.73(4)                                    & 24.5$^{+1.3}_{-1.2}$                      & CV  & p-v \\
16              & 00:24:08.300  & -72:04:35.79   & 0.345 & 0.05/0.6     &  21.97(2) &  22.35(2)   & - & -     & -                                                                          &13.3$^{+1.0}_{-0.9}$                        & CV  & c-v \\
23$^1$      & 00:24:07.804  & -72:04:41.53   & 0.246 & 0.03/0.4     &  23.03(2) &  22.45(2)  & 22.27(2) & 20.11(2)  & 20.47(1)                                         & $\leq$41.8$^{+1.7}_{-1.6}$      & CV  & n-v \\
24              & 00:24:07.380  &  -72:04:49.65  & 0.138 & 0.06/0.7     &  22.94(1) &  24.32(4)  &  24.36(8)  &  -   & -                                                               & $\leq$31.8$^{+1.7}_{-1.6}$       & CV? & n \\
25              & 00:24:07.142  & -72:05:45.75   & 0.891 & 0.03/0.4     &  20.702(8) &  21.568(7)  &  20.44(3)  & 18.89(2) & 19.62(3)                                   & 56.8$^{+1.9}_{-1.7}$                  & CV   & p-v \\
27$^1$      & 00:24:06.373  & -72:04:43.00   & 0.169 & 0.02/0.3     &  20.943(4) &  22.80(1)  &  21.99(2)  & 20.20(3) &  21.01(5)                                    & 259.3$^{+4.2}_{-4.1}$          & CV   & p-v \\
30              & 00:24:06.002  & -72:04:56.16   & 0.061 & 0.03/0.4     &  19.74(4) &  20.51(1)  &  21.69(1)  & 20.10(3)  & 20.62(2)                                      & 51.0$^{+1.8}_{-1.7}$                   & CV   & p-v \\ 
33              & 00:24:05.381  & -72:04:21.67   & 0.517 & 0.15/1.6     &  22.392(8) &  23.28(1)  &  23.33(6)  & 21.24(3) &  22.21(4)                                     & 8.3$^{+0.8}_{-0.7}$                   & CV   & p-v \\
35              & 00:24:05.024  & -72:05:06.30   & 0.230 & 0.4/2.5       &  22.59(3) &  21.87(3)  &  20.96(5) & -    &  -                                                               & $\leq$1.0$^{+0.4}_{-0.2}$           & CV   & p-v\\
36              & 00:24:04.924  & -72:04:55.45   & 0.075 & 0.09/1.1     &  18.97(3) &  18.6(2)  &  18.12(1) & 16.66(1) & 16.98(1)                                           & 50.0$^{+2.2}_{-2.0}$                 & CV   & p-v \\
44              & 00:24:03.688  & -72:04:58.98   & 0.187 & 0.08/ 1.0    &  22.98(3) &  24.28(6)  &  - & -    & -                                                                          & 19.6$^{+1.2}_{-1.1}$                & CV   & p-v*\\
45              & 00:24:03.764  & -72:04:22.97   & 0.517 & 0.03/0.3     &   20.83(2) &  22.06(2)  &  22.06(3)  & 20.29(2) & 21.39(2)                                      & 72.8$^{+2.1}_{-2.0}$                  & CV   & p-v \\
49              & 00:24:03.102  & -72:04:47.34   & 0.219 & 0.07/0.2     &   23.54(2) &  24.25(3)  &  25.5(2)  &  -  &  -                                                             &4.0$^{+0.8}_{-0.6}$            & CV  & p-v \\
51              & 00:24:02.814  & -72:04:49.10   & 0.232  & 0.1/1.6     & 23.81(3)    & 23.37(2)  &  23.13(7)  & 20.65(5) & 21.41(6)                                   & 82.9 $^{+2.3}_{-2.2}$                   & CV  & p-v \\
53              & 00:24:02.533  & -72:05:11.26   & 0.394  &  0.03/0.3     &  24.49(3)  & 24.12(2)  &  23.52(4)  & 20.93(2) & -                                   & 36.8$^{+1.5}_{-1.4}$                    & CV   & p-v* \\
55              & 00:24:02.185  & -72:04:50.28   & 0.273  &  0.3/2.5       &  22.24(2) &  23.48(3)  &  24.04(6) & -    & -                                                          & 2.4$^{+0.5}_{-0.4}$                     & CV?   & c \\
56              & 00:24:02.125  &  -72:05:42.02  & 0.866  &  0.01/0.1     &  20.33(2) &  21.27(1) &  21.41(2)  &  19.55(3) & 20.56(1)                                   & 143.5$^{+2.9}_{-2.8}$                    & CV  & p-v \\
75             & 00:24:06.425    &  -72:04:52.35  & 0.055  & 0.3/3.7       &  21.391(5) &  23.09(2) &  23.54(9) & -  & -                                                          &$\leq$ 5.6$^{+0.8}_{-0.6}$               & CV?  & n \\
98             & 00:24:05.108  &  -72:05:02.92  & 0.176  &  0.1/1.3        &   22.75(1)  &  23.16(2)  &  -     & -    & -                                                                  & 2.5$^{+0.5}_{-0.4}$                          & CV? & n\\ 
100-a$^2$ & 00:24:01.318  & -72:04:26.99   & 0.545  & 0.25/1.5        & 24.62(3)  &  25.29(4)  &  26.3(2)     & -   & -                                  &1.7$^{+0.4}_{-0.3}$              & CV? & n\\
100-b$^2$ & 00:24:01.281  &  -72:04:27.11  & 0.545  & 0.18/ 1.0       & 25.76(7)  &  25.56(6)  &  -     & -    & -                                                                 &" "              & CV? & n\\
120            & 00:24:11.097   &  -72:06:20.03 &  1.514   &  -  & -           & -  & 23.40(1)  & 22.67(3)  & 23.07(2)                                               & 15.0$^{+1.1}_{-1.0}$                        & CV & p\\
122            &00:24:03.849   & -72:06:21.60	& 1.488    &  -  &-            & -  & 23.25(2)  & 21.48(1)  & 22.36(1)                                               & 20.2$^{+1.2}_{-1.1}$                      & CV & p\\
140            & 00:24:09.213  &  -72:05:43.87  & 0.894  &  0.04/0.4   &  22.62(1)  & 23.86(2) &  24.4(1)  &  -   & 24.2(3)                                                  & 7.7$^{+0.8}_{-0.7}$                     & CV   & c-v \\
187            & 00:23:56.897  &  -72:03:41.33  & 1.369  &  0.3/2.2     &  24.09(1) &  25.21(3)   & -  & - & -                                                                        & 1.9$^{+0.5}_{-0.3}$                      & CV? & n\\
200            & 00:24:07.125  &  -72:05:06.81  & 0.259  &  0.2/2.0     &   22.226(8) & 23.78(3)  & 23.37(5)  & -& 24.35(2)                                                  & $\leq$4.1$^{+0.6}_{-0.5}$             & CV & n-v?\\
202            & 00:24:04.946   &  -72:04:43.71  & 0.161  &  0.3/2.7    &   21.472(5) &  23.11(4)   & 23.16(5)     &  -   &  -                                                    &$\leq$ 3.6$^{+0.6}_{-0.5}$             & CV? & n\\
223            & 00:24:20.916   & -72:05:18.62   & 1.246  &  0.3/1.5    &   24.63(3) &  25.84(7)  & -      &  -   &  -                                                                & 0.6$^{+0.3}_{-0.2}$                     & CV? & n \\
229            & 00:24:18.511   & -72:03:34.79   & 1.629  &  0.1/0.8    &    25.6(2) &$ < 26.5$ & -      &  -   & -                                                                                  & 1.2$^{+0.4}_{-0.3}$                     & CV? & n \\    
235            & 00:24:15.986   & -72:04:12.52   & 1.036  &  0.3/1.4    &   24.59(3) &  25.65(4)  & 25.60(8)     & -  & -                                             & 0.6$^{+0.3}_{-0.2}$                        & CV? & n \\
256           & 00:24:10.047   & -72:04:26.64 & 0.547    &  0.6/1.7  &  24.20(2) & 25.23(5) & 25.6(1)   & - & -                                                               & $\leq$0.7$^{+0.4}_{-0.2}$              & CV & n-v? \\ 
292           & 00:24:04.972   & -72:03:30.14 & 1.377    &  0.3/1.7  &  23.55(2) & 24.49(2) &  - & - & -                                                                         & 1.9$^{+0.5}_{-0.3}$                        &  CV & n-v \\
295           & 00:24:04.657   & -72:05:09.46 & 0.291    &  0.09/ 0.6&  22.56(1) & 23.63(3)  &  23.7(1)     &  21.5(1) & 21.90(7)                                  & 1.1$^{+0.5}_{-0.3}$                        &  CV & n-v \\
299           & 00:24:04.136   & -72:05:12.32 & 0.348    &  0.1/ 1.2   & 23.72(6) & 24.3(1)  &  23.7(1)   & 21.59(8) & 22.06(9)                                    & $\leq$ 1.8$^{+0.5}_{-0.3}$              &  CV & n-v? \\
\\

\vspace{10cm}\\
\multicolumn{13}{c}{Continued}\\

W source &R.A. & Decl. & $r$ ($\arcmin$)& $d_X$ &U$_{300}$ & B$_{390}$ & B$_{435}$ & H$\alpha$& R$_{625}$ & $L_{X (0.5-6 \ keV)}$   & Type & Notes  \\
counterpart & (J2000)  & (J2000)  & &$(\arcsec/\sigma_X)$ & & & & & & (10$^{30}$ erg s$^{-1}$)  &\\
\hline
302    & 00:24:02.976   &  -72:04:43.80 & 0.258  &  0.3/0.8  & 23.07(1)  &  24.38(3) & 24.7(1)            &  -            &  -                                   & $\leq$  1.2$^{+0.4}_{-0.3}$            & CV?  & n \\
304    & 00:24:02.047   & -72:05:01.75 & 0.319   &  0.4/2.4  &  23.11(1) & 23.83(2)  &  25.3(3)   & -   &  -                                                      &$\leq$ 0.9$^{+0.4}_{-0.2}$              &  CV& n-v? \\  
309    & 00:24:01.070   & -72:05:30.48 & 0.724    & 0.3/1.9  & 23.32(1)   & 22.585(7)  & -  & - &-                                                                 & 1.4$^{+0.4}_{-0.3}$                      &  CV & n-v \\ 
311    & 00:24:00.631   & -72:04:12.72 & 0.772    & 0.3/1.7  &  23.29(2) & 24.08(9)  &  24.9(1)   &  -   & -                                                    & 0.8$^{+0.5}_{-0.2}$                          &  CV & n-v \\
324    & 00:23:57.691   & -72:04:14.27 & 0.889    & 0.2/1.1  &  21.882(6) & 22.84(1) & -   & - & -                               & 0.8$^{+0.4}_{-0.2}$                           &  CV?& n \\
329    & 00:23:55.025   & -72:04:34.22 & 0.878    & 0.5/2.2  &  25.02(6) & 25.9(1)  & - & -  & -                                                                     & 0.5$^{+0.3}_{-0.2}$                         &  CV?& n \\
335    & 00:23:52.053   & -72:04:37.11 & 1.082    & 0.4/2.4  &  24.30(2) & 24.94(3)   & 25.8(2)     &  - & -                                                    &$\leq$ 1.1$^{+0.4}_{-0.2}$               &  CV?& n \\
\hline
21   & 00:24:07.766  &  -72:05:27.28  & 0.597     & 0.06/0.7   &    21.381(7)  & 21.43(1)  &21.17(1)  & 20.29(2)  &      20.57(2)                               &14.2$^{+1.0}_{-0.9}$                  & ?  & r-v \\
62   & 00:23:59.340  &  -72:04:48.25  & 0.049     & 0.1/1.0     &  19.651(2)  & 19.71(3)  & 19.54(2) &   19.04(5) &   19.16(6)                                & 2.6$^{+0.5}_{-0.4}$                &  AM CVn?  & n-v \\

\hline
\label{table:results}
\vspace{-0.5cm}\\
\end{longtable}
NOTES.-  
Astrometric positions correspond to the counterparts associated to the respective W source. 
Units of right ascension are hours, minutes, and seconds, and units of declination are degrees, arcminutes, and arcseconds.
The photometric values are calibrated to the Vega-mag system. Photometric errors are given in parenthesis. 
The projected distance from the cluster center ($r$) was determined using the coordinates of the center of 47 
as given by \citet{2010-Golds}. The offsets between the X-ray and NUV positions ($d_X$)
were determined using the positions from \heinke. 
The X-ray luminosity at 4.85 kpc given in column 11 was taken from \heinke. 
These luminosity values reduce by 6\% using a more recent distance estimate of 4.69 kpc. Column 12 indicates the kind of
object identified. The letters in last column indicate if the object was confirmed in this work (c), 
reported for the first time as CV or CV candidate in this paper (n), a reclassified CV (r) or if it was previously classified by other authors (p). 
The letters v and v? means that the object shows variability at $99\%$ or $95\%$ confidence level, respectively.
The letter v* means that the object shows variability in the I or F300W band according to \edmondsb.\\
$^1$ Mean magnitudes.\\
$^2$ For this object the correct counterpart is ambiguous. \\
\end{landscape}
\twocolumn

\bibliographystyle{mn2e}
\bibliography{ref_cv}

\begin{thebibliography}{}

\bibitem[\protect\citeauthoryear{{Albrow}, {Gilliland}, {Brown}, {Edmonds},
  {Guhathakurta} \& {Sarajedini}}{{Albrow} et~al.}{2001}]{albrow01}
{Albrow} M.~D.,  {Gilliland} R.~L.,  {Brown} T.~M.,  {Edmonds} P.~D.,
  {Guhathakurta} P.,    {Sarajedini} A.,  2001, \apj, 559, 1060

\bibitem[\protect\citeauthoryear{{Augusteijn}, {van der Hooft}, {de Jong} \&
  {van Paradijs}}{{Augusteijn} et~al.}{1996}]{1996-augus}
{Augusteijn} T.,  {van der Hooft} F.,  {de Jong} J.~A.,    {van Paradijs} J.,
  1996, \aap, 311, 889

\bibitem[\protect\citeauthoryear{{Auriere}, {Koch-Miramond} \&
  {Ortolani}}{{Auriere} et~al.}{1989}]{1989ortolani}
{Auriere} M.,  {Koch-Miramond} L.,    {Ortolani} S.,  1989, \aap, 214, 113

\bibitem[\protect\citeauthoryear{{Bahramian}, {Heinke}, {Sivakoff} \&
  {Gladstone}}{{Bahramian} et~al.}{2013}]{2013-bara}
{Bahramian} A.,  {Heinke} C.~O.,  {Sivakoff} G.~R.,    {Gladstone} J.~C.,
  2013, \apj, 766, 136

\bibitem[\protect\citeauthoryear{{Baraffe}, {Chabrier}, {Allard} \&
  {Hauschildt}}{{Baraffe} et~al.}{1997}]{1997_Bara}
{Baraffe} I.,  {Chabrier} G.,  {Allard} F.,    {Hauschildt} P.~H.,  1997, \aap,
  327, 1054

\bibitem[\protect\citeauthoryear{{Bassa}, {van Kerkwijk}, {Koester} \&
  {Verbunt}}{{Bassa} et~al.}{2006}]{2006-bassa}
{Bassa} C.~G.,  {van Kerkwijk} M.~H.,  {Koester} D.,    {Verbunt} F.,  2006,
  \aap, 456, 295

\bibitem[\protect\citeauthoryear{{Beccari}, {De Marchi}, {Panagia} \&
  {Pasquini}}{{Beccari} et~al.}{2014}]{2014-beccari}
{Beccari} G.,  {De Marchi} G.,  {Panagia} N.,    {Pasquini} L.,  2014, \mnras,
  437, 2621

\bibitem[\protect\citeauthoryear{{Belloni}, {Giersz}, {Askar}, {Leigh} \&
  {Hypki}}{{Belloni} et~al.}{2016}]{2016-belloni}
{Belloni} D.,  {Giersz} M.,  {Askar} A.,  {Leigh} N.,    {Hypki} A.,  2016,
  \mnras, 462, 2950

\bibitem[\protect\citeauthoryear{{Belloni}, {Giersz}, {Rocha-Pinto}, {Leigh} \&
  {Askar}}{{Belloni} et~al.}{2017}]{2017-belloni}
{Belloni} D.,  {Giersz} M.,  {Rocha-Pinto} H.~J.,  {Leigh} N.~W.~C.,    {Askar}
  A.,  2017, \mnras, 464, 4077

\bibitem[\protect\citeauthoryear{{Belloni}, {Zorotovic}, {Schreiber}, {Leigh},
  {Giersz} \& {Askar}}{{Belloni} et~al.}{2017}]{2017-bellonib}
{Belloni} D.,  {Zorotovic} M.,  {Schreiber} M.~R.,  {Leigh} N.~W.~C.,  {Giersz}
  M.,    {Askar} A.,  2017, \mnras, 468, 2429

\bibitem[\protect\citeauthoryear{{Bildsten}, {Townsley}, {Deloye} \&
  {Nelemans}}{{Bildsten} et~al.}{2006}]{bil}
{Bildsten} L.,  {Townsley} D.~M.,  {Deloye} C.~J.,    {Nelemans} G.,  2006,
  \apj, 640, 466

\bibitem[\protect\citeauthoryear{{Bogdanov}, {Grindlay}, {Heinke}, {Camilo},
  {Freire} \& {Becker}}{{Bogdanov} et~al.}{2006}]{2006-bogda}
{Bogdanov} S.,  {Grindlay} J.~E.,  {Heinke} C.~O.,  {Camilo} F.,  {Freire}
  P.~C.~C.,    {Becker} W.,  2006, \apj, 646, 1104

\bibitem[\protect\citeauthoryear{{Bogdanov}, {van den Berg}, {Heinke}, {Cohn},
  {Lugger} \& {Grindlay}}{{Bogdanov} et~al.}{2010}]{2010-bogda}
{Bogdanov} S.,  {van den Berg} M.,  {Heinke} C.~O.,  {Cohn} H.~N.,  {Lugger}
  P.~M.,    {Grindlay} J.~E.,  2010, \apj, 709, 241

\bibitem[\protect\citeauthoryear{{Bressan}, {Marigo}, {Girardi}, {Salasnich},
  {Dal Cero}, {Rubele} \& {Nanni}}{{Bressan} et~al.}{2012}]{2012-bressan}
{Bressan} A.,  {Marigo} P.,  {Girardi} L.,  {Salasnich} B.,  {Dal Cero} C.,
  {Rubele} S.,    {Nanni} A.,  2012, \mnras, 427, 127

\bibitem[\protect\citeauthoryear{{Cadelano}, {Pallanca}, {Ferraro}, {Stairs},
  {Ransom}, {Dalessandro}, {Lanzoni}, {Hessels} \& {Freire}}{{Cadelano}
  et~al.}{2015}]{cadelano71}
{Cadelano} M.,  {Pallanca} C.,  {Ferraro} F.~R.,  {Stairs} I.,  {Ransom} S.~M.,
   {Dalessandro} E.,  {Lanzoni} B.,  {Hessels} J.~W.~T.,    {Freire} P.~C.~C.,
  2015, \apj, 807, 91

\bibitem[\protect\citeauthoryear{{Carter}, {Marsh}, {Steeghs}, {Groot},
  {Nelemans}, {Levitan}, {Rau}, {Copperwheat}, {Kupfer} \& {Roelofs}}{{Carter}
  et~al.}{2013}]{carter}
{Carter} P.~J.,  {Marsh} T.~R.,  {Steeghs} D.,  {Groot} P.~J.,  {Nelemans} G.,
  {Levitan} D.,  {Rau} A.,  {Copperwheat} C.~M.,  {Kupfer} T.,    {Roelofs}
  G.~H.~A.,  2013, \mnras, 429, 2143

\bibitem[\protect\citeauthoryear{{Cohn}, {Lugger}, {Couch}, {Anderson}, {Cool},
  {van den Berg}, {Bogdanov}, {Heinke} \& {Grindlay}}{{Cohn}
  et~al.}{2010}]{cohn}
{Cohn} H.~N.,  {Lugger} P.~M.,  {Couch} S.~M.,  {Anderson} J.,  {Cool} A.~M.,
  {van den Berg} M.,  {Bogdanov} S.,  {Heinke} C.~O.,    {Grindlay} J.~E.,
  2010, \apj, 722, 20

\bibitem[\protect\citeauthoryear{{Cool}, {Haggard}, {Arias}, {Brochmann},
  {Dorfman}, {Gafford}, {White} \& {Anderson}}{{Cool} et~al.}{2013}]{cool}
{Cool} A.~M.,  {Haggard} D.,  {Arias} T.,  {Brochmann} M.,  {Dorfman} J.,
  {Gafford} A.,  {White} V.,    {Anderson} J.,  2013, \apj, 763, 126

\bibitem[\protect\citeauthoryear{{Davies}}{{Davies}}{1997}]{1997-davies}
{Davies} M.~B.,  1997, \mnras, 288, 117

\bibitem[\protect\citeauthoryear{{Di Stefano} \& {Rappaport}}{{Di Stefano} \&
  {Rappaport}}{1994}]{1994-stefano}
{Di Stefano} R.,  {Rappaport} S.,  1994, \apj, 423, 274

\bibitem[\protect\citeauthoryear{{Dieball}, {Long}, {Knigge}, {Thomson} \&
  {Zurek}}{{Dieball} et~al.}{2010}]{2010-die}
{Dieball} A.,  {Long} K.~S.,  {Knigge} C.,  {Thomson} G.~S.,    {Zurek} D.~R.,
  2010, \apj, 710, 332

\bibitem[\protect\citeauthoryear{{Dolphin}}{{Dolphin}}{2000}]{dolphin}
{Dolphin} A.~E.,  2000, \pasp, 112, 1383

\bibitem[\protect\citeauthoryear{{Edmonds}, {Gilliland}, {Camilo}, {Heinke} \&
  {Grindlay}}{{Edmonds} et~al.}{2002}]{2002-Edmonds}
{Edmonds} P.~D.,  {Gilliland} R.~L.,  {Camilo} F.,  {Heinke} C.~O.,
  {Grindlay} J.~E.,  2002, \apj, 579, 741

\bibitem[\protect\citeauthoryear{{Edmonds}, {Gilliland}, {Heinke} \&
  {Grindlay}}{{Edmonds} et~al.}{2003a}]{2003-edmonds1}
{Edmonds} P.~D.,  {Gilliland} R.~L.,  {Heinke} C.~O.,    {Grindlay} J.~E.,
  2003a, \apj, 596, 1177 (E03a)

\bibitem[\protect\citeauthoryear{{Edmonds}, {Gilliland}, {Heinke} \&
  {Grindlay}}{{Edmonds} et~al.}{2003b}]{2003-edmonds2}
{Edmonds} P.~D.,  {Gilliland} R.~L.,  {Heinke} C.~O.,    {Grindlay} J.~E.,
  2003b, \apj, 596, 1197 (E03b)

\bibitem[\protect\citeauthoryear{{Ferraro}, {Lapenna}, {Mucciarelli},
  {Lanzoni}, {Dalessandro}, {Pallanca} \& {Massari}}{{Ferraro}
  et~al.}{2016}]{2016-ferraro}
{Ferraro} F.~R.,  {Lapenna} E.,  {Mucciarelli} A.,  {Lanzoni} B.,
  {Dalessandro} E.,  {Pallanca} C.,    {Massari} D.,  2016, \apj, 816, 70

\bibitem[\protect\citeauthoryear{{Forestell}, {Heinke}, {Cohn}, {Lugger},
  {Sivakoff}, {Bogdanov}, {Cool} \& {Anderson}}{{Forestell}
  et~al.}{2014}]{2014-forestell}
{Forestell} L.~M.,  {Heinke} C.~O.,  {Cohn} H.~N.,  {Lugger} P.~M.,  {Sivakoff}
  G.~R.,  {Bogdanov} S.,  {Cool} A.~M.,    {Anderson} J.,  2014, \mnras, 441,
  757

\bibitem[\protect\citeauthoryear{{Fregeau}, {G{\"u}rkan}, {Joshi} \&
  {Rasio}}{{Fregeau} et~al.}{2003}]{2003-Frege}
{Fregeau} J.~M.,  {G{\"u}rkan} M.~A.,  {Joshi} K.~J.,    {Rasio} F.~A.,  2003,
  \apj, 593, 772

\bibitem[\protect\citeauthoryear{{G{\"a}nsicke}, {Dillon}, {Southworth},
  {Thorstensen}, {Rodr{\'i}guez-Gil}, {Aungwerojwit}, {Marsh}, {Szkody},
  {Barros}, {Casares}, {de Martino}, {Groot}, {Hakala}, {Kolb}, {Littlefair} \&
  et al.}{{G{\"a}nsicke} et~al.}{2009}]{2009-gan}
{G{\"a}nsicke} B.~T.,  {Dillon} M.,  {Southworth} J.,  {Thorstensen} J.~R.,
  {Rodr{\'i}guez-Gil} P.,  {Aungwerojwit} A.,  {Marsh} T.~R.,  {Szkody} P.,
  {Barros} S.~C.~C.,  {Casares} J.,  {de Martino} D.,  {Groot} P.~J.,  {Hakala}
  P.,  {Kolb} U.,  {Littlefair} S.~P.,    et al. 2009, \mnras, 397, 2170

\bibitem[\protect\citeauthoryear{{Goldsbury}, {Richer}, {Anderson}, {Dotter},
  {Sarajedini} \& {Woodley}}{{Goldsbury} et~al.}{2010}]{2010-Golds}
{Goldsbury} R.,  {Richer} H.~B.,  {Anderson} J.,  {Dotter} A.,  {Sarajedini}
  A.,    {Woodley} K.,  2010, \aj, 140, 1830

\bibitem[\protect\citeauthoryear{{Grindlay}, {Cool}, {Callanan}, {Bailyn},
  {Cohn} \& {Lugger}}{{Grindlay} et~al.}{1995}]{1995-grin}
{Grindlay} J.~E.,  {Cool} A.~M.,  {Callanan} P.~J.,  {Bailyn} C.~D.,  {Cohn}
  H.~N.,    {Lugger} P.~M.,  1995, \apjl, 455, L47

\bibitem[\protect\citeauthoryear{{Grindlay}, {Heinke}, {Edmonds} \&
  {Murray}}{{Grindlay} et~al.}{2001}]{2001-grindlay}
{Grindlay} J.~E.,  {Heinke} C.,  {Edmonds} P.,    {Murray} S.~S.,  2001,
  Science, 292, 2290

\bibitem[\protect\citeauthoryear{{Grindlay}, {Heinke}, {Edmonds}, {Murray} \&
  {Cool}}{{Grindlay} et~al.}{2001}]{2001-grinb}
{Grindlay} J.~E.,  {Heinke} C.~O.,  {Edmonds} P.~D.,  {Murray} S.~S.,    {Cool}
  A.~M.,  2001, \apjl, 563, L53

\bibitem[\protect\citeauthoryear{{Harlaftis} \& {Horne}}{{Harlaftis} \&
  {Horne}}{1999}]{1999emilios}
{Harlaftis} E.~T.,  {Horne} K.,  1999, \mnras, 305, 437

\bibitem[\protect\citeauthoryear{{Harris}}{{Harris}}{1996}]{Harris}
{Harris} W.~E.,  1996, \aj, 112, 1487

\bibitem[\protect\citeauthoryear{{Heggie} \& {Giersz}}{{Heggie} \&
  {Giersz}}{2009}]{2009-heggie}
{Heggie} D.~C.,  {Giersz} M.,  2009, \mnras, 397, L46

\bibitem[\protect\citeauthoryear{{Heinke}, {Grindlay}, {Edmonds}, {Cohn},
  {Lugger}, {Camilo}, {Bogdanov} \& {Freire}}{{Heinke} et~al.}{2005}]{heinke}
{Heinke} C.~O.,  {Grindlay} J.~E.,  {Edmonds} P.~D.,  {Cohn} H.~N.,  {Lugger}
  P.~M.,  {Camilo} F.,  {Bogdanov} S.,    {Freire} P.~C.,  2005, \apj, 625, 796
  (H05)

\bibitem[\protect\citeauthoryear{{Heyl}, {Richer}, {Anderson}, {Fahlman},
  {Dotter}, {Hurley}, {Kalirai}, {Rich}, {Shara}, {Stetson}, {Woodley} \&
  {Zurek}}{{Heyl} et~al.}{2012}]{Heyl}
{Heyl} J.~S.,  {Richer} H.,  {Anderson} J.,  {Fahlman} G.,  {Dotter} A.,
  {Hurley} J.,  {Kalirai} J.,  {Rich} R.~M.,  {Shara} M.,  {Stetson} P.,
  {Woodley} K.~A.,    {Zurek} D.,  2012, \apj, 761, 51

\bibitem[\protect\citeauthoryear{{Hong}, {Vesperini}, {Belloni} \&
  {Giersz}}{{Hong} et~al.}{2017}]{hong}
{Hong} J.,  {Vesperini} E.,  {Belloni} D.,    {Giersz} M.,  2017, \mnras, 464,
  2511

\bibitem[\protect\citeauthoryear{{Howell}, {Guhathakurta} \&
  {Gilliland}}{{Howell} et~al.}{2000}]{2000-howell}
{Howell} J.~H.,  {Guhathakurta} P.,    {Gilliland} R.~L.,  2000, \pasp, 112,
  1200

\bibitem[\protect\citeauthoryear{{Howell}, {Nelson} \& {Rappaport}}{{Howell}
  et~al.}{2001}]{2001-howell}
{Howell} S.~B.,  {Nelson} L.~A.,    {Rappaport} S.,  2001, \apj, 550, 897

\bibitem[\protect\citeauthoryear{{Huang}, {Becker}, {Edmonds}, {Elsner},
  {Heinke} \& {Hsieh}}{{Huang} et~al.}{2010}]{2010-huang}
{Huang} R.~H.~H.,  {Becker} W.,  {Edmonds} P.~D.,  {Elsner} R.~F.,  {Heinke}
  C.~O.,    {Hsieh} B.~C.,  2010, \aap, 513, A16

\bibitem[\protect\citeauthoryear{{Hurley} \& {Shara}}{{Hurley} \&
  {Shara}}{2012}]{2012Hurley}
{Hurley} J.~R.,  {Shara} M.~M.,  2012, \mnras, 425, 2872

\bibitem[\protect\citeauthoryear{{Israel}, {Esposito}, {Rodr{\'{\i}}guez
  Castillo} \& {Sidoli}}{{Israel} et~al.}{2016}]{2016-israel}
{Israel} G.~L.,  {Esposito} P.,  {Rodr{\'{\i}}guez Castillo} G.~A.,    {Sidoli}
  L.,  2016, \mnras, 462, 4371

\bibitem[\protect\citeauthoryear{{Israel}, {Hummel}, {Covino}, {Campana},
  {Appenzeller}, {G{\"a}ssler}, {Mantel}, {Marconi}, {Mauche}, {Munari},
  {Negueruela}, {Nicklas}, {Rupprecht}, {Smart}, {Stahl} \& {Stella}}{{Israel}
  et~al.}{2002}]{2002-israel}
{Israel} G.~L.,  {Hummel} W.,  {Covino} S.,  {Campana} S.,  {Appenzeller} I.,
  {G{\"a}ssler} W.,  {Mantel} K.-H.,  {Marconi} G.,  {Mauche} C.~W.,  {Munari}
  U.,  {Negueruela} I.,  {Nicklas} H.,  {Rupprecht} G.,  {Smart} R.~L.,
  {Stahl} O.,    {Stella} L.,  2002, \aap, 386, L13

\bibitem[\protect\citeauthoryear{{Ivanova}, {Heinke}, {Rasio}, {Taam},
  {Belczynski} \& {Fregeau}}{{Ivanova} et~al.}{2006}]{2006-ivanova}
{Ivanova} N.,  {Heinke} C.~O.,  {Rasio} F.~A.,  {Taam} R.~E.,  {Belczynski} K.,
     {Fregeau} J.,  2006, \mnras, 372, 1043

\bibitem[\protect\citeauthoryear{{Kimmig}, {Seth}, {Ivans}, {Strader},
  {Caldwell}, {Anderton} \& {Gregersen}}{{Kimmig} et~al.}{2015}]{kimming}
{Kimmig} B.,  {Seth} A.,  {Ivans} I.~I.,  {Strader} J.,  {Caldwell} N.,
  {Anderton} T.,    {Gregersen} D.,  2015, \aj, 149, 53

\bibitem[\protect\citeauthoryear{{King}}{{King}}{1966}]{king}
{King} I.~R.,  1966, \aj, 71, 64

\bibitem[\protect\citeauthoryear{{Knigge}, {Baraffe} \& {Patterson}}{{Knigge}
  et~al.}{2011}]{2011-knigge}
{Knigge} C.,  {Baraffe} I.,    {Patterson} J.,  2011, \apjs, 194, 28

\bibitem[\protect\citeauthoryear{{Knigge}, {Dieball}, {Ma{\'{\i}}z
  Apell{\'a}niz}, {Long}, {Zurek} \& {Shara}}{{Knigge} et~al.}{2008}]{knigge08}
{Knigge} C.,  {Dieball} A.,  {Ma{\'{\i}}z Apell{\'a}niz} J.,  {Long} K.~S.,
  {Zurek} D.~R.,    {Shara} M.~M.,  2008, \apj, 683, 1006

\bibitem[\protect\citeauthoryear{{Knigge}, {Zurek}, {Shara} \& {Long}}{{Knigge}
  et~al.}{2002}]{knigge02-1}
{Knigge} C.,  {Zurek} D.~R.,  {Shara} M.~M.,    {Long} K.~S.,  2002, \apj, 579,
  752

\bibitem[\protect\citeauthoryear{{Knigge}, {Zurek}, {Shara}, {Long} \&
  {Gilliland}}{{Knigge} et~al.}{2003}]{2003-knigge}
{Knigge} C.,  {Zurek} D.~R.,  {Shara} M.~M.,  {Long} K.~S.,    {Gilliland}
  R.~L.,  2003, \apj, 599, 1320

\bibitem[\protect\citeauthoryear{{Kuiper}}{{Kuiper}}{1960}]{kuiper}
{Kuiper} N.~H.,  1960, Proc. of the Koninkl. Nederl. Akad. van Wetenschappen,
  Ser. A., 63, 38

\bibitem[\protect\citeauthoryear{{Lu}, {Kong}, {Verbunt}, {Lewin}, {Anderson}
  \& {Pooley}}{{Lu} et~al.}{2011}]{2011-lu}
{Lu} T.-N.,  {Kong} A.~K.~H.,  {Verbunt} F.,  {Lewin} W.~H.~G.,  {Anderson}
  S.~F.,    {Pooley} D.,  2011, \apj, 736, 158

\bibitem[\protect\citeauthoryear{{Lugger}, {Cohn}, {Cool}, {Heinke} \&
  {Anderson}}{{Lugger} et~al.}{2017}]{2017Lugger}
{Lugger} P.~M.,  {Cohn} H.~N.,  {Cool} A.~M.,  {Heinke} C.~O.,    {Anderson}
  J.,  2017, \apj, 841, 53

\bibitem[\protect\citeauthoryear{{Lugger}, {Cohn} \& {Grindlay}}{{Lugger}
  et~al.}{1995}]{1995-Lugger}
{Lugger} P.~M.,  {Cohn} H.~N.,    {Grindlay} J.~E.,  1995, \apj, 439, 191

\bibitem[\protect\citeauthoryear{{Lugger}, {Cohn}, {Heinke}, {Grindlay} \&
  {Edmonds}}{{Lugger} et~al.}{2007}]{2007-lugger}
{Lugger} P.~M.,  {Cohn} H.~N.,  {Heinke} C.~O.,  {Grindlay} J.~E.,    {Edmonds}
  P.~D.,  2007, \apj, 657, 286

\bibitem[\protect\citeauthoryear{{Majewski}, {Patterson}, {Dinescu}, {Johnson},
  {Ostheimer}, {Kunkel} \& {Palma}}{{Majewski} et~al.}{2000}]{2000-omega}
{Majewski} S.~R.,  {Patterson} R.~J.,  {Dinescu} D.~I.,  {Johnson} W.~Y.,
  {Ostheimer} J.~C.,  {Kunkel} W.~E.,    {Palma} C.,  2000, in {Noels} A.,
  {Magain} P.,  {Caro} D.,  {Jehin} E.,  {Parmentier} G.,   {Thoul} A.~A.,
  eds, Liege International Astrophysical Colloquia Vol.~35 of Liege
  International Astrophysical Colloquia, {{$\omega$} Centauri : Nucleus of a
  milky way dwarf spheroidal?}.
p.~619

\bibitem[\protect\citeauthoryear{{Maxwell}, {Lugger}, {Cohn}, {Heinke},
  {Grindlay}, {Budac}, {Drukier} \& {Bailyn}}{{Maxwell}
  et~al.}{2012}]{2012-max}
{Maxwell} J.~E.,  {Lugger} P.~M.,  {Cohn} H.~N.,  {Heinke} C.~O.,  {Grindlay}
  J.~E.,  {Budac} S.~A.,  {Drukier} G.~A.,    {Bailyn} C.~D.,  2012, \apj, 756,
  147

\bibitem[\protect\citeauthoryear{{Miller-Jones}, {Strader}, {Heinke},
  {Maccarone}, {van den Berg}, {Knigge}, {Chomiuk}, {Noyola}, {Russell}, {Seth}
  \& {Sivakoff}}{{Miller-Jones} et~al.}{2015}]{2015-x9}
{Miller-Jones} J.~C.~A.,  {Strader} J.,  {Heinke} C.~O.,  {Maccarone} T.~J.,
  {van den Berg} M.,  {Knigge} C.,  {Chomiuk} L.,  {Noyola} E.,  {Russell}
  T.~D.,  {Seth} A.~C.,    {Sivakoff} G.~R.,  2015, \mnras, 453, 3918

\bibitem[\protect\citeauthoryear{{Mukai}}{{Mukai}}{2017}]{2017Mukai}
{Mukai} K.,  2017, \pasp, 129, 062001

\bibitem[\protect\citeauthoryear{{Nelemans}}{{Nelemans}}{2005}]{2005-nelemans}
{Nelemans} G.,  2005, in {Hameury} J.-M.,  {Lasota} J.-P.,  eds, The
  Astrophysics of Cataclysmic Variables and Related Objects Vol.~330 of
  Astronomical Society of the Pacific Conference Series, {AM CVn stars}.
p.~27

\bibitem[\protect\citeauthoryear{{Nelson}, {Rappaport} \& {Joss}}{{Nelson}
  et~al.}{1986}]{1986Nelson}
{Nelson} L.~A.,  {Rappaport} S.~A.,    {Joss} P.~C.,  1986, \apj, 304, 231

\bibitem[\protect\citeauthoryear{{Nice} \& {Thorsett}}{{Nice} \&
  {Thorsett}}{1992}]{1992-nice}
{Nice} D.~J.,  {Thorsett} S.~E.,  1992, \apj, 397, 249

\bibitem[\protect\citeauthoryear{{Noyola}, {Gebhardt} \& {Bergmann}}{{Noyola}
  et~al.}{2008}]{2008-eva}
{Noyola} E.,  {Gebhardt} K.,    {Bergmann} M.,  2008, \apj, 676, 1008

\bibitem[\protect\citeauthoryear{{Pan}, {Hobbs}, {Li}, {Ridolfi}, {Wang} \&
  {Freire}}{{Pan} et~al.}{2016}]{2016-pan}
{Pan} Z.,  {Hobbs} G.,  {Li} D.,  {Ridolfi} A.,  {Wang} P.,    {Freire} P.,
  2016, \mnras, 459, L26

\bibitem[\protect\citeauthoryear{{Paresce} \& {de Marchi}}{{Paresce} \& {de
  Marchi}}{1994}]{1994-Pare}
{Paresce} F.,  {de Marchi} G.,  1994, \apjl, 427, L33

\bibitem[\protect\citeauthoryear{{Paresce}, {de Marchi} \& {Ferraro}}{{Paresce}
  et~al.}{1992}]{1992-pare}
{Paresce} F.,  {de Marchi} G.,    {Ferraro} F.~R.,  1992, \nat, 360, 46

\bibitem[\protect\citeauthoryear{{Parzen}}{{Parzen}}{1962}]{parzen}
{Parzen} E.,  1962, Ann. Math. Statist., 33, 1065

\bibitem[\protect\citeauthoryear{{Pooley}, {Lewin}, {Homer}, {Verbunt},
  {Anderson}, {Gaensler}, {Margon}, {Miller}, {Fox}, {Kaspi} \& {van der
  Klis}}{{Pooley} et~al.}{2002}]{2002-pooley}
{Pooley} D.,  {Lewin} W.~H.~G.,  {Homer} L.,  {Verbunt} F.,  {Anderson} S.~F.,
  {Gaensler} B.~M.,  {Margon} B.,  {Miller} J.~M.,  {Fox} D.~W.,  {Kaspi}
  V.~M.,    {van der Klis} M.,  2002, \apj, 569, 405

\bibitem[\protect\citeauthoryear{{Pretorius}, {Knigge} \&
  {Schwope}}{{Pretorius} et~al.}{2013}]{2013-Preto}
{Pretorius} M.~L.,  {Knigge} C.,    {Schwope} A.~D.,  2013, \mnras, 432, 570

\bibitem[\protect\citeauthoryear{{Ramsay}, {Groot}, {Marsh}, {Nelemans},
  {Steeghs} \& {Hakala}}{{Ramsay} et~al.}{2006}]{2006-ramsay}
{Ramsay} G.,  {Groot} P.~J.,  {Marsh} T.,  {Nelemans} G.,  {Steeghs} D.,
  {Hakala} P.,  2006, \aap, 457, 623

\bibitem[\protect\citeauthoryear{{Reynolds}, {Callanan}, {Fruchter}, {Torres},
  {Beer} \& {Gibbons}}{{Reynolds} et~al.}{2007}]{2007Rey}
{Reynolds} M.~T.,  {Callanan} P.~J.,  {Fruchter} A.~S.,  {Torres} M.~A.~P.,
  {Beer} M.~E.,    {Gibbons} R.~A.,  2007, \mnras, 379, 1117

\bibitem[\protect\citeauthoryear{{Rivera Sandoval}, {Hernandez Santisteban},
  {Degenaar}, {Wijnands}, {Knigge}, {Miller}, {Reynolds}, {Altamirano}, {van
  den Berg} \& {Hill}}{{Rivera Sandoval} et~al.}{2017}]{2017muv}
{Rivera Sandoval} L.~E.,  {Hernandez Santisteban} J.~V.,  {Degenaar} N.,
  {Wijnands} R.,  {Knigge} C.,  {Miller} J.~M.,  {Reynolds} M.,  {Altamirano}
  D.,  {van den Berg} M.,    {Hill} A.,  2017, ArXiv e-prints

\bibitem[\protect\citeauthoryear{{Rivera-Sandoval}, {van den Berg}, {Heinke},
  {Cohn}, {Lugger}, {Freire}, {Anderson}, {Serenelli}, {Althaus}, {Cool},
  {Grindlay}, {Edmonds}, {Wijnands} \& {Ivanova}}{{Rivera-Sandoval}
  et~al.}{2015}]{rivera}
{Rivera-Sandoval} L.~E.,  {van den Berg} M.,  {Heinke} C.~O.,  {Cohn} H.~N.,
  {Lugger} P.~M.,  {Freire} P.,  {Anderson} J.,  {Serenelli} A.~M.,  {Althaus}
  L.~G.,  {Cool} A.~M.,  {Grindlay} J.~E.,  {Edmonds} P.~D.,  {Wijnands} R.,
  {Ivanova} N.,  2015, \mnras, 453, 2707

\bibitem[\protect\citeauthoryear{{Rodr{\'{\i}}guez-Gil}, {G{\"a}nsicke},
  {Hagen}, {Nogami}, {Torres}, {Lehto}, {Aungwerojwit}, {Littlefair},
  {Araujo-Betancor} \& {Engels}}{{Rodr{\'{\i}}guez-Gil}
  et~al.}{2005}]{2005-rodriguez}
{Rodr{\'{\i}}guez-Gil} P.,  {G{\"a}nsicke} B.~T.,  {Hagen} H.-J.,  {Nogami} D.,
   {Torres} M.~A.~P.,  {Lehto} H.,  {Aungwerojwit} A.,  {Littlefair} S.,
  {Araujo-Betancor} S.,    {Engels} D.,  2005, \aap, 440, 701

\bibitem[\protect\citeauthoryear{{Romani}, {Filippenko} \& {Cenko}}{{Romani}
  et~al.}{2014}]{2014-romani}
{Romani} R.~W.,  {Filippenko} A.~V.,    {Cenko} S.~B.,  2014, \apjl, 793, L20

\bibitem[\protect\citeauthoryear{{Salaris}, {Held}, {Ortolani}, {Gullieuszik}
  \& {Momany}}{{Salaris} et~al.}{2007}]{2007-salaris}
{Salaris} M.,  {Held} E.~V.,  {Ortolani} S.,  {Gullieuszik} M.,    {Momany} Y.,
   2007, \aap, 476, 243

\bibitem[\protect\citeauthoryear{{Serenelli}, {Althaus}, {Rohrmann} \&
  {Benvenuto}}{{Serenelli} et~al.}{2002}]{2002-serenelli}
{Serenelli} A.~M.,  {Althaus} L.~G.,  {Rohrmann} R.,    {Benvenuto} O.~G.,
  2002, \mnras, 337, 1091

\bibitem[\protect\citeauthoryear{{Servillat}, {Dieball}, {Webb}, {Knigge},
  {Cornelisse}, {Barret}, {Long}, {Shara} \& {Zurek}}{{Servillat}
  et~al.}{2008}]{2008-servi}
{Servillat} M.,  {Dieball} A.,  {Webb} N.~A.,  {Knigge} C.,  {Cornelisse} R.,
  {Barret} D.,  {Long} K.~S.,  {Shara} M.~M.,    {Zurek} D.~R.,  2008, \aap,
  490, 641

\bibitem[\protect\citeauthoryear{{Shara}, {Bergeron}, {Gilliland}, {Saha} \&
  {Petro}}{{Shara} et~al.}{1996}]{1996-shara}
{Shara} M.~M.,  {Bergeron} L.~E.,  {Gilliland} R.~L.,  {Saha} A.,    {Petro}
  L.,  1996, \apj, 471, 804

\bibitem[\protect\citeauthoryear{{Shara} \& {Hurley}}{{Shara} \&
  {Hurley}}{2006}]{2006-shara}
{Shara} M.~M.,  {Hurley} J.~R.,  2006, \apj, 646, 464

\bibitem[\protect\citeauthoryear{{Silverman}}{{Silverman}}{1986}]{silverman}
{Silverman} B.~W.,  1986, {Density estimation for statistics and data analysis}

\bibitem[\protect\citeauthoryear{{Sirianni}, {Jee}, {Ben{\'{\i}}tez},
  {Blakeslee}, {Martel}, {Meurer}, {Clampin}, {De Marchi}, {Ford}, {Gilliland},
  {Hartig}, {Illingworth}, {Mack} \& {McCann}}{{Sirianni}
  et~al.}{2005}]{siriea05}
{Sirianni} M.,  {Jee} M.~J.,  {Ben{\'{\i}}tez} N.,  {Blakeslee} J.~P.,
  {Martel} A.~R.,  {Meurer} G.,  {Clampin} M.,  {De Marchi} G.,  {Ford} H.~C.,
  {Gilliland} R.,  {Hartig} G.~F.,  {Illingworth} G.~D.,  {Mack} J.,
  {McCann} W.~J.,  2005, \pasp, 117, 1049

\bibitem[\protect\citeauthoryear{{Solheim}}{{Solheim}}{2010}]{Sole}
{Solheim} J.-E.,  2010, \pasp, 122, 1133

\bibitem[\protect\citeauthoryear{{Steeghs}, {Knigge}, {Drew}, {Unruh} \&
  {Greimel}}{{Steeghs} et~al.}{2008}]{2008stee}
{Steeghs} D.,  {Knigge} C.,  {Drew} J.,  {Unruh} Y.,    {Greimel} R.,  2008,
  The Astronomer's Telegram, 1653

\bibitem[\protect\citeauthoryear{{Szkody}, {Anderson}, {Brooks},
  {G{\"a}nsicke}, {Kronberg}, {Riecken}, {Ross}, {Schmidt}, {Schneider},
  {Ag{\"u}eros}, {Gomez-Moran}, {Knapp}, {Schreiber} \& {Schwope}}{{Szkody}
  et~al.}{2011}]{2011-szkody}
{Szkody} P.,  {Anderson} S.~F.,  {Brooks} K.,  {G{\"a}nsicke} B.~T.,
  {Kronberg} M.,  {Riecken} T.,  {Ross} N.~P.,  {Schmidt} G.~D.,  {Schneider}
  D.~P.,  {Ag{\"u}eros} M.~A.,  {Gomez-Moran} A.~N.,  {Knapp} G.~R.,
  {Schreiber} M.~R.,    {Schwope} A.~D.,  2011, \aj, 142, 181

\bibitem[\protect\citeauthoryear{{Thomson}, {Knigge}, {Dieball}, {Maccarone},
  {Dolphin}, {Zurek}, {Long}, {Shara} \& {Sarajedini}}{{Thomson}
  et~al.}{2012}]{2012-thom}
{Thomson} G.~S.,  {Knigge} C.,  {Dieball} A.,  {Maccarone} T.~J.,  {Dolphin}
  A.,  {Zurek} D.,  {Long} K.~S.,  {Shara} M.,    {Sarajedini} A.,  2012,
  \mnras, 423, 2901

\bibitem[\protect\citeauthoryear{{Thorstensen}, {Fenton}, {Patterson}, {Kemp},
  {Krajci} \& {Baraffe}}{{Thorstensen} et~al.}{2002}]{2002-Thor}
{Thorstensen} J.~R.,  {Fenton} W.~H.,  {Patterson} J.~O.,  {Kemp} J.,  {Krajci}
  T.,    {Baraffe} I.,  2002, \apjl, 567, L49

\bibitem[\protect\citeauthoryear{{Trager}, {King} \& {Djorgovski}}{{Trager}
  et~al.}{1995}]{trager}
{Trager} S.~C.,  {King} I.~R.,    {Djorgovski} S.,  1995, \aj, 109, 218

\bibitem[\protect\citeauthoryear{{van den Berg}, {Hong} \& {Grindlay}}{{van den
  Berg} et~al.}{2009}]{2009-vandenberg}
{van den Berg} M.,  {Hong} J.~S.,    {Grindlay} J.~E.,  2009, \apj, 700, 1702

\bibitem[\protect\citeauthoryear{{van Kerkwijk}, {Bergeron} \& {Kulkarni}}{{van
  Kerkwijk} et~al.}{1996}]{1996-van}
{van Kerkwijk} M.~H.,  {Bergeron} P.,    {Kulkarni} S.~R.,  1996, \apjl, 467,
  L89

\bibitem[\protect\citeauthoryear{{VandenBerg}, {Brogaard}, {Leaman} \&
  {Casagrande}}{{VandenBerg} et~al.}{2013}]{VandenBerg}
{VandenBerg} D.~A.,  {Brogaard} K.,  {Leaman} R.,    {Casagrande} L.,  2013,
  \apj, 775, 134

\bibitem[\protect\citeauthoryear{{Verbunt} \& {Meylan}}{{Verbunt} \&
  {Meylan}}{1988}]{1988-ver}
{Verbunt} F.,  {Meylan} G.,  1988, \aap, 203, 297

\bibitem[\protect\citeauthoryear{{Warner}}{{Warner}}{1987}]{1987-Warner}
{Warner} B.,  1987, \mnras, 227, 23

\bibitem[\protect\citeauthoryear{{Wijnen}, {Zorotovic} \& {Schreiber}}{{Wijnen}
  et~al.}{2015}]{2015_wijnen}
{Wijnen} T.~P.~G.,  {Zorotovic} M.,    {Schreiber} M.~R.,  2015, \aap, 577,
  A143

\bibitem[\protect\citeauthoryear{{Witham}, {Knigge}, {G{\"a}nsicke},
  {Aungwerojwit}, {Corradi}, {Drew}, {Greimel}, {Groot}, {Morales-Rueda},
  {Rodriguez-Flores}, {Rodriguez-Gil} \& {Steeghs}}{{Witham}
  et~al.}{2006}]{2006-whil}
{Witham} A.~R.,  {Knigge} C.,  {G{\"a}nsicke} B.~T.,  {Aungwerojwit} A.,
  {Corradi} R.~L.~M.,  {Drew} J.~E.,  {Greimel} R.,  {Groot} P.~J.,
  {Morales-Rueda} L.,  {Rodriguez-Flores} E.~R.,  {Rodriguez-Gil} P.,
  {Steeghs} D.,  2006, \mnras, 369, 581

\bibitem[\protect\citeauthoryear{{Woodley}, {Goldsbury}, {Kalirai}, {Richer},
  {Tremblay}, {Anderson}, {Bergeron}, {Dotter}, {Esteves}, {Fahlman}, {Hansen},
  {Heyl}, {Hurley}, {Rich}, {Shara} \& {Stetson}}{{Woodley}
  et~al.}{2012}]{2012-woodley}
{Woodley} K.~A.,  {Goldsbury} R.,  {Kalirai} J.~S.,  {Richer} H.~B.,
  {Tremblay} P.-E.,  {Anderson} J.,  {Bergeron} P.,  {Dotter} A.,  {Esteves}
  L.,  {Fahlman} G.~G.,  {Hansen} B.~M.~S.,  {Heyl} J.,  {Hurley} J.,  {Rich}
  R.~M.,  {Shara} M.~M.,    {Stetson} P.~B.,  2012, \aj, 143, 50

\bibitem[\protect\citeauthoryear{{Zamanov}, {Latev}, {Stoyanov}, {Boeva},
  {Spassov} \& {Tsvetkova}}{{Zamanov} et~al.}{2012}]{2012-zama}
{Zamanov} R.~K.,  {Latev} G.~Y.,  {Stoyanov} K.~A.,  {Boeva} S.,  {Spassov} B.,
     {Tsvetkova} S.~V.,  2012, Astronomische Nachrichten, 333, 736

\bibitem[\protect\citeauthoryear{{Zorotovic}, {Schreiber} \&
  {G{\"a}nsicke}}{{Zorotovic} et~al.}{2011}]{2011_zoro}
{Zorotovic} M.,  {Schreiber} M.~R.,    {G{\"a}nsicke} B.~T.,  2011, \aap, 536,
  A42

\bibitem[\protect\citeauthoryear{{Zurek}, {Knigge}, {Maccarone}, {Pooley},
  {Dieball}, {Long}, {Shara} \& {Sarajedini}}{{Zurek} et~al.}{2016}]{2016Zurek}
{Zurek} D.~R.,  {Knigge} C.,  {Maccarone} T.~J.,  {Pooley} D.,  {Dieball} A.,
  {Long} K.~S.,  {Shara} M.,    {Sarajedini} A.,  2016, \mnras, 460, 3660

\end{thebibliography}

\appendix
\label{sec:appendix}

\begin{table*}
\section{The core of 47 Tuc in NUV}
\label{stacked}
We show a stacked image in the filter U$_{300}$ which shows the portion of 47 Tuc covered in our study.
\end{table*}

\begin{figure*}
\centering
\includegraphics[width=19cm]{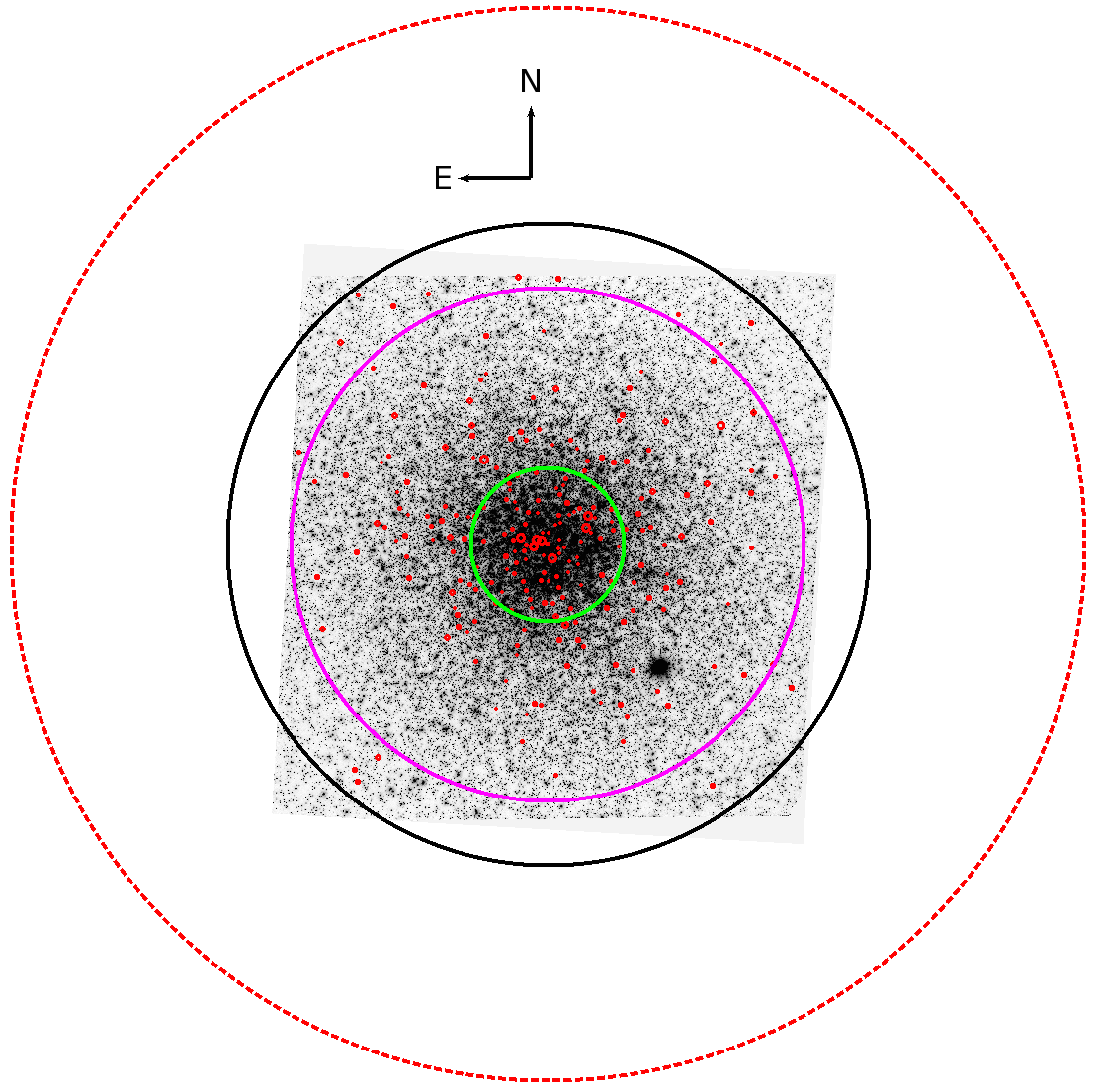}
\caption{Stacked \textit{HST} U$_{300}$ image of 47 Tuc with $3\sigma$ \textit{Chandra} source error circles (small red circles). 
The red dashed circle indicates the r$_h$ of the cluster ($2\farcm79$). The green circle shows the core radius ($24\arcsec$). 
The black circle has a radius of $100\arcsec$, and the magenta one indicates the largest circle that can be inscribed in our NUV images ($80\arcsec$). 
There are 88 X-ray sources within the core radius out of the 238 sources detected in our FOV. }
\label{fig:mosaic}
\vspace{2cm}
\end{figure*}

\begin{table*}
\section{Finding charts}
\label{finding_charts}
We present finding charts in the NUV for some CVs and CV candidates discussed in section \ref{results}. 
For more finding charts see \edmondsa, \edmondsb, \citet{knigge08} and \citet{2014-beccari}. 
\end{table*}

\begin{figure*}
\centering
  \begin{tabular}{@{}cccc@{}}
    \includegraphics[width=.23\textwidth]{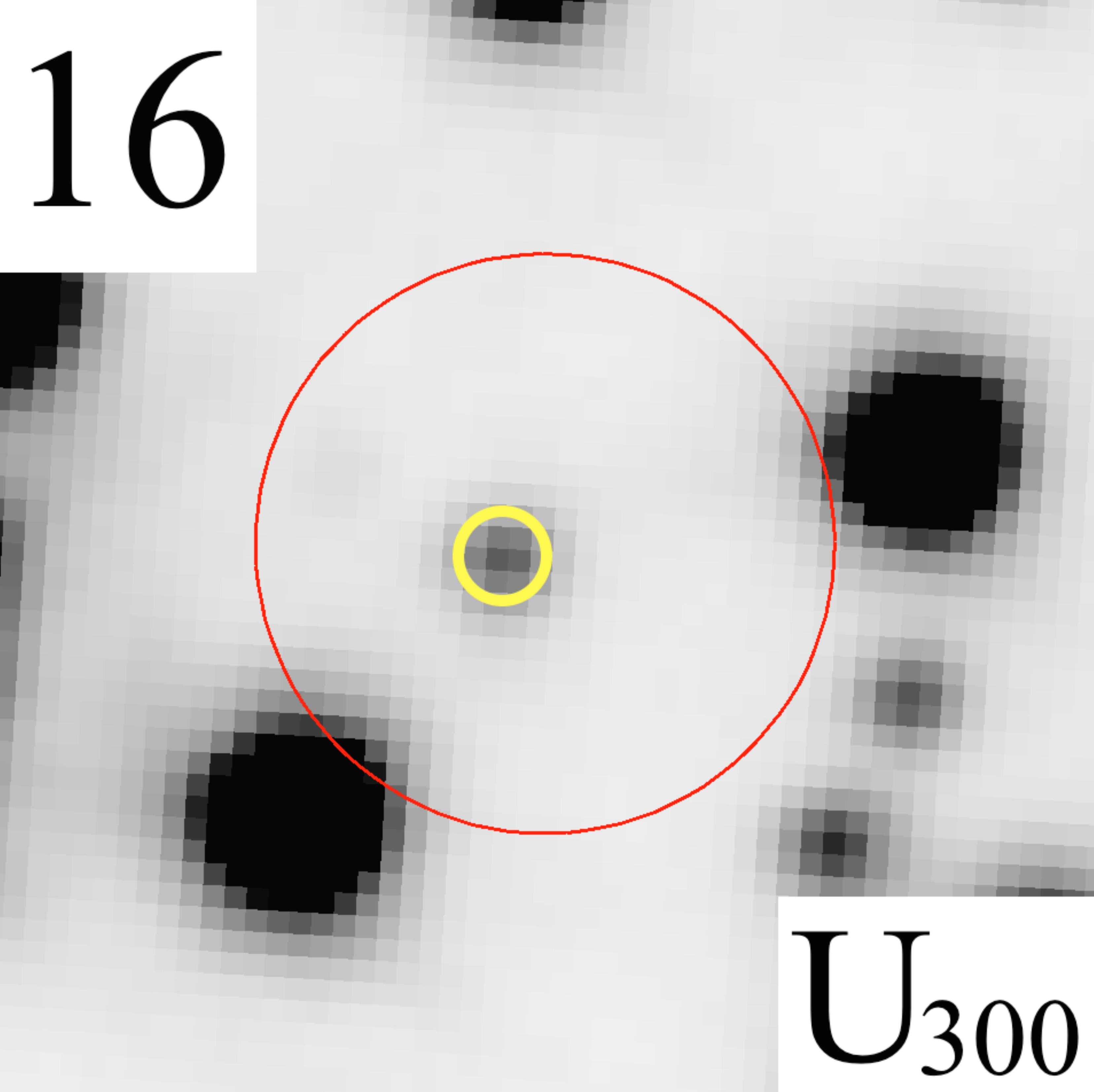} &
    \includegraphics[width=.23\textwidth]{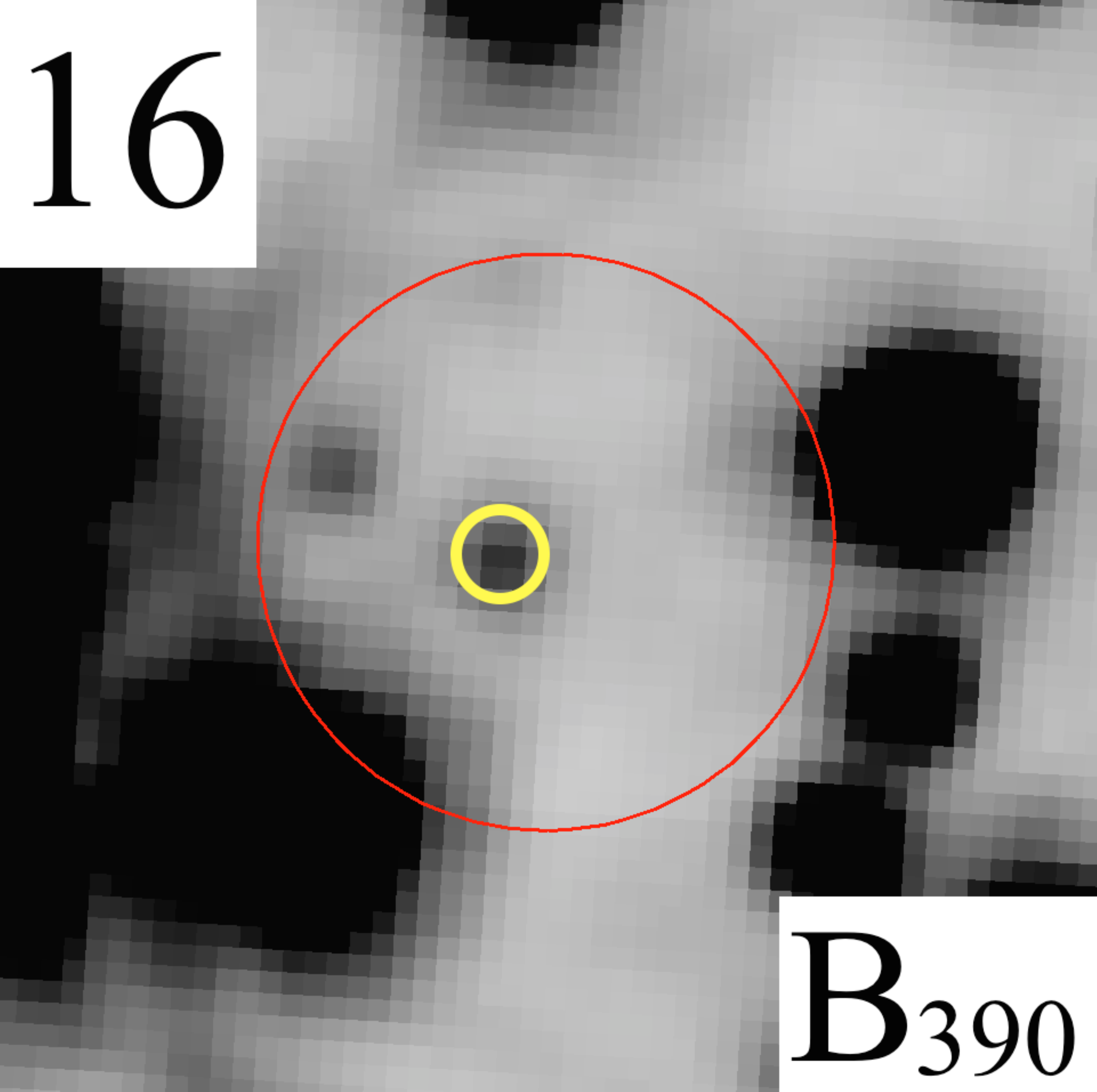} &
    \includegraphics[width=.23\textwidth]{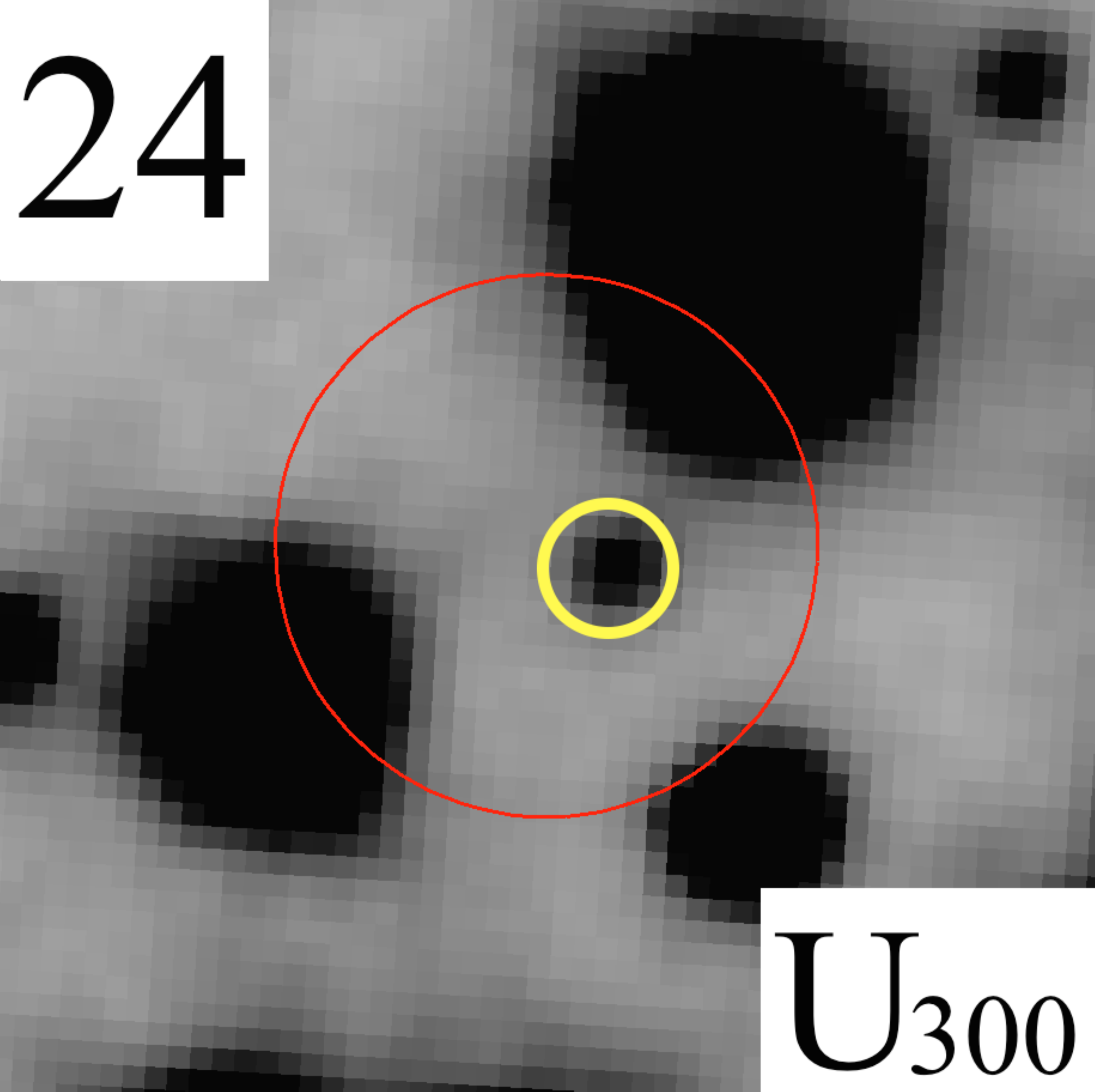} &
    \includegraphics[width=.23\textwidth]{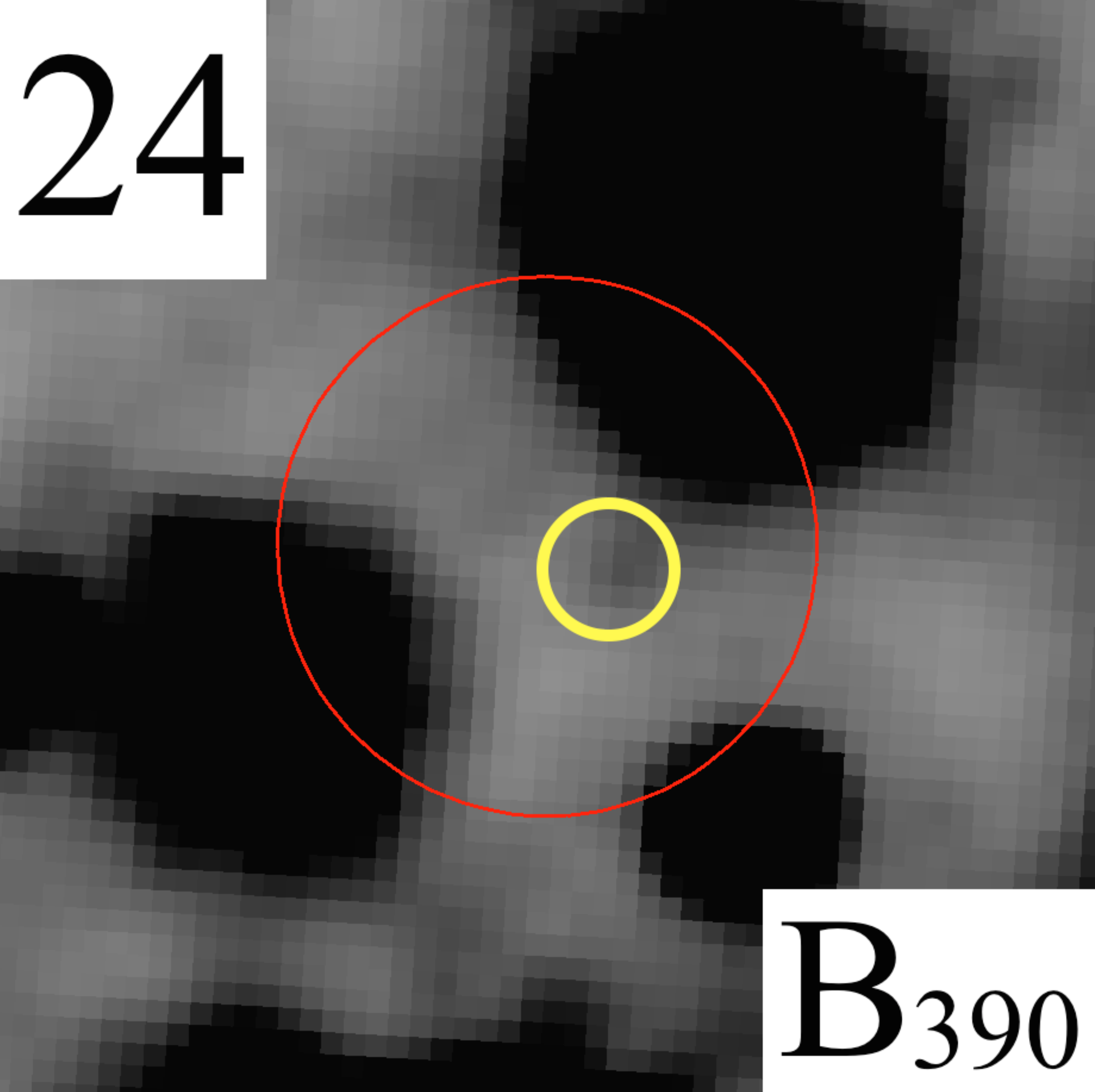}   \\
    \includegraphics[width=.23\textwidth]{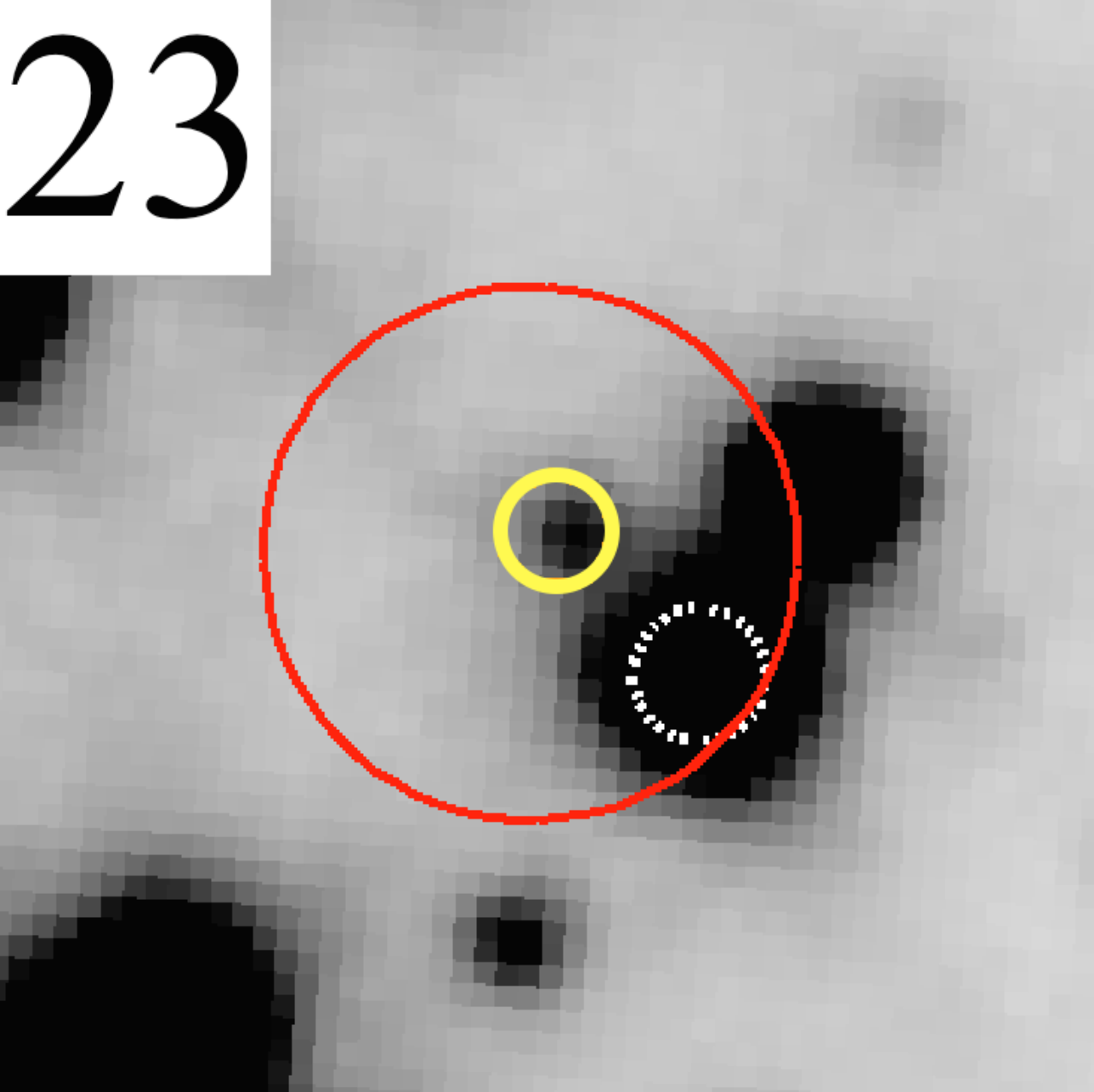} &
    \includegraphics[width=.23\textwidth]{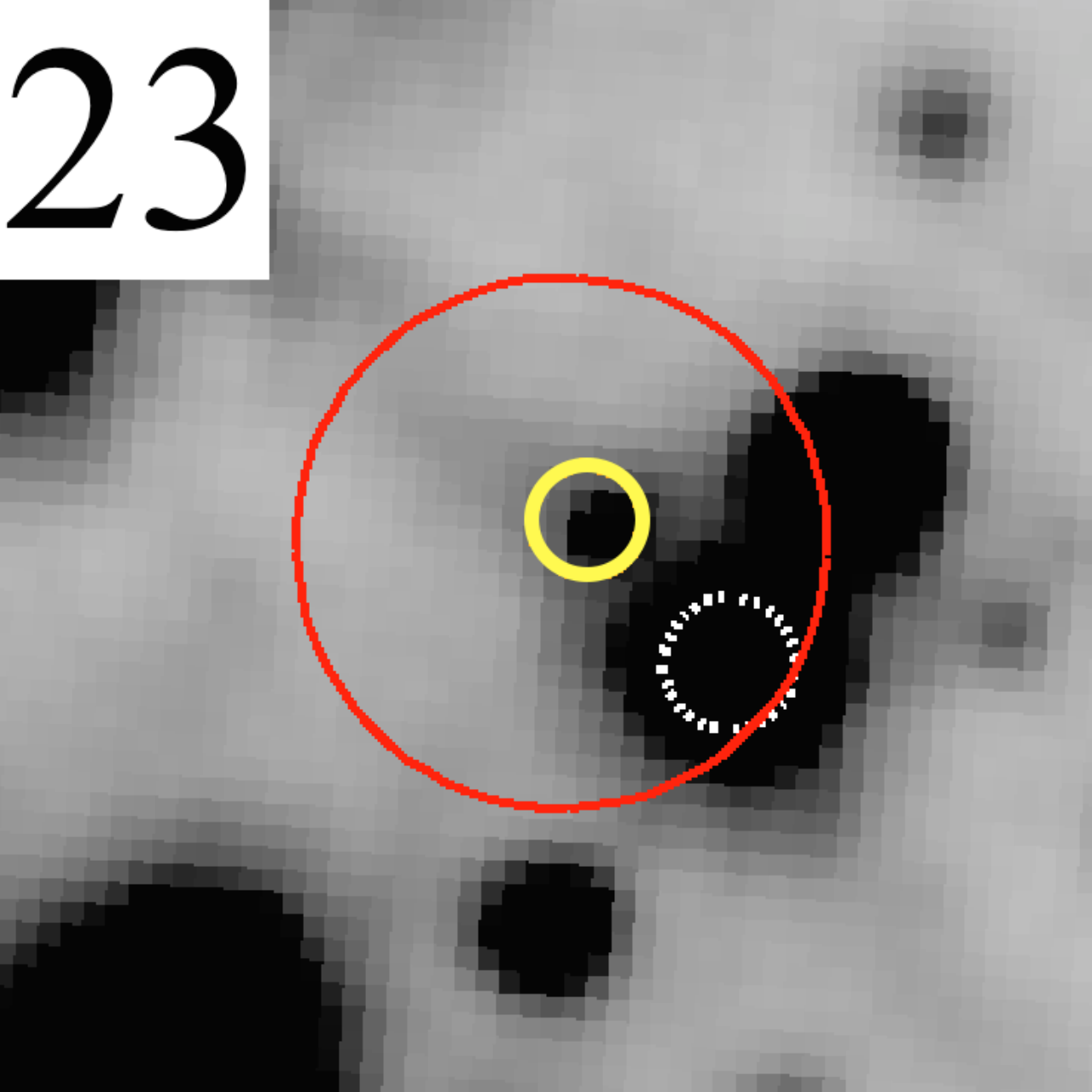} &
    \includegraphics[width=.23\textwidth]{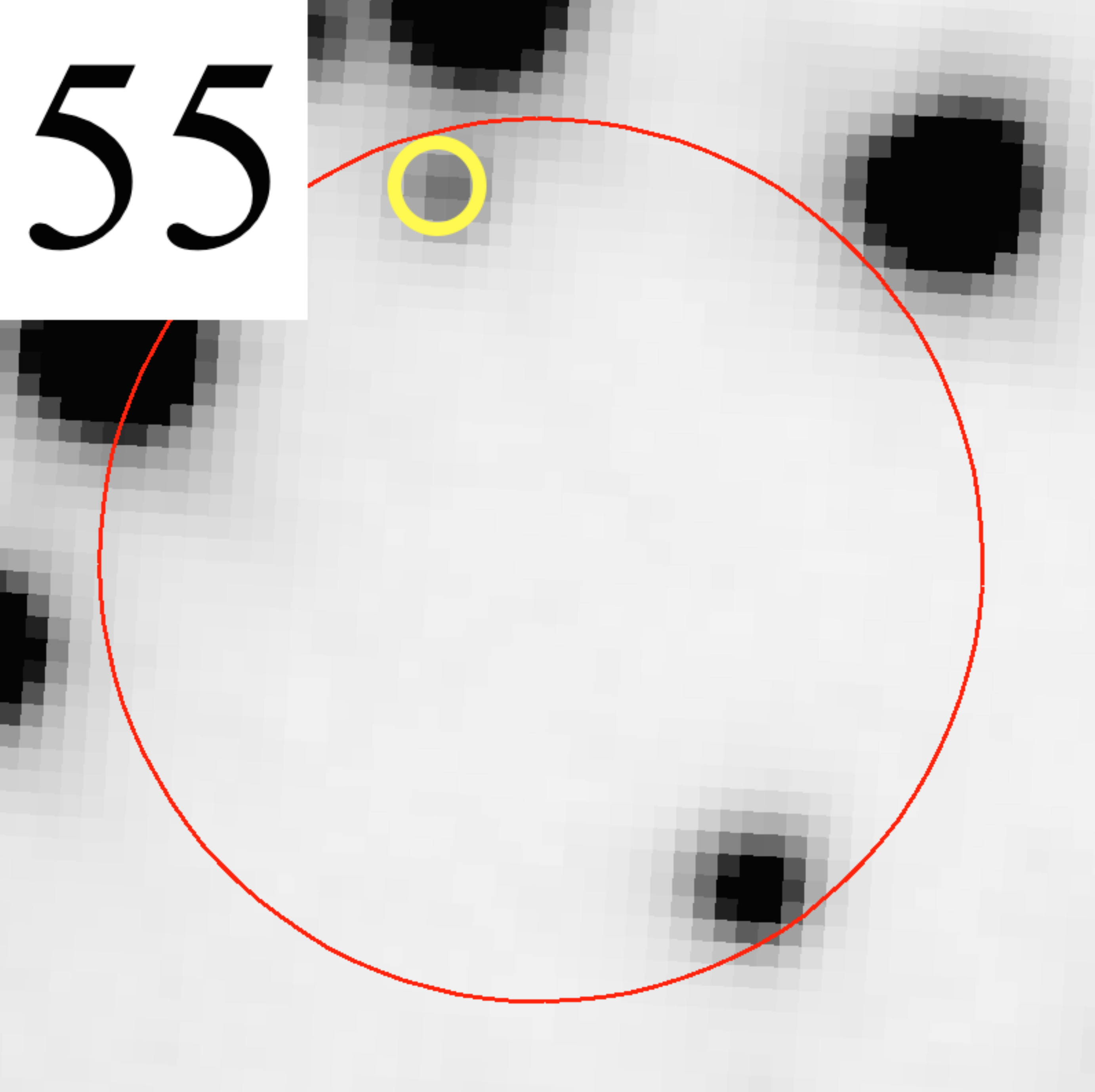} &
    \includegraphics[width=.23\textwidth]{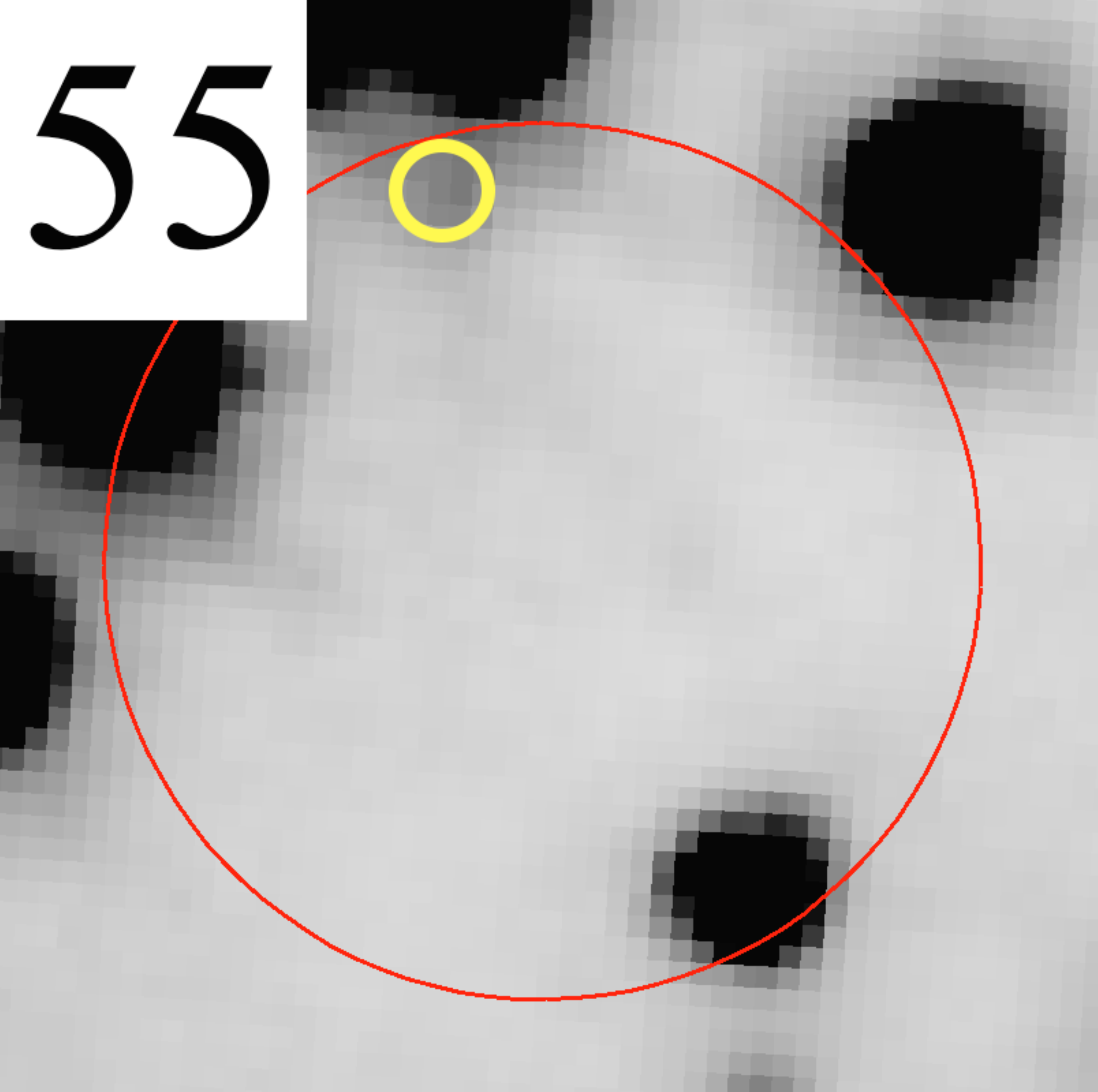}   \\
    \includegraphics[width=.23\textwidth]{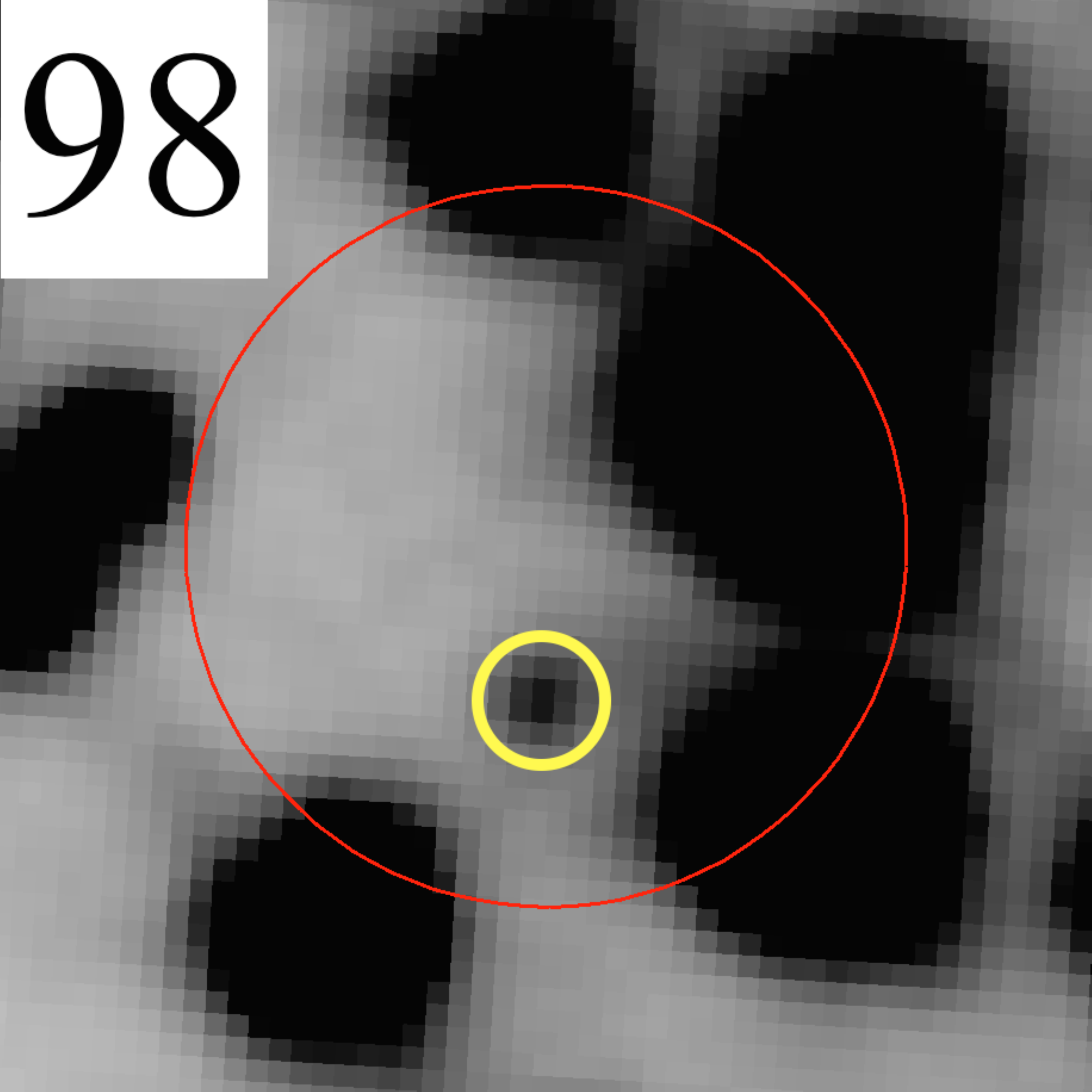} &
    \includegraphics[width=.23\textwidth]{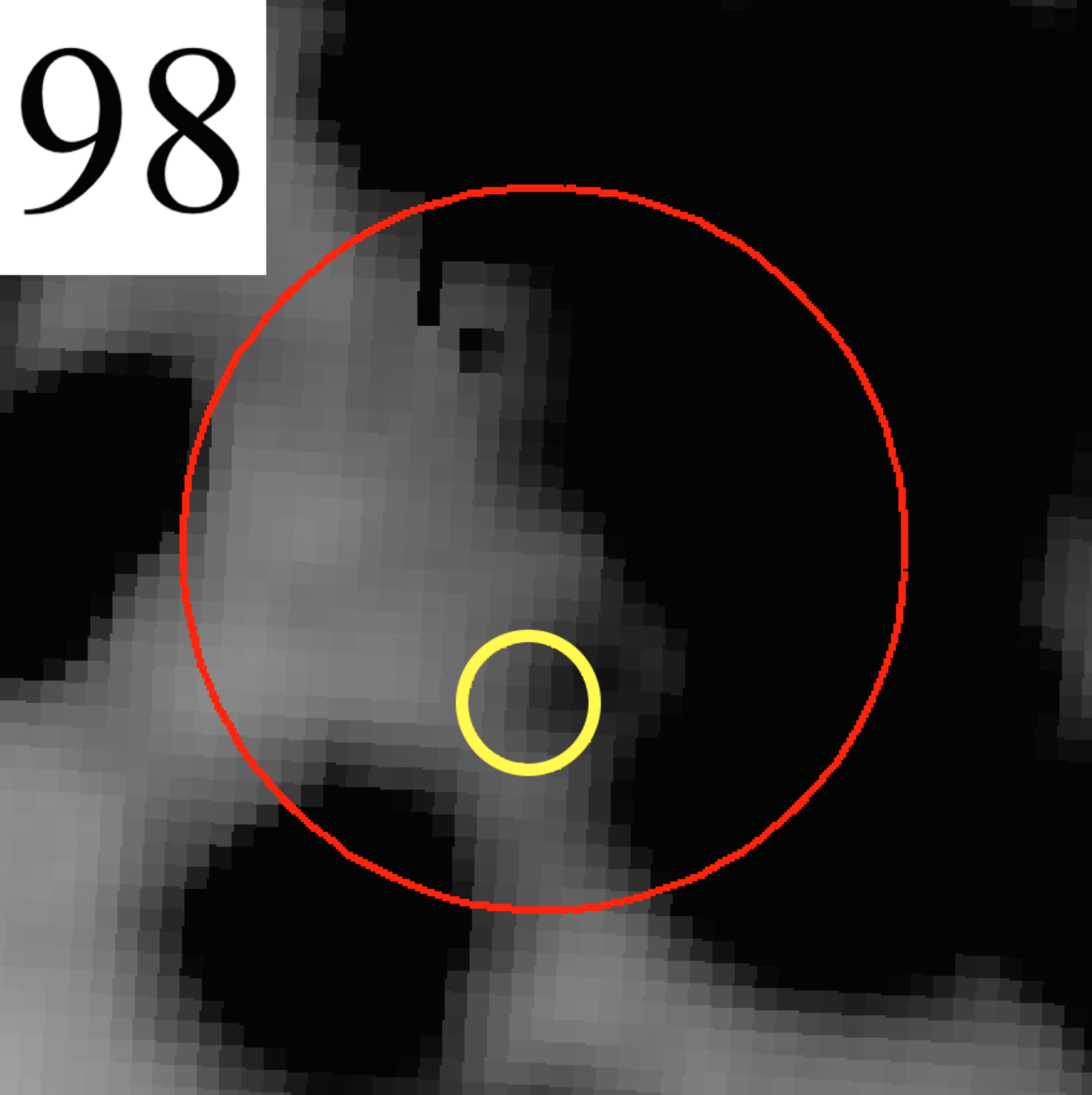} &
    \includegraphics[width=.23\textwidth]{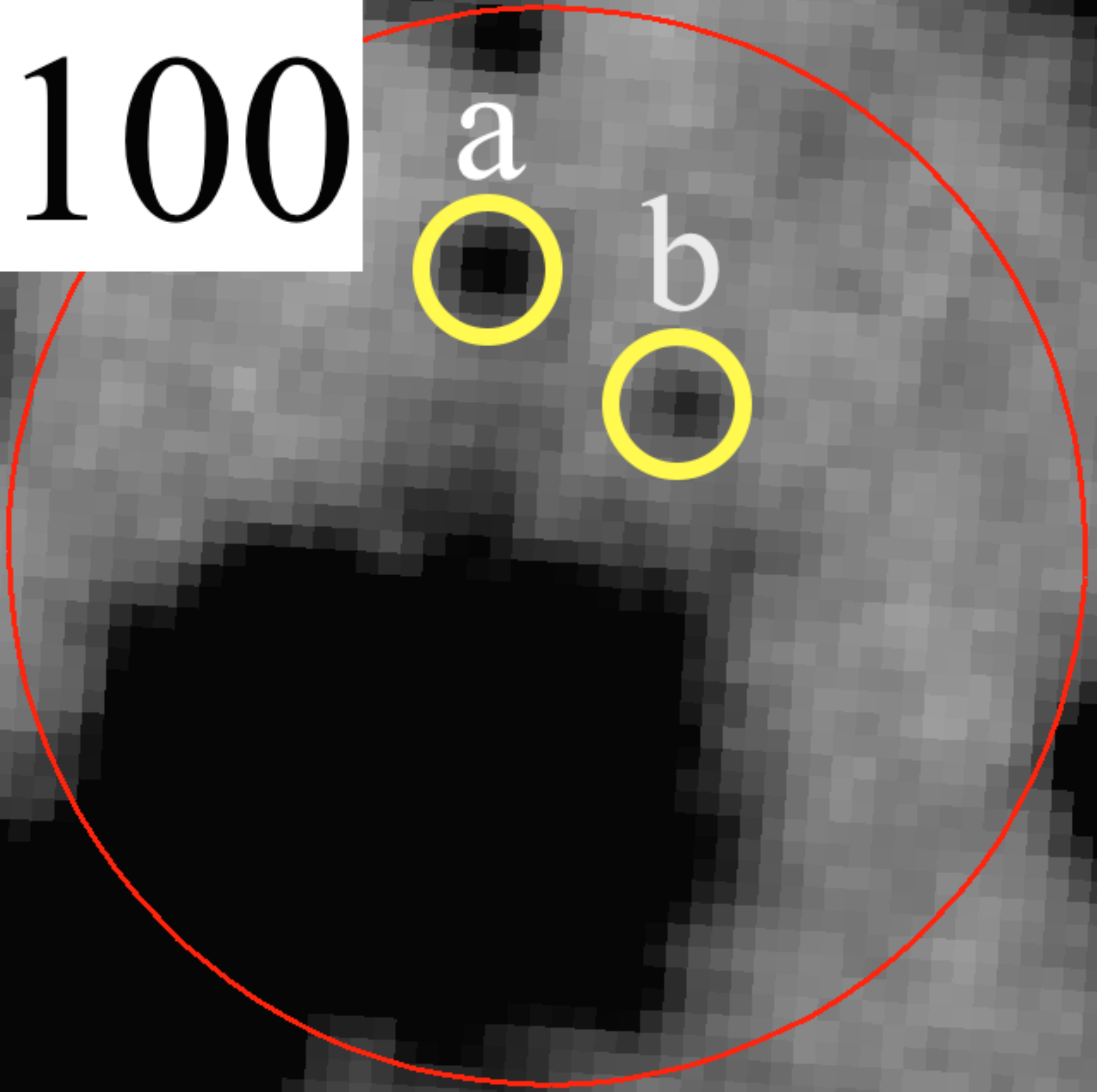} &
    \includegraphics[width=.23\textwidth]{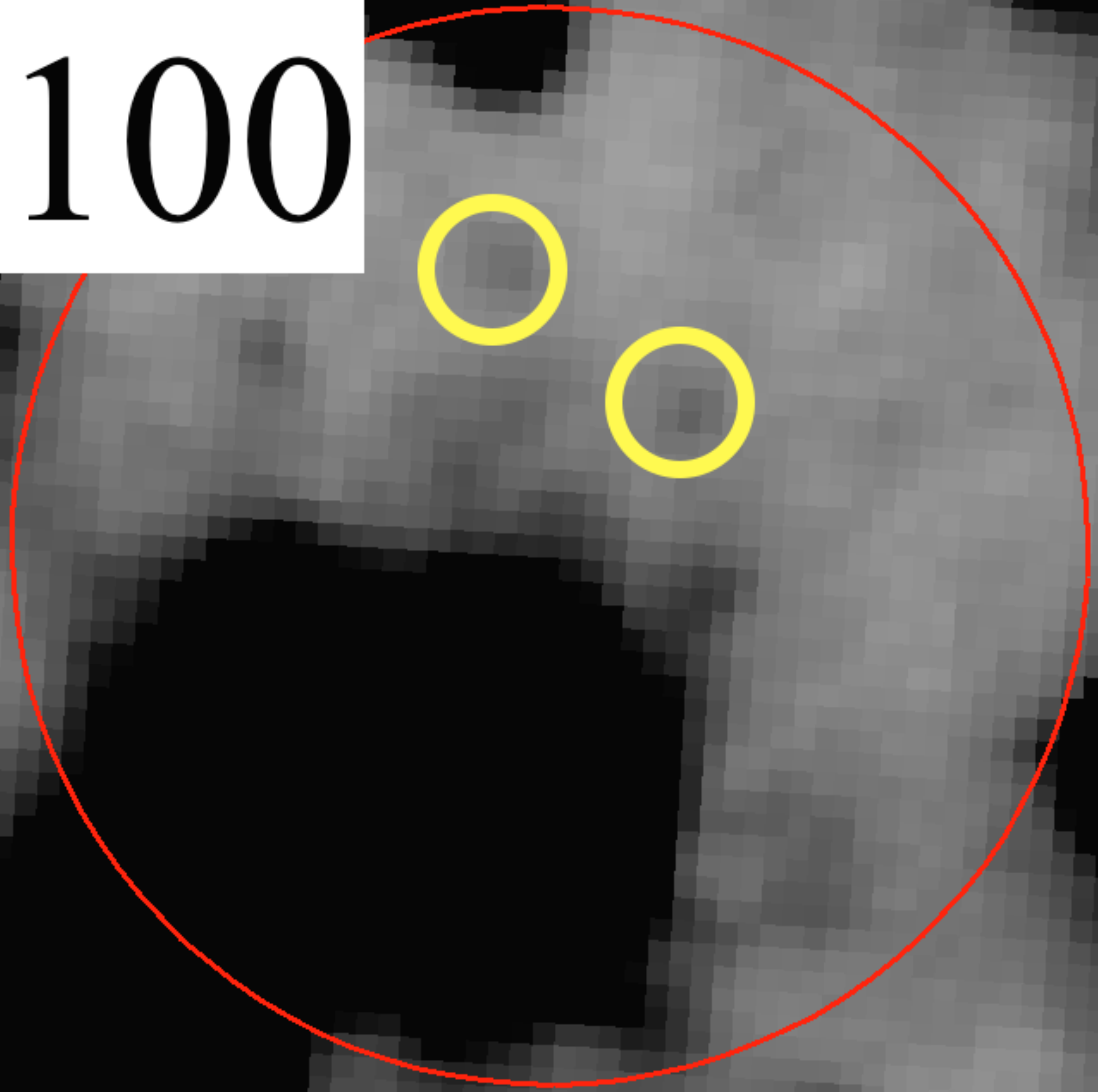}   \\
    \includegraphics[width=.23\textwidth]{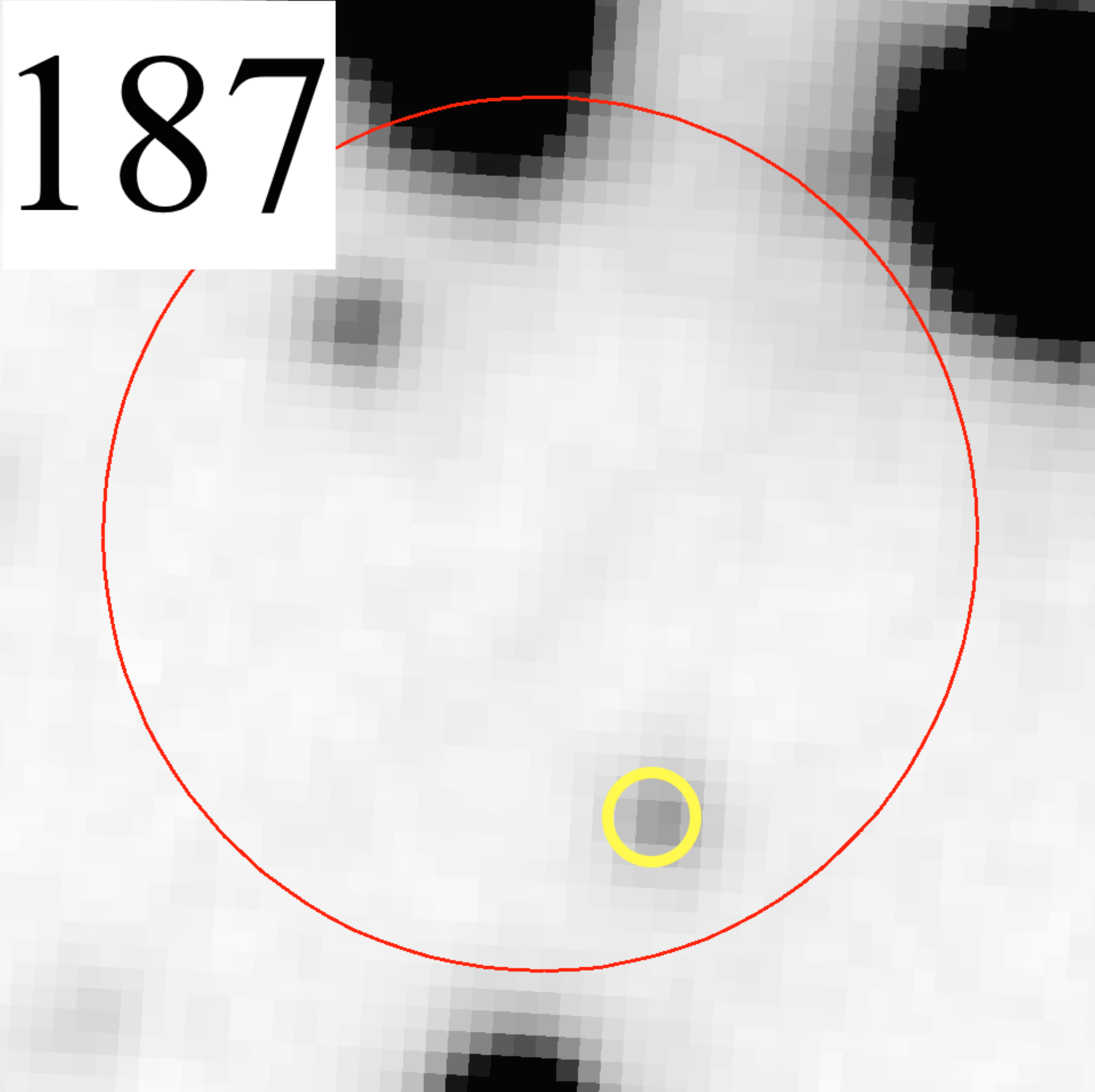} &
    \includegraphics[width=.23\textwidth]{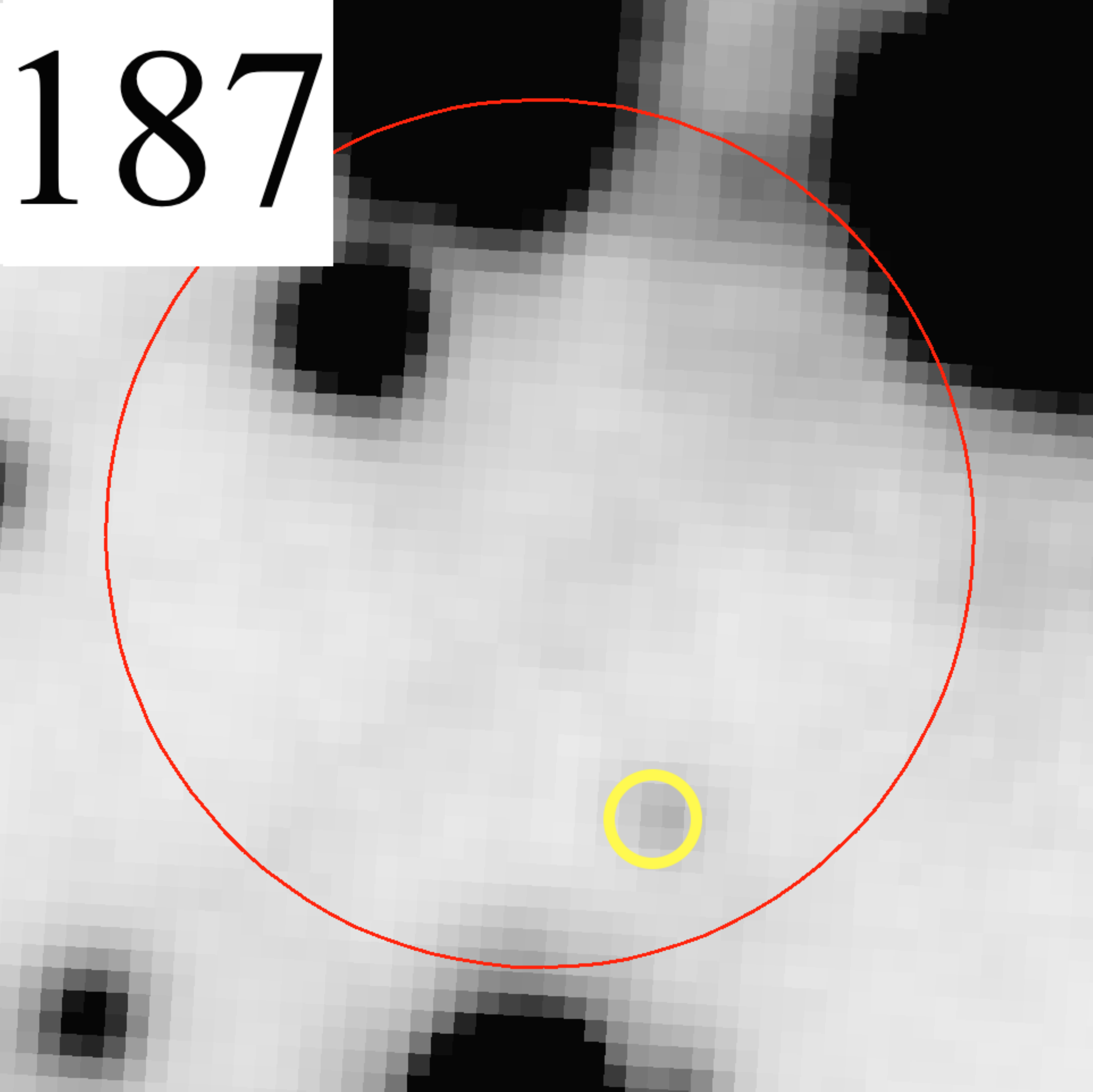} &
    \includegraphics[width=.23\textwidth]{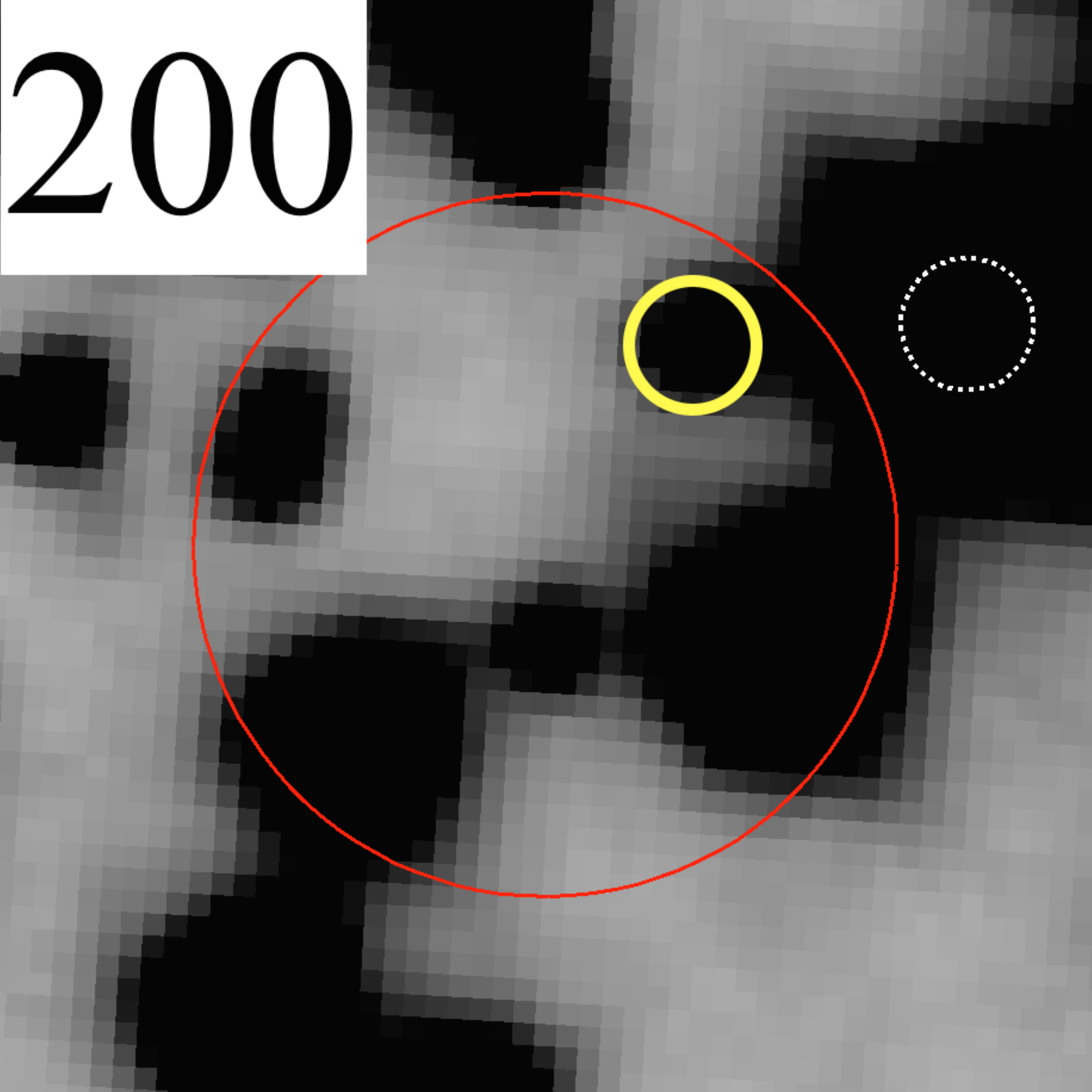} &
    \includegraphics[width=.23\textwidth]{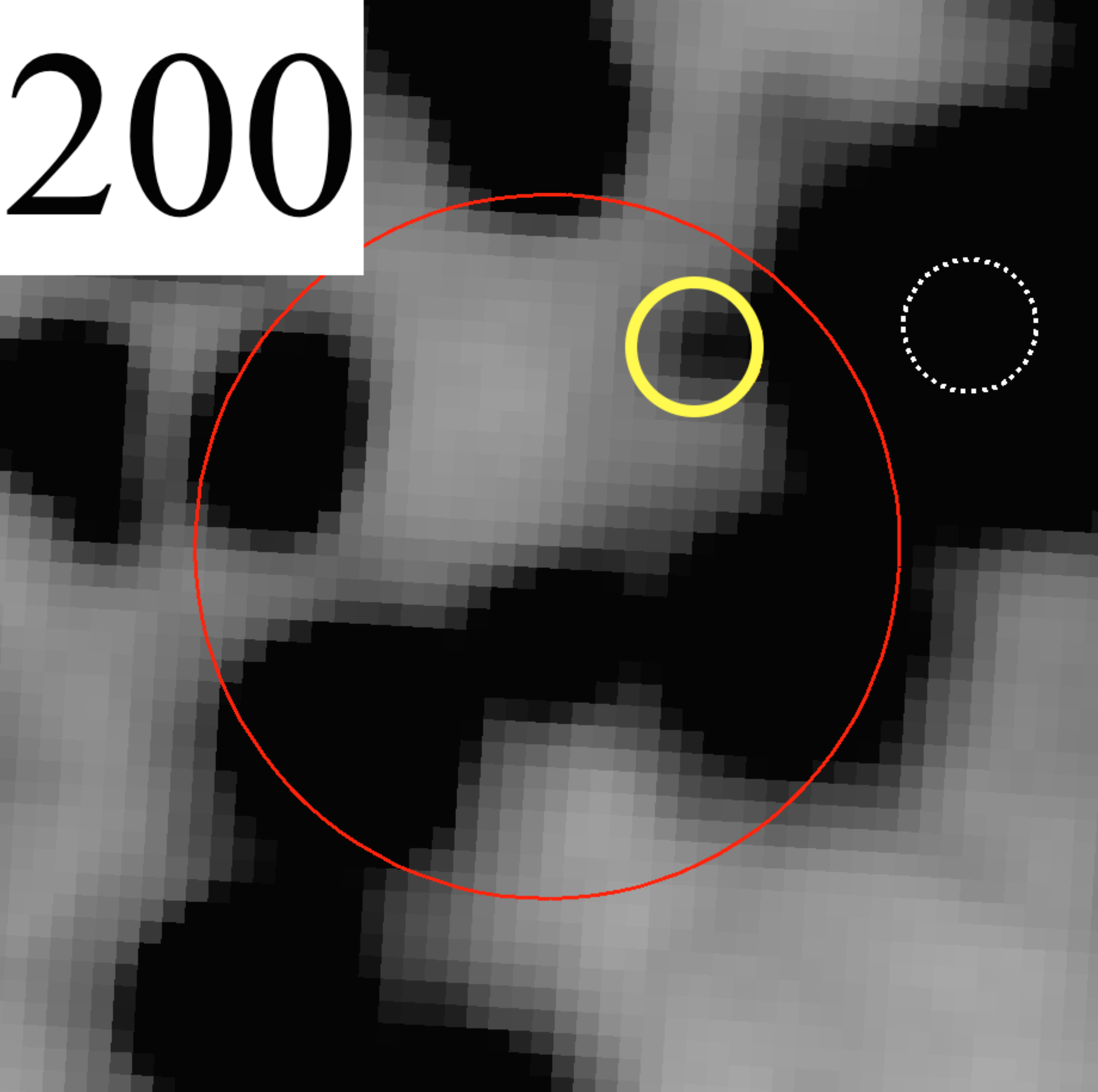}   \\
  \end{tabular}
  \caption{Finding charts in the filters U$_{300}$ and B$_{390}$ for some CVs and CV candidates in 47 Tuc. 
The solid red circles are the 3$\sigma$ match circles of \textit{Chandra} sources. The CVs and CV candidates are indicated with a solid yellow circle. 
The dashed white circles indicate possible additional X-ray counterparts (see section \ref{sec:multiple}). Each image is $1\arcsec \times 1\arcsec$.
In each column the images correspond to the same filter that is indicated by the top image of the column. The pixel scale is $0\farcs02$.}
\label{fig:A1}
\end{figure*}

\begin{figure*}
\centering
  \begin{tabular}{@{}cccc@{}}
   \includegraphics[width=.23\textwidth]{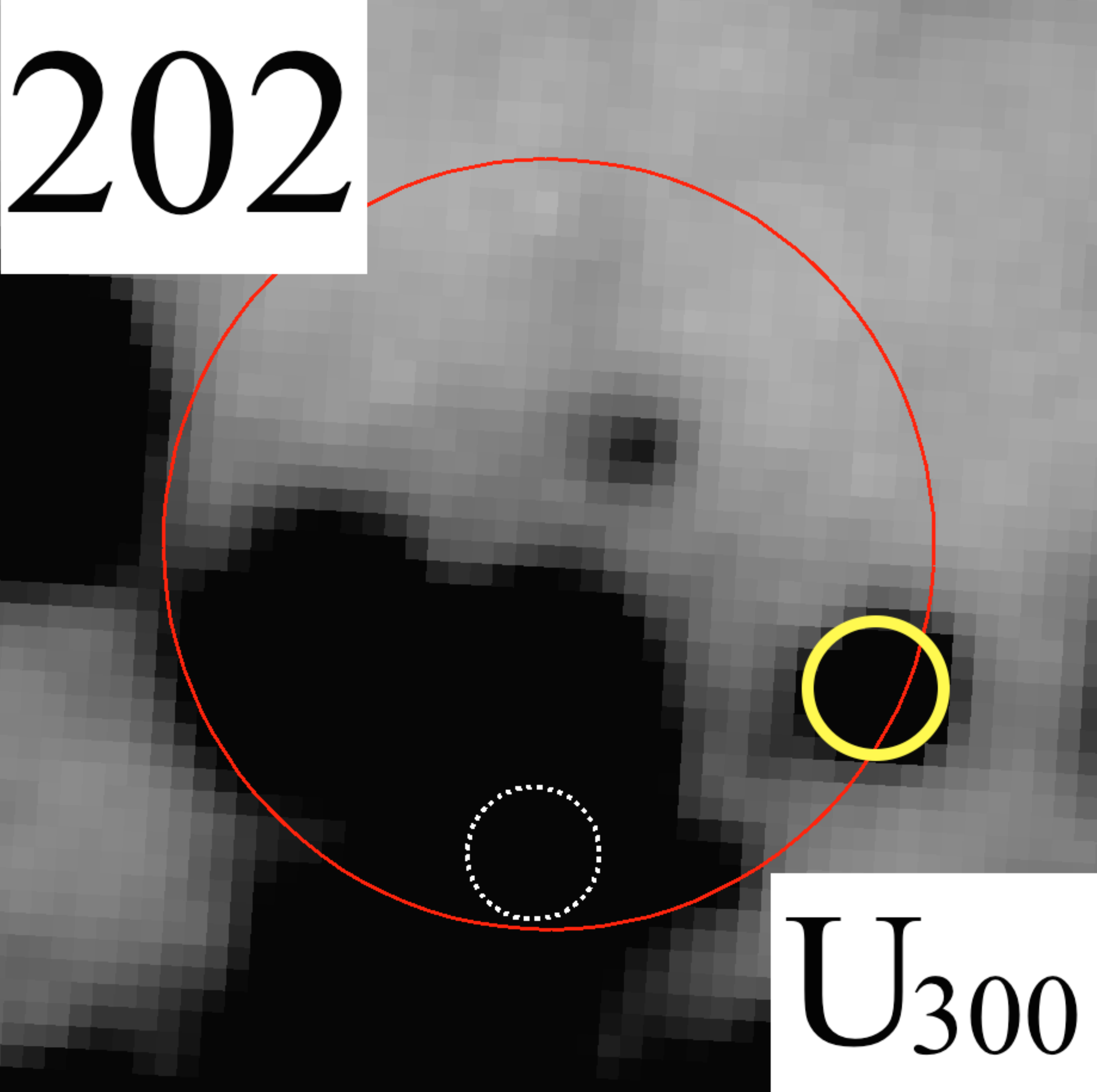} &
    \includegraphics[width=.23\textwidth]{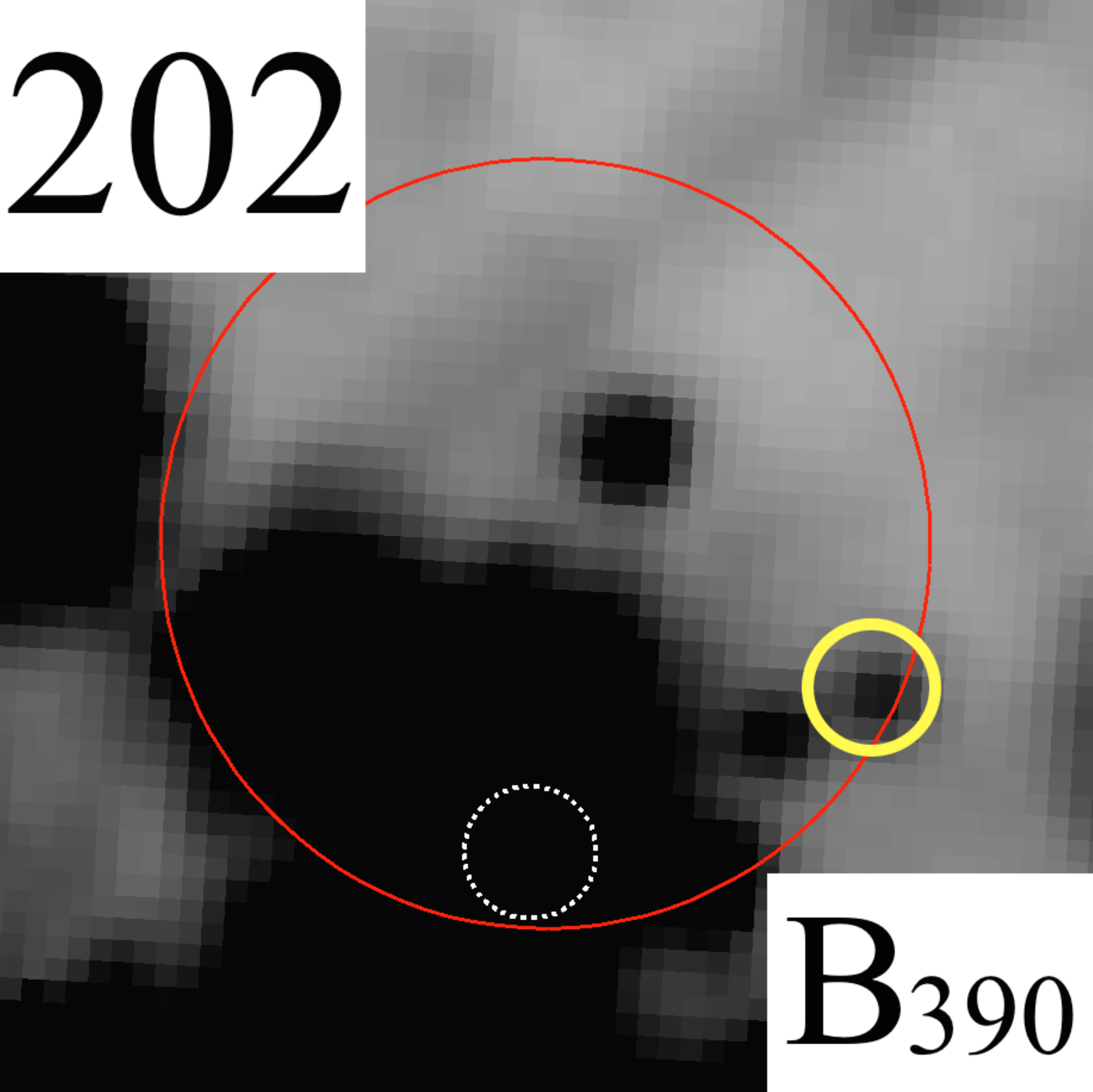} &
    \includegraphics[width=.23\textwidth]{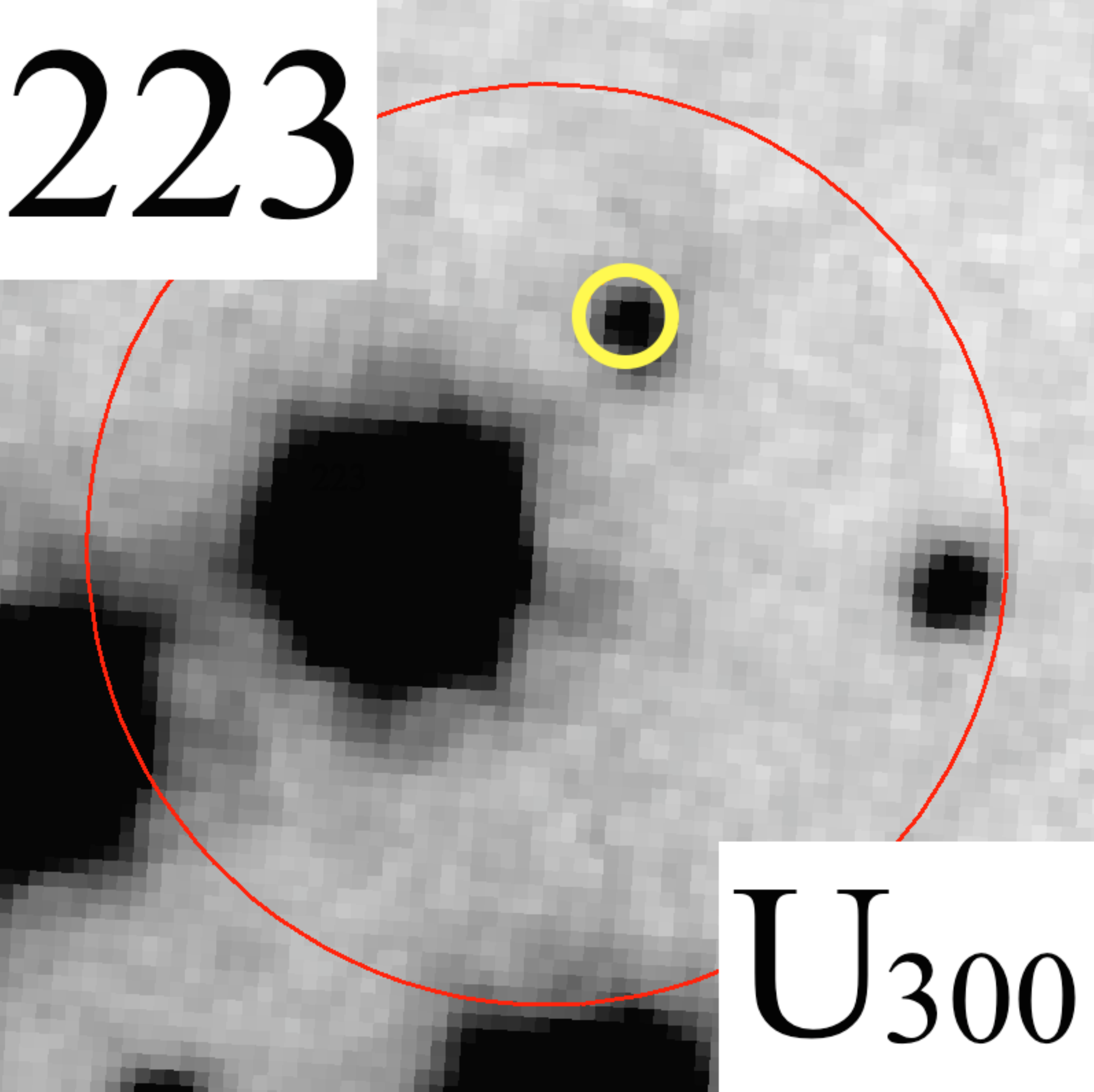} &
    \includegraphics[width=.23\textwidth]{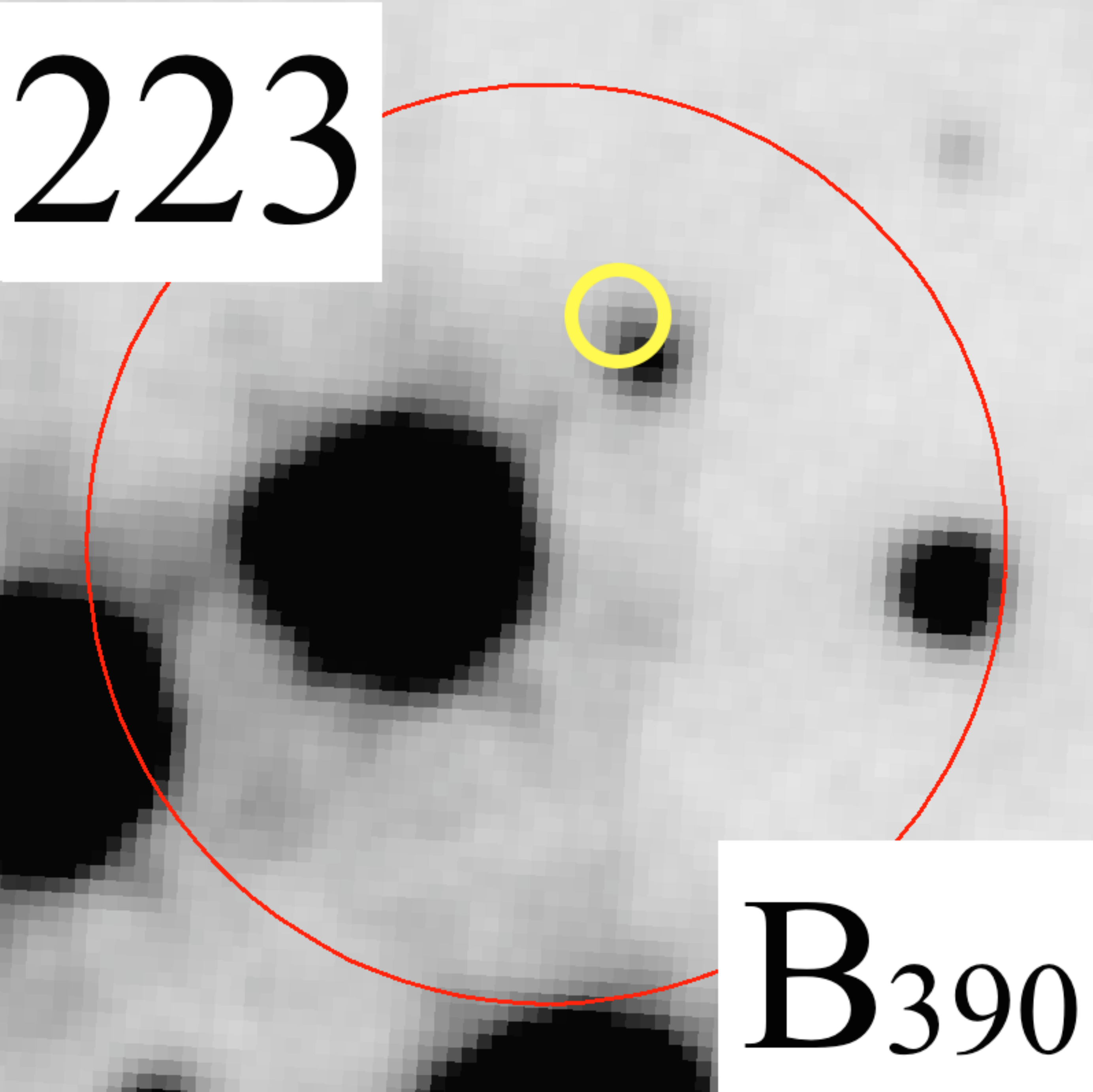}   \\
    \includegraphics[width=.23\textwidth]{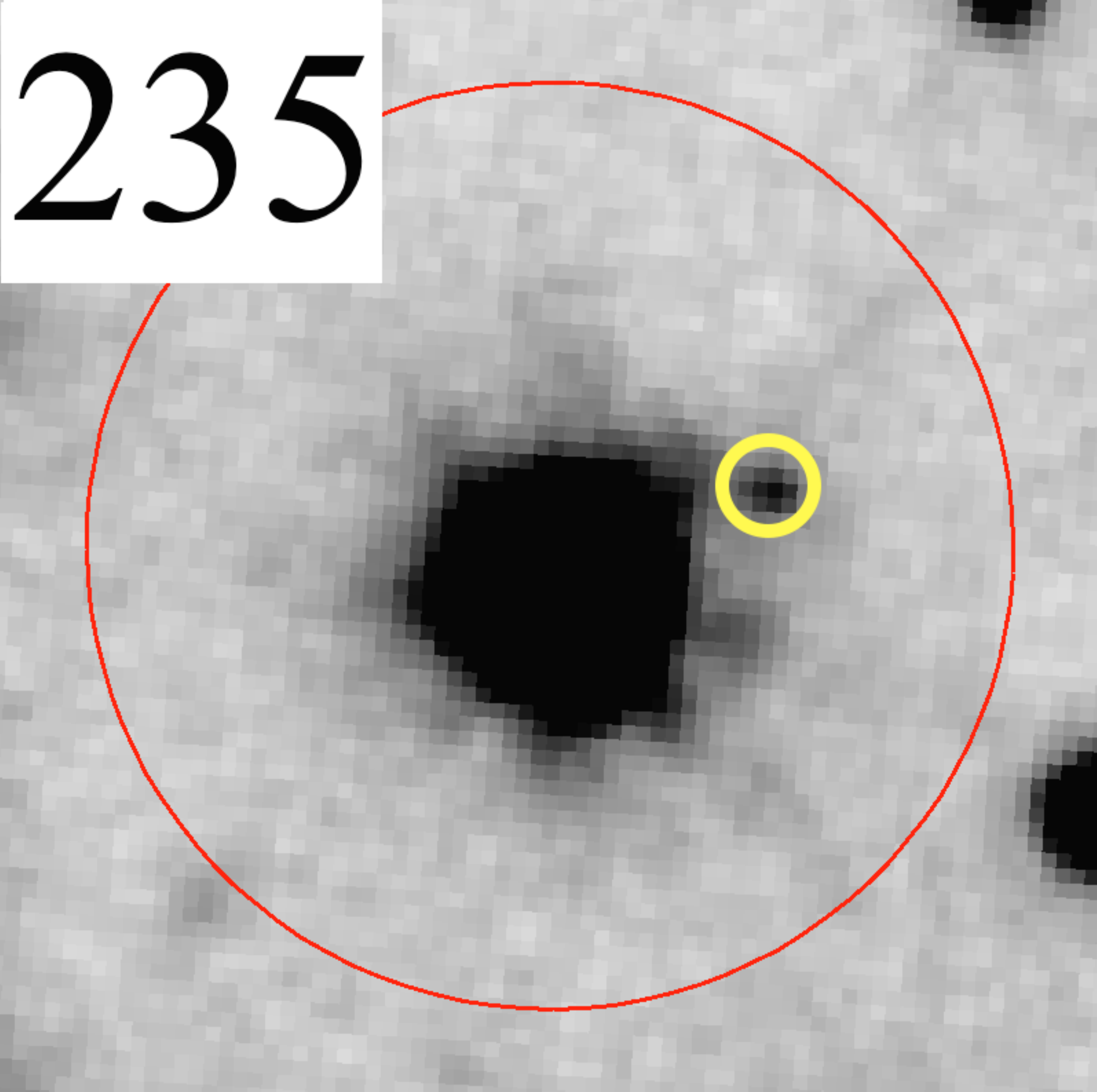} &
    \includegraphics[width=.23\textwidth]{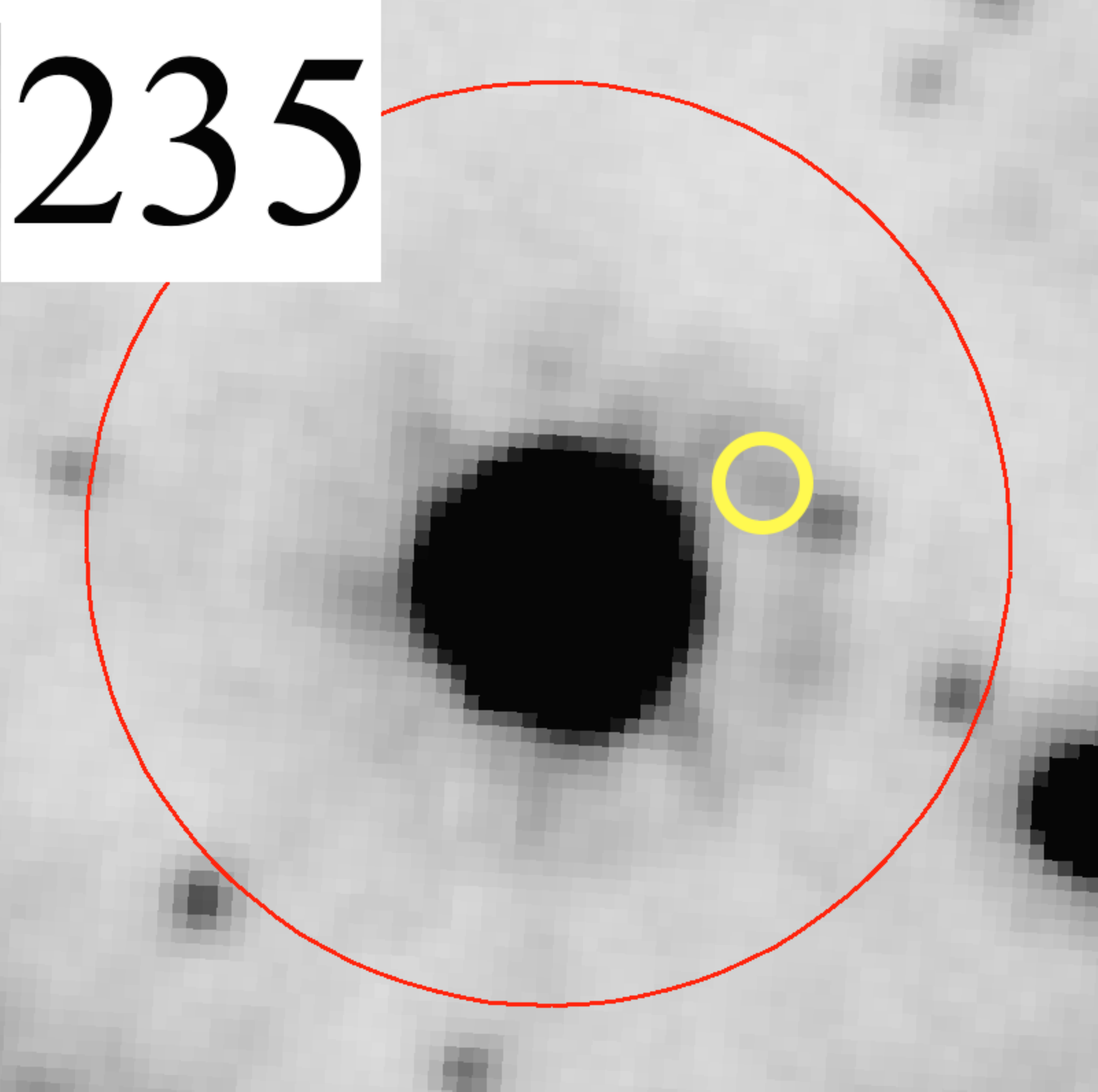} &
    \includegraphics[width=.23\textwidth]{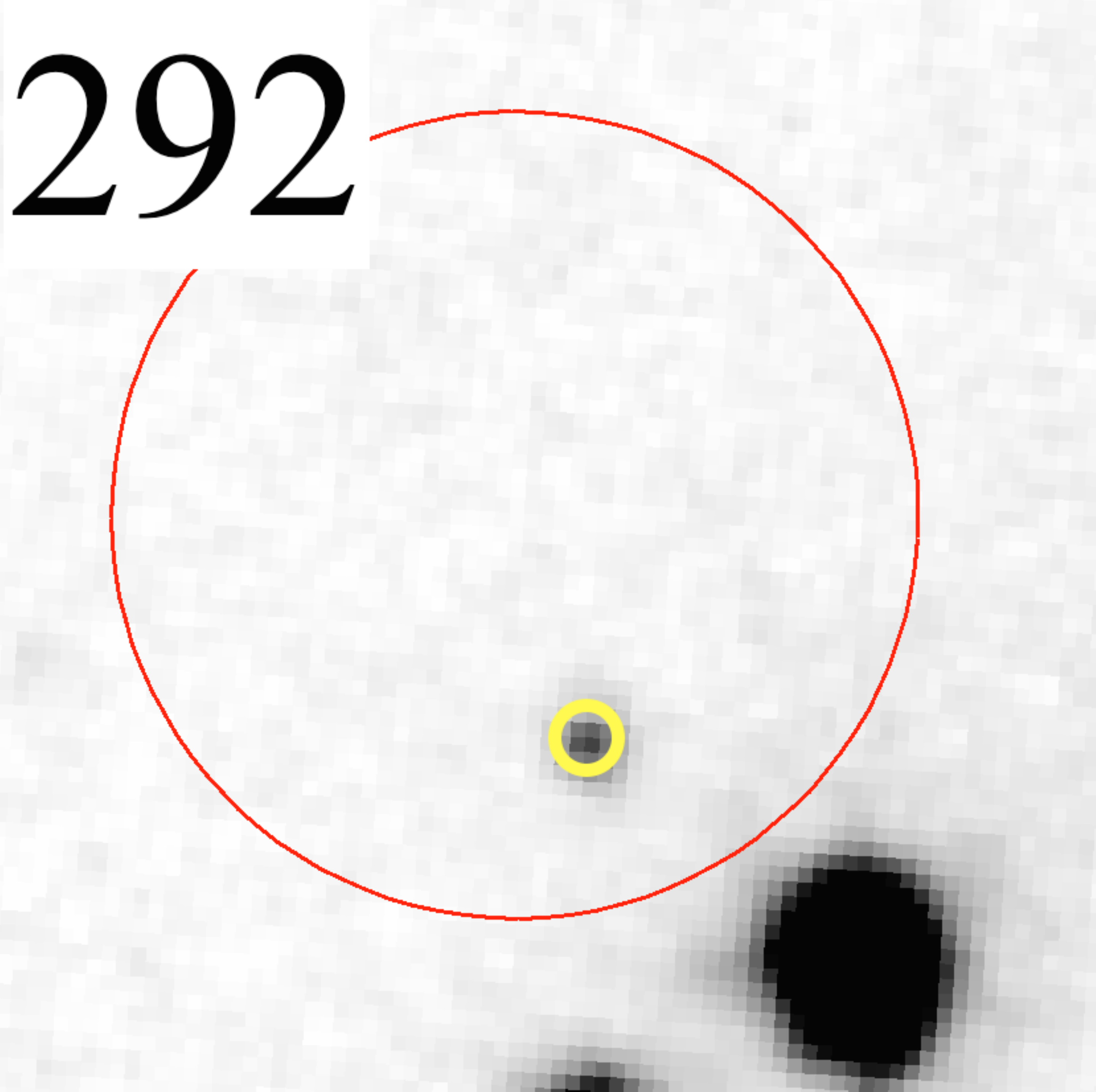} &
    \includegraphics[width=.23\textwidth]{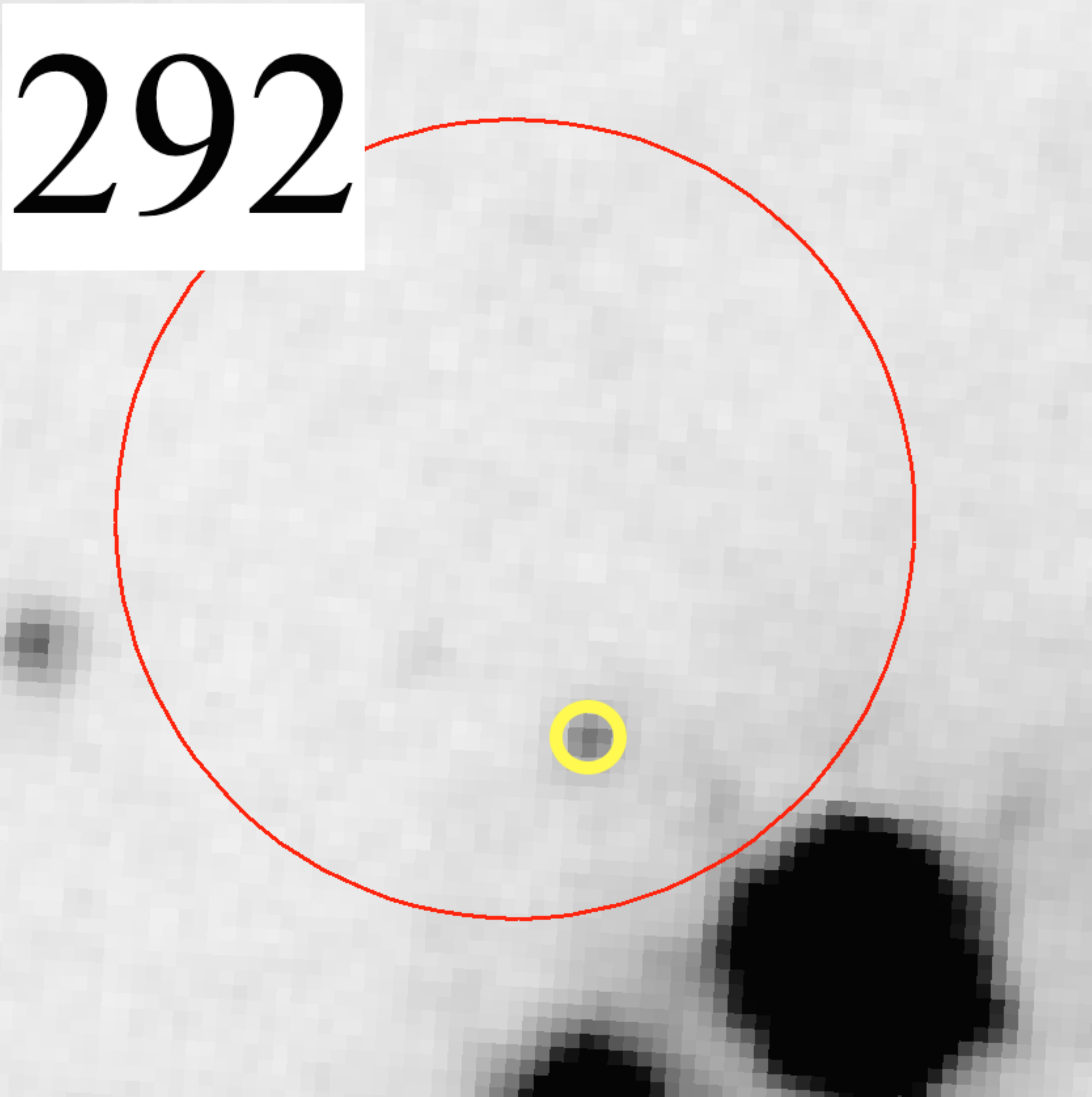}  \\
    \includegraphics[width=.23\textwidth]{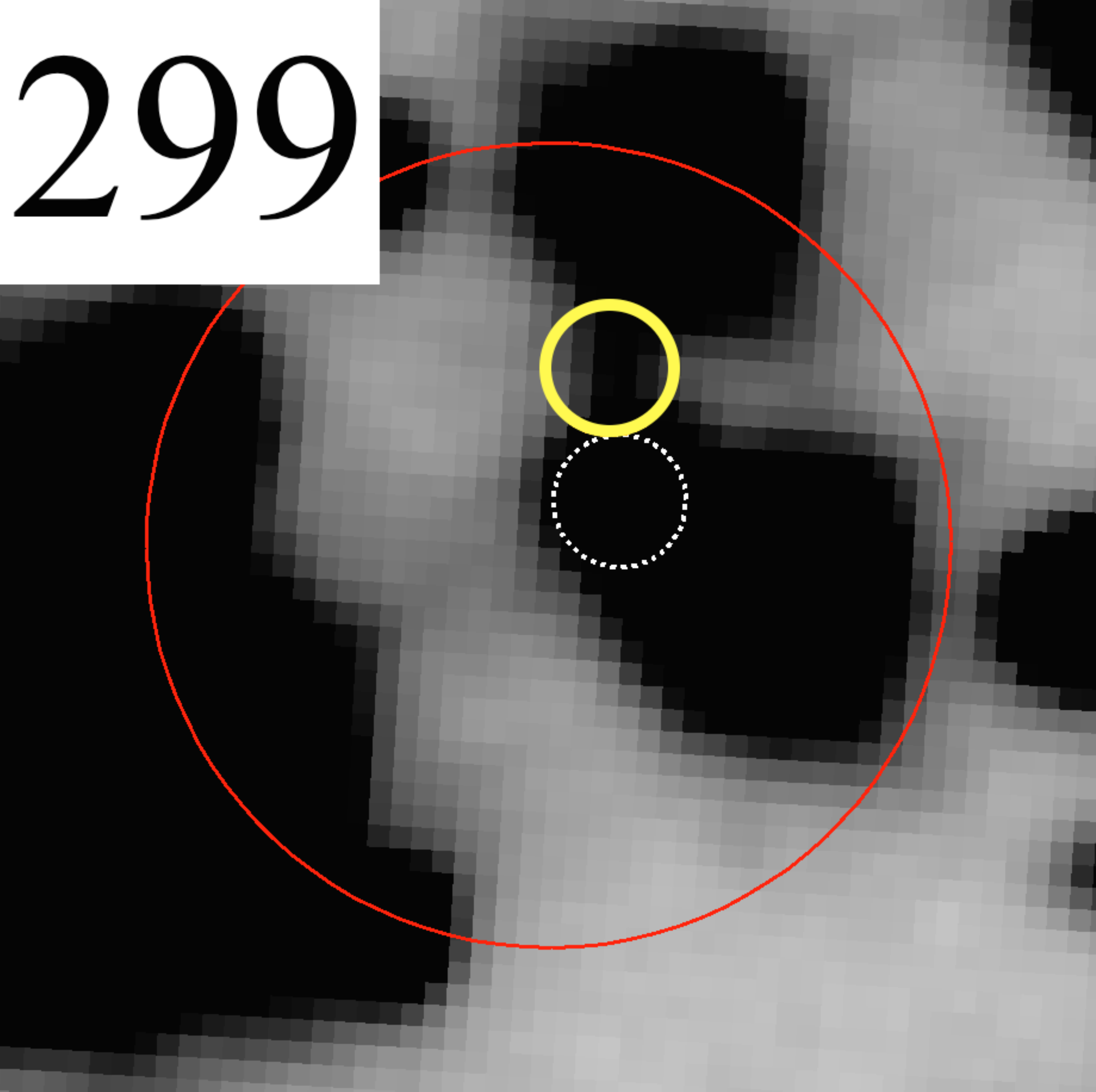} &
    \includegraphics[width=.23\textwidth]{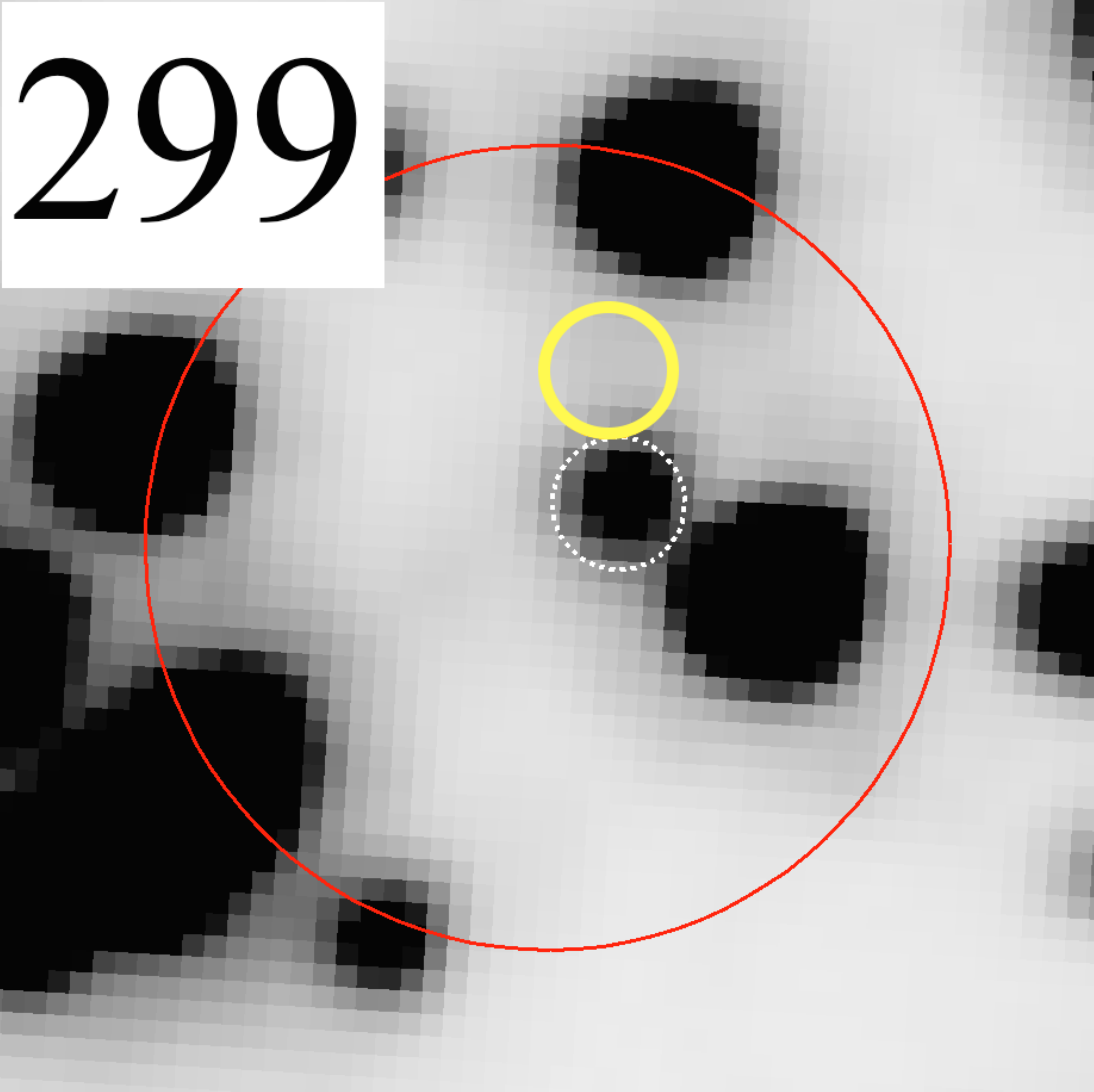} &
    \includegraphics[width=.23\textwidth]{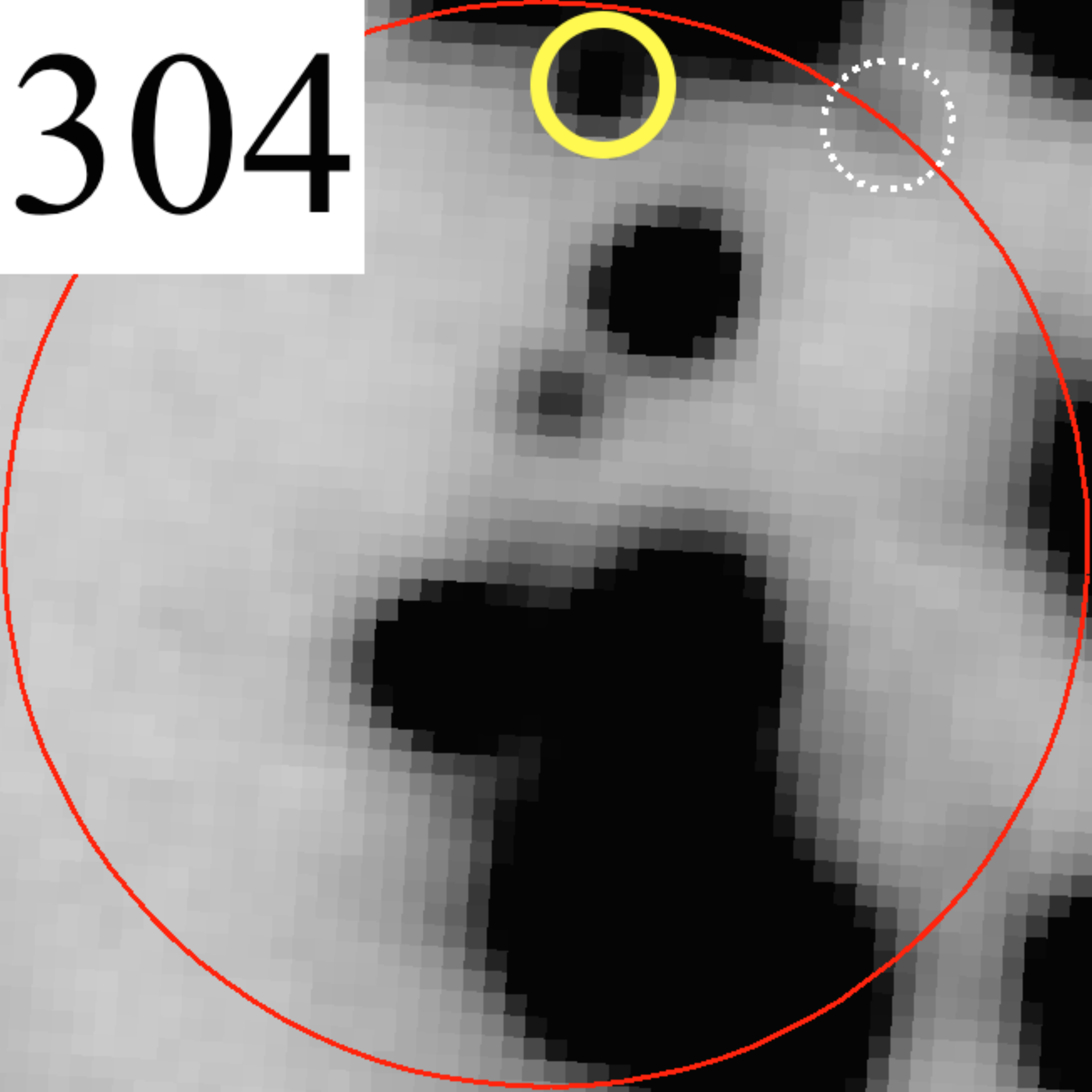} &
    \includegraphics[width=.23\textwidth]{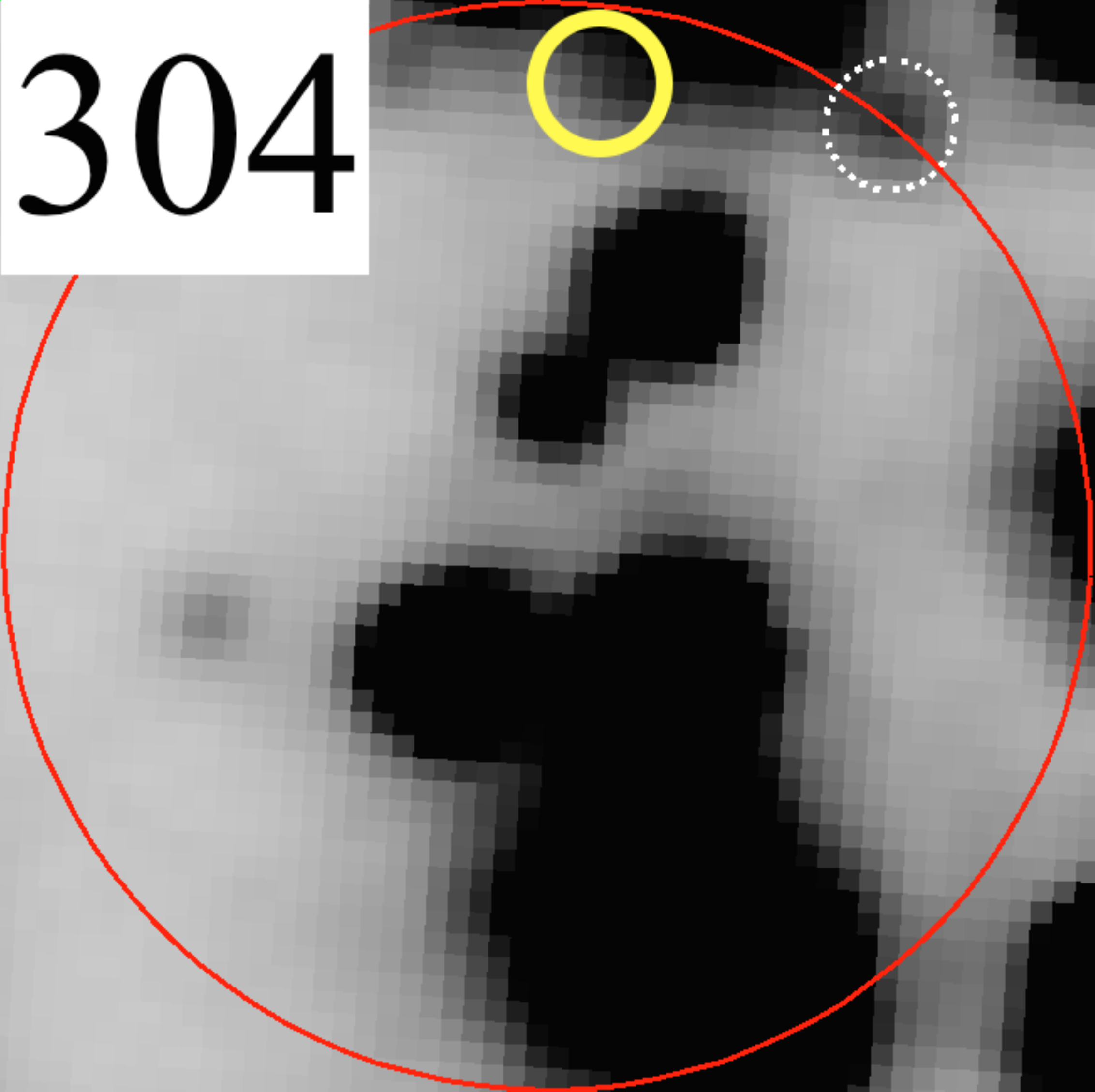}   \\
    \includegraphics[width=.23\textwidth]{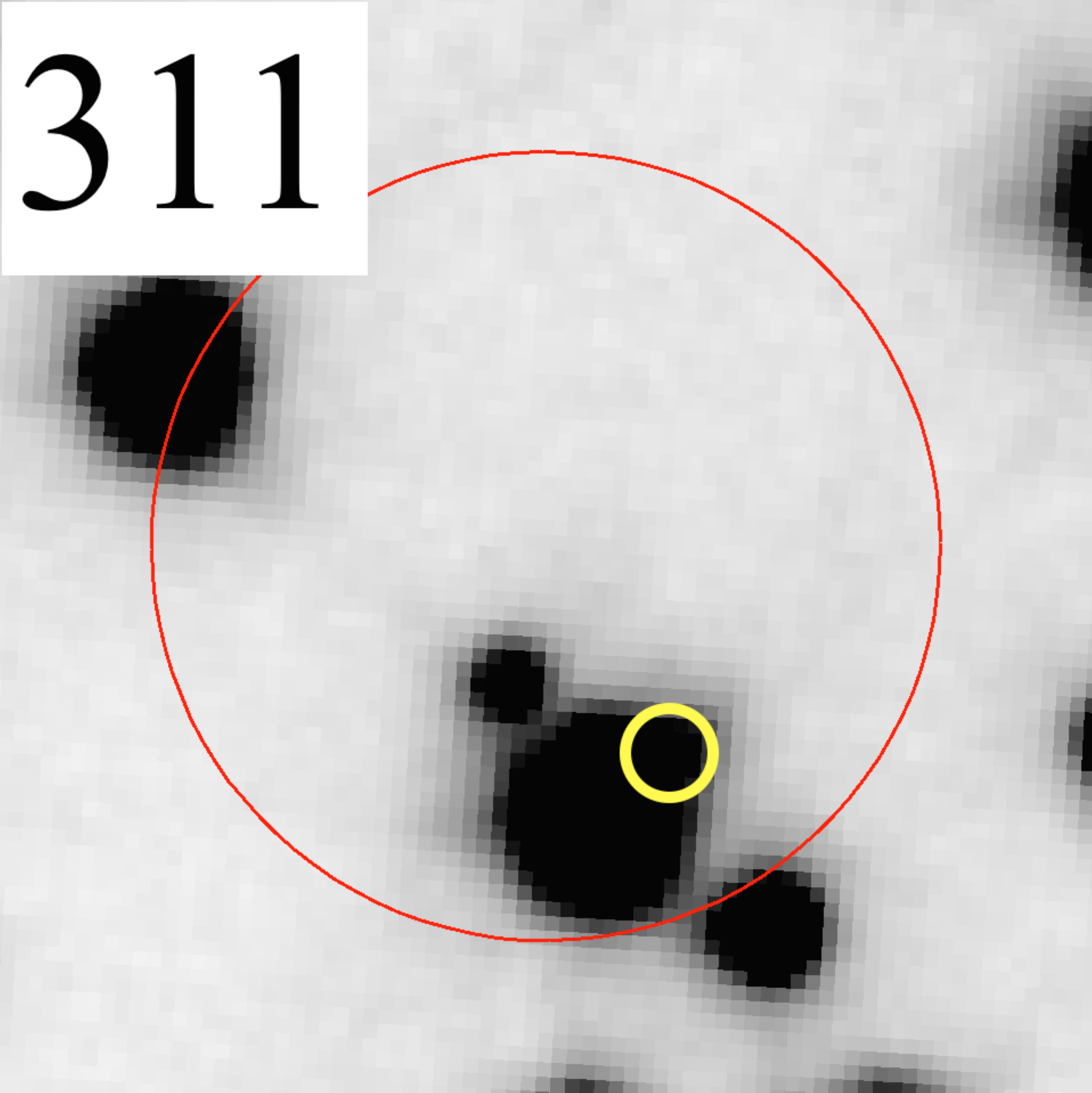} &
    \includegraphics[width=.23\textwidth]{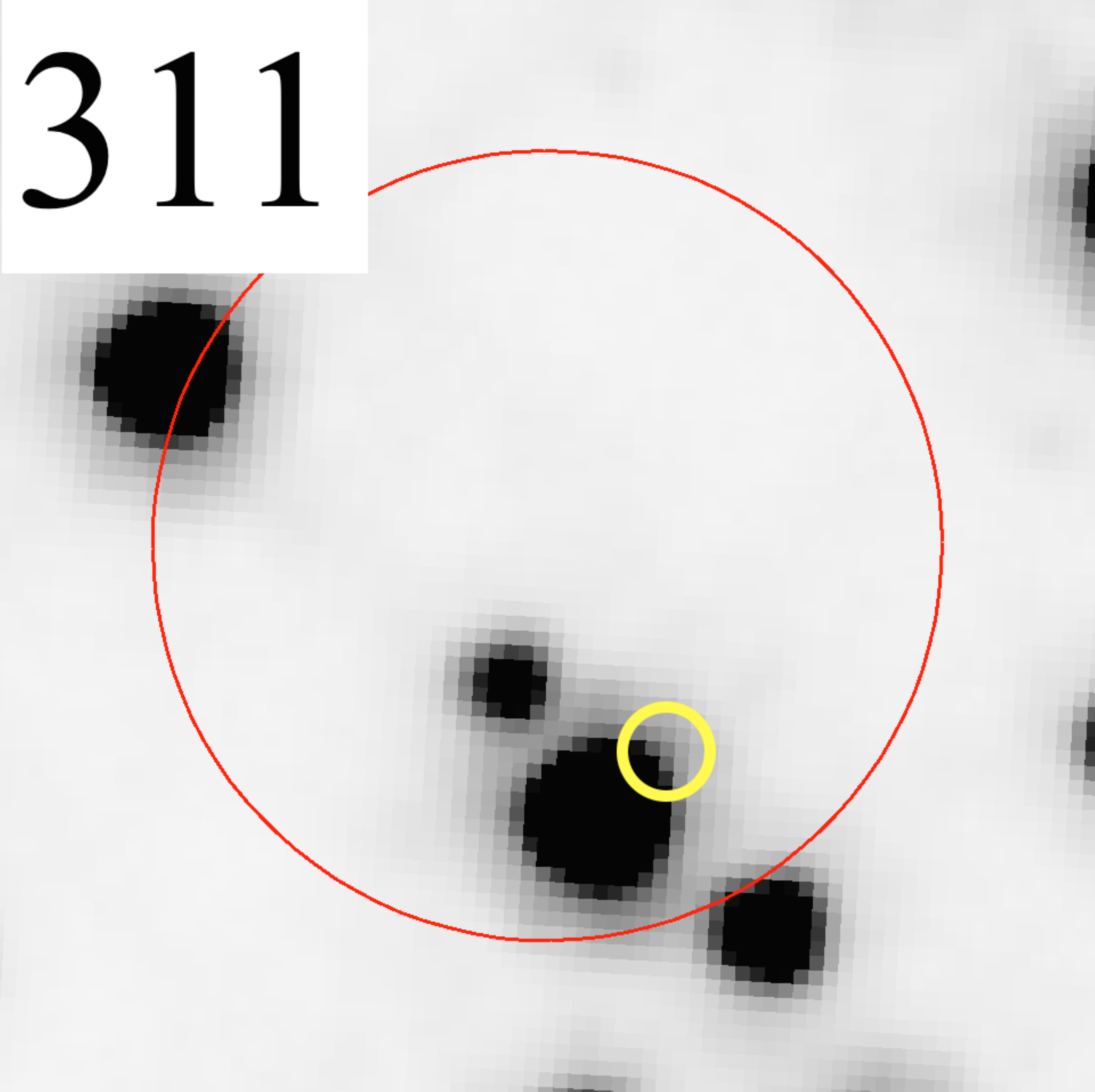} &
    \includegraphics[width=.23\textwidth]{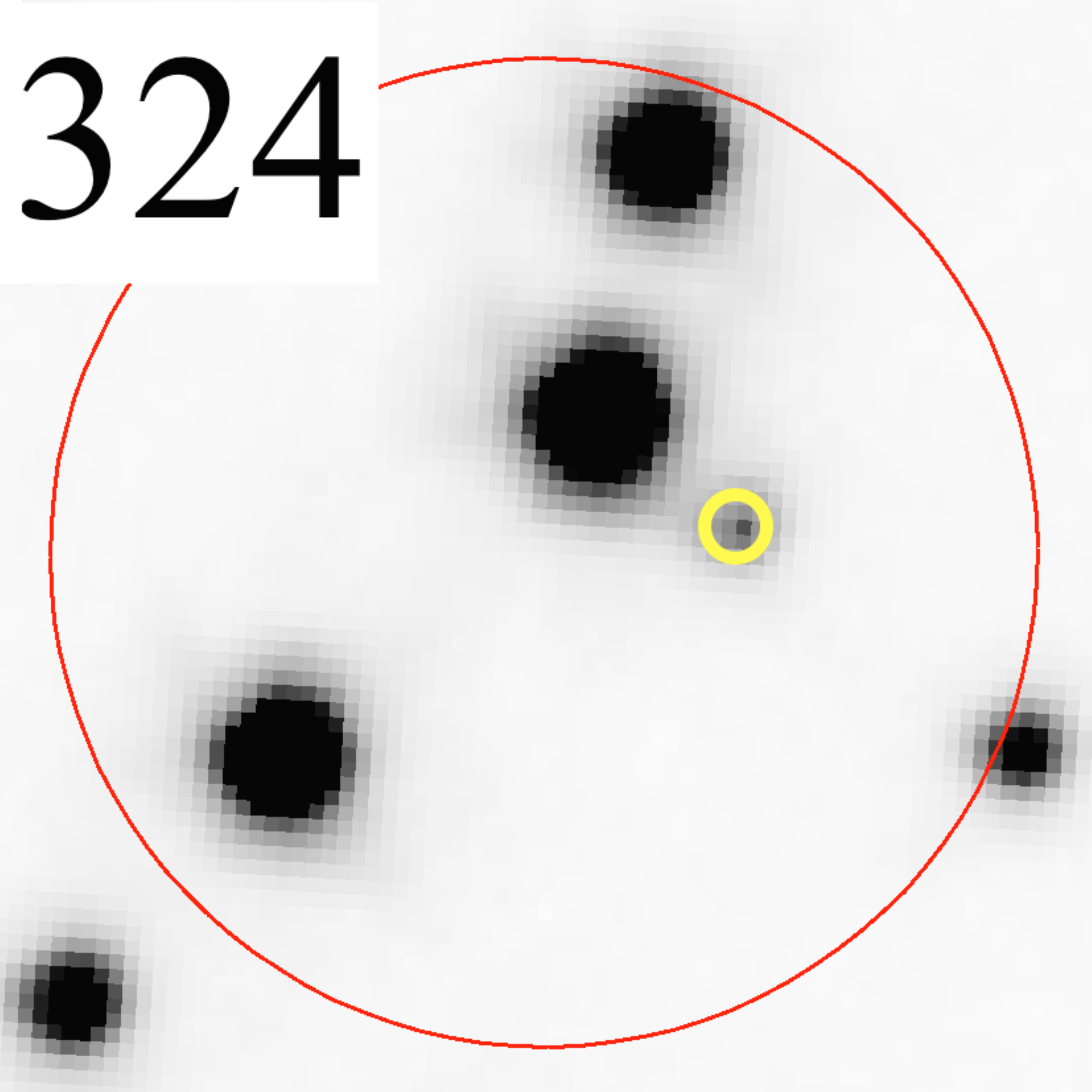} &
    \includegraphics[width=.23\textwidth]{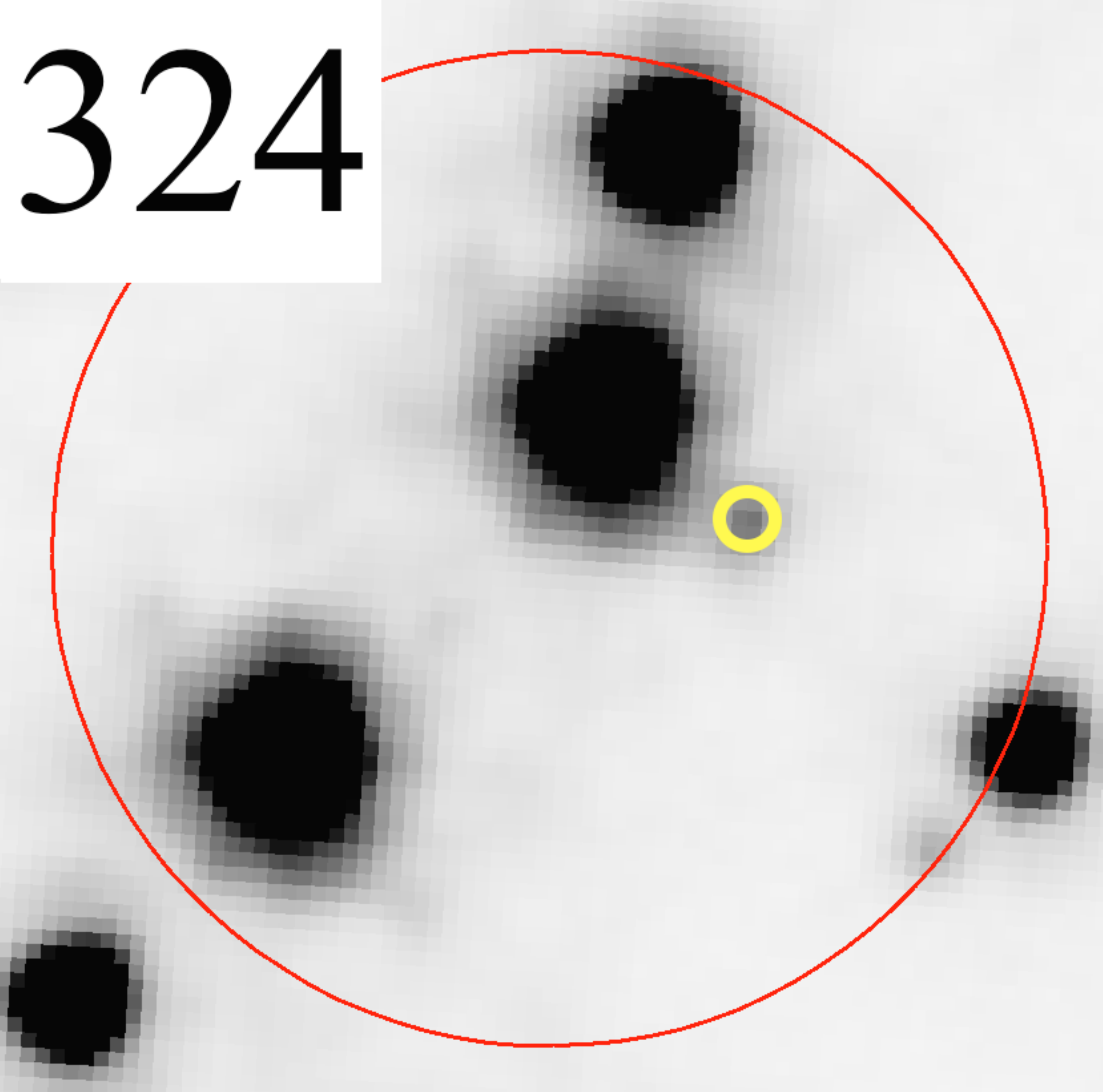}   \\
    \includegraphics[width=.23\textwidth]{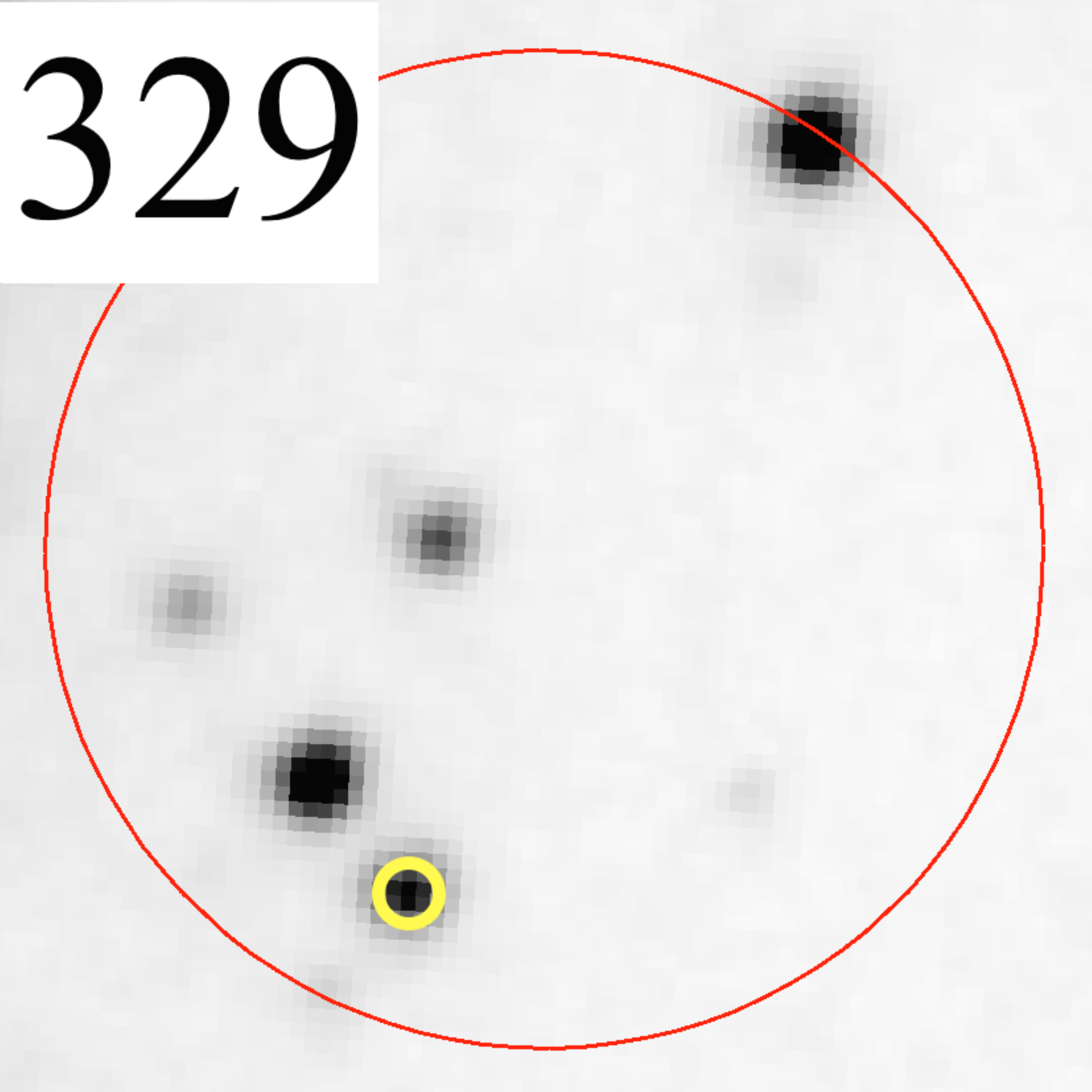} &
    \includegraphics[width=.23\textwidth]{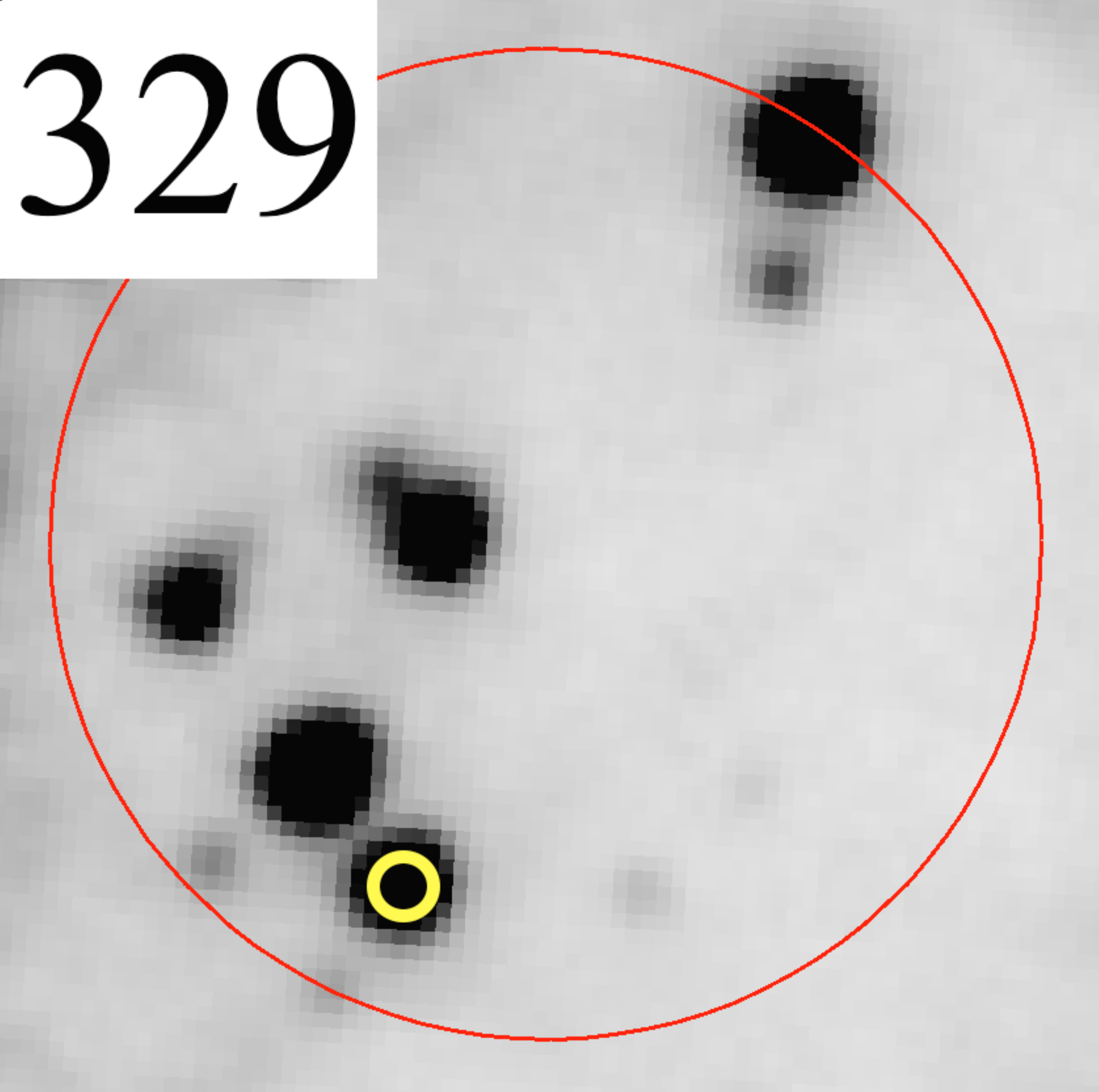} &
    \includegraphics[width=.23\textwidth]{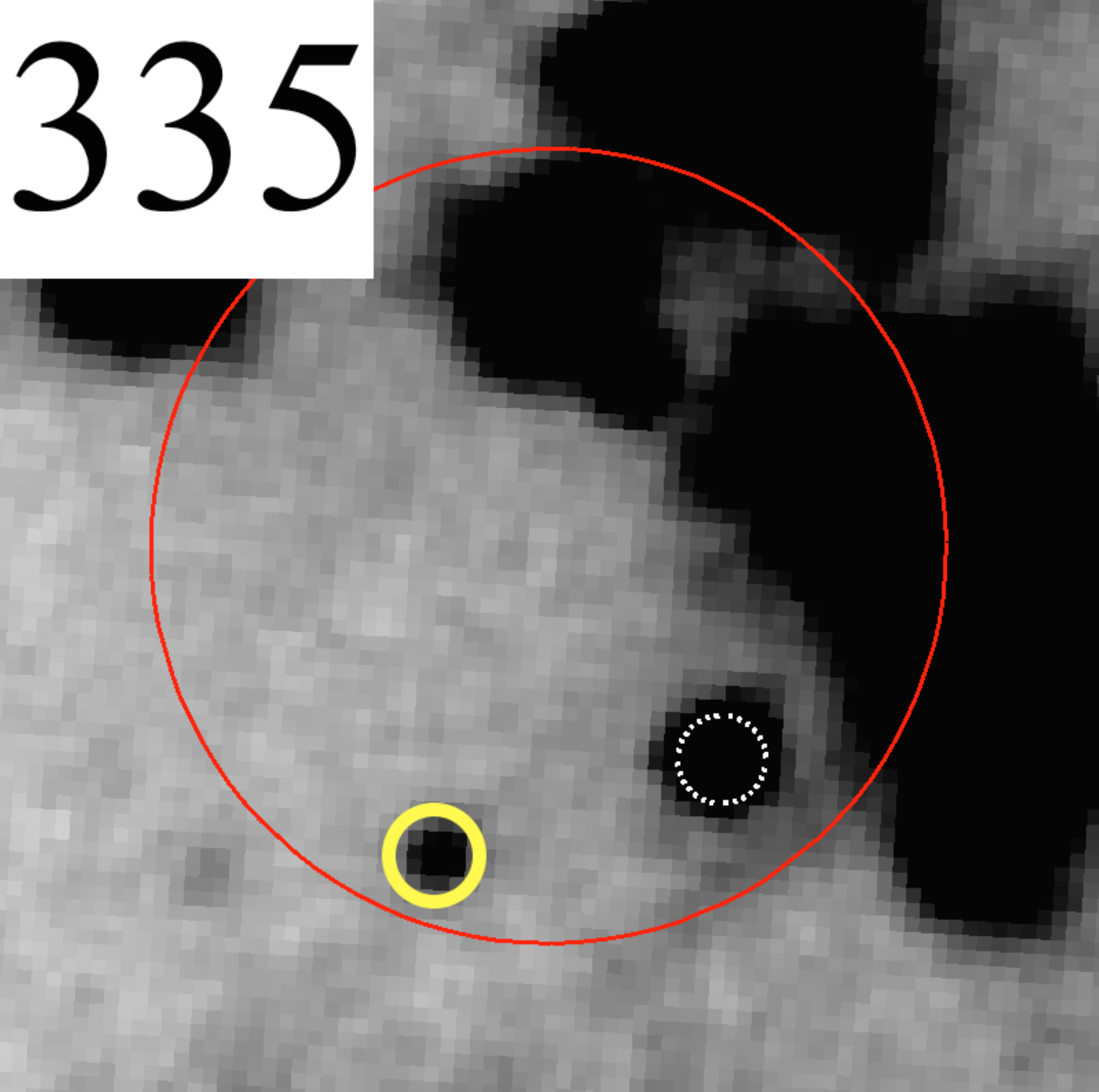} &
    \includegraphics[width=.23\textwidth]{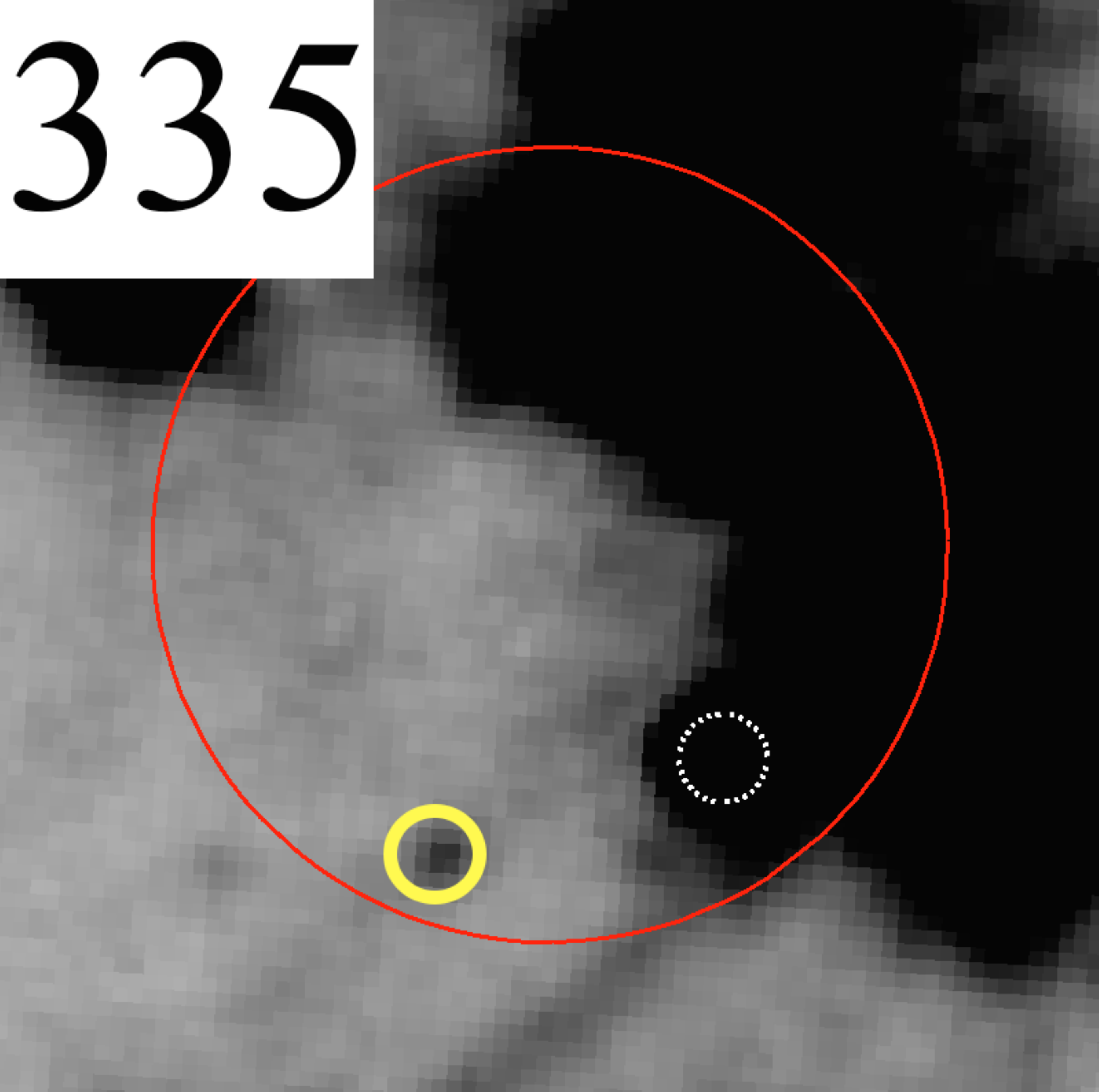}   \\
  \end{tabular}
  \caption{Finding charts in the filters U$_{300}$ and B$_{390}$ for some CVs and CV candidates in 47 Tuc. 
The solid red circles are the 3$\sigma$ match circles of \textit{Chandra} sources. The CVs and CV candidates are indicated with a solid yellow circle. 
The dashed white circles indicate possible additional X-ray counterparts (see section \ref{sec:multiple}). Each image is $1.5\arcsec \times 1.5\arcsec$,
except for W202, W299 and W304 which images are $1\arcsec \times 1\arcsec$.
In each column the images correspond to the same filter that is indicated by the top image of the column. The pixel scale is $0\farcs02$.}
\label{fig:A2}
\end{figure*}

\clearpage

\begin{table*}
\section{Cataclysmic variables in other globular clusters}
\label{fCVs_other}

In this appendix we show the CV and CV candidates in the core-collapsed clusters NGC 6397 and NGC 6752 
that were used in the comparison with the CV population in 47 Tuc (Section \ref{sec:comparison}). 
\end{table*}

\begin{table*}
 \centering
  \caption{CVs and CV candidates in NGC 6397 which have proper motions that are consistent with that of the cluster. 
The information was taken from \citet{cohn} and \citet{2010-bogda}.}
\label{table:otherclusterscvs_1}
  \begin{tabular}{c c c  c c c}
\hline
\hline
NGC 6397& & &  & & \\
\hline
Source & R.A. & Decl (J2000)& r (') & $L_{X (0.5-6 \ keV)}$* &  Type\\
            &          &                    &         & (10$^{30}$ ergs s$^{-1}$) &\\
\hline

U7 - CV10  & 17:40:52.832 & --53:41:21.77 &1.81 & 4.8 & CV\\ 
U10 - CV6  & 17:40:48.978 &--53:39:48.62 &1.21 & 22.5 & CV\\
U11 - CV7  & 17:40:45.781 & --53:40:41.52 &0.58 &1.5 & CV\\
U13 - CV8  &17:40:44.084 & --53:40:39.17 &0.33 & 2.9 & CV\\
U17 - CV3  & 17:40:42.651 & --53:40:19.30 & 0.17 & 141.8 & CV\\ 
U19 - CV2  &17:40:42.306 & --53:40:28.70 & 0.02 & 111.2 & CV\\
U21 - CV4  &17:40:41.830 & --53:40:21.37 & 0.13 & 52.0 & CV\\
U22 - CV5  & 17:40:41.701 & --53:40:29.00 & 0.07 & 2.9 & CV\\
U23 - CV1  & 17:40:41.597 & --53:40:19.30 &  0.18 & 92.8 & CV\\
U25 - CV13 & 17:40:41.237 & --53:40:25.79 &  1.0 & 1.3 & CV?\\  
U60 - CV9   & 17:40:47.807 & --53:41:28.40 & 1.30 & 0.25 & CV\\
U61 - CV12 & 17:40:45.223 & --53:40:28.60  &0.45 & 2.0 & CV?\\
U80 - CV14 & 17:40:46.455 &  --53:41:56.50 & 1.60& 0.4 & CV?\\
U83 - CV15 & 17:40:49.615 & --53:40:43.02 & 1.13 & 0.1 & CV?\\
\hline
\end{tabular}
{\flushleft
\begin{flushleft}
* Luminosities in the 0.5-6 keV band were calculated by taking the fluxes from \citet{2010-bogda},
 and assuming a distance to NGC 6397 of 2.2 kpc. Errors are not provided by the authors.
\end{flushleft}
}
\end{table*}

\begin{table*}
 \centering
  \caption{CVs and CV candidates in NGC 6752 which have proper motions that are consistent with that of the cluster, 
the information was taken from Lugger et al. \citet{2017Lugger} and \citet{2014-forestell}. }
\label{table:otherclusterscvs_2}
  \begin{tabular}{c c c c c c }
\hline
\hline
NGC 6752 \\
\hline
Source & R.A. & Decl (J2000)& r (') &  $L_{X (0.5-6 \ keV)}$* & Type\\
            &          &                    &         & (10$^{30}$ ergs s$^{-1}$) &\\
\hline
CX1 & 19:10:51.138 & --59:59:11.92 & 0.17 & 259.8$^{+17.4}_{-17.4}$ &CV\\
CX2 & 19:10:56.005 & --59:59:37.33 & 0.73 & 126.4$^{+12.1}_{-11.9} $  &CV\\
CX4 & 19:10:51.586 & --59:59:01.73 & 0.08 & 88.6$^{+11.5}_{-11.2} $  & CV\\
CX5 & 19:10:51.414 & --59:59:05.18 & 0.09 & 101.6$^{+12.4}_{-12.1} $  & CV \\
CX6 & 19:10:51.505 & --59:59:27.10 & 0.38 & 42.3$^{+6.5}_{-6.2} $  & CV?\\
CX7 & 19:10:51.511 & --59:58:56.85 & 0.15 & 68.7$^{+9.4}_{-9.2} $  & CV\\
CX9 & 19:10:51.766 & --59:58:59.25 & 0.10 & 21.5$^{+6.2}_{-5.0} $  & CV?\\
CX13 & 19:10:40.610 & --60:00:05.91 & 1.76 & 5.3$^{+1.2}_{-1.0} $  & CV\\
CX21 & 19:10:49.516 & --59:59:43.16 & 0.72 & 2.9$^{+0.9}_{-0.8} $  & CV?\\
CX23 & 19:10:52.546 & --59:59:04.38 & 0.06 & 2.1$^{+0.8}_{-0.7} $  & CV?\\
CX24 & 19:10:52.670 & --59:59:03.21 & 0.07 & 2.1$^{+0.8}_{-0.7} $  & CV?\\
CX25 & 19:10:51.957 & --59:58:40.55 & 0.40 & 1.6$^{+0.8}_{-0.6} $  & CV?\\
CX32 & 19:10:54.137 & --59:59:11.04 & 0.28 & 1.05$^{+0.7}_{-0.5} $  &  CV?\\
CX35 & 19:10:52.165 & --59:59:16.73 & 0.20 & 0.8$^{+0.6}_{-0.4} $  & CV?\\
CX36 & 19:10:49.585 & --59:58:26.41 & 0.71 & 0.7$^{+0.6}_{-0.4} $  & CV\\
\hline
\end{tabular}
\flushleft
\begin{flushleft}
* Luminosities in the 0.5-6 keV band were calculated by multiplying the 0.5-8 keV band luminosities given in
\citet{2014-forestell} by the appropriate factor (0.95) assuming a 2 keV bremsstrahlung model.
\end{flushleft}
\end{table*}

\end{document}